\documentclass[a4paper,11pt]{article}
\usepackage[DIV12]{typearea}

\usepackage[T1]{fontenc}
\usepackage[utf8]{inputenc}
\usepackage{authblk}

\usepackage[font=small,labelfont=bf]{caption}
\usepackage{graphicx}
\usepackage{multirow}
\usepackage{amsmath,amssymb,amsfonts}
\usepackage{verbatim}
\usepackage{bm}
\usepackage{color}
\usepackage{ulem}
\usepackage{mathtools}
\usepackage{cite}
\usepackage{colortbl}
\definecolor{light-gray}{gray}{0.85}
\usepackage[pdftex,colorlinks,pdfpagelabels]{hyperref}
\hypersetup{
   bookmarksnumbered,
   citecolor={blue},
   linkcolor={blue},
   urlcolor={blue},
   filecolor={blue}
} 
\newcommand{\eVdist}{\kern-0.06em}

\newcommand{\gev}{\:\text{Ge\eVdist V}}
\newcommand{\gv}{\:\text{G\eVdist V}}
\newcommand{\kpc}{\:\text{kpc}}
\newcommand{\mb}{\:\text{mb}}
\newcommand{\cm}{\:\text{cm}}
\newcommand{\pbar}{{$\bar{\text{p}}$}}
\newcommand{\tev}{\:\text{Te\eVdist V}}
\newcommand{\s}{\:\text{s}}
\newcommand{\pl}{p_{\text{\tiny{L}}}}
\newcommand{\pt}{p_{\text{\tiny{T}}}}
\newcommand{\mt}{m_{\text{\tiny{T}}}}
\newcommand{\iso}{\ensuremath{\Delta_{\text{IS}}}}
\newcommand{\hyp}{\ensuremath{\Delta_{\Lambda}}}
\newcommand{\R}{\mathcal{R}}

\newcommand{\AddrBonn}{%
\textit{Bethe Center for Theoretical Physics and Physikalisches Institut der 
Universit\"at Bonn, Nu{\ss}allee 12, 
 53115 Bonn, Germany}
}
\newcommand{\AddrStockholm}{
\textit{Nordita, KTH Royal Institute of Technology and Stockholm University, Roslagstullsbacken 23, 10 691 Stockholm, Sweden}
}

\usepackage{fancyhdr}
 
\fancypagestyle{plain}{%
    \fancyhead[R]{NORDITA-2017-127\\BONN-TH-2017-12}
    
}
\date{}
\title{\Large\bf A Precision Search for WIMPs with Charged Cosmic Rays}
\author[1]{Annika Reinert\thanks{areinert@th.physik.uni-bonn.de}}
\author[2]{Martin Wolfgang Winkler\thanks{martin.winkler@su.se}}
\affil[1]{\AddrBonn}
\affil[2]{\AddrStockholm }

\begin{document}
\maketitle
\vspace*{0mm}
\begin{abstract}
AMS-02 has reached the sensitivity to probe canonical thermal WIMPs by their annihilation into antiprotons. Due to the high precision of the data, uncertainties in the astrophysical background have become the most limiting factor for indirect dark matter detection. In this work we systematically quantify and -- where possible -- reduce uncertainties in the antiproton background. We constrain the propagation of charged cosmic rays through the combination of antiproton, B/C and positron data. Cross section uncertainties are determined from a wide collection of accelerator data and are -- for the first time ever -- fully taken into account. This allows us to robustly constrain even subdominant dark matter signals through their spectral properties. For a standard NFW dark matter profile we are able to exclude thermal WIMPs with masses up to 570~GeV which annihilate into bottom quarks. While we confirm a reported excess compatible with dark matter of mass around 80 GeV, its local (global) significance only reaches 2.2$\,\sigma$ (1.1$\,\sigma$) in our analysis.
\end{abstract}
\clearpage

\section{Introduction}

The last decade has seen dramatic progress in the measurement of charged cosmic ray fluxes. As the experiments entered new territory in energy and precision, a number of surprises came along. Most strikingly, the positron flux failed to show the strong decrease with energy which would have established it as a secondary, i.e. one that is produced by scattering of protons or nuclear cosmic rays on the interstellar matter. When the rise of the positron fraction was unambiguously proven by PAMELA~\cite{Adriani:2008zr}, it seemed that dark matter discovery was within reach.
The wave of excitement prevailed until gamma ray~\cite{Abdo:2010dk,Ackermann:2011wa} and CMB data~\cite{Galli:2009zc,Slatyer:2009yq,Huetsi:2009ex} put increasingly strong pressure on this interpretation which required annihilation rates far beyond those of Weakly Interacting Massive Particles (WIMPs). Skepticism was later also raised on whether dark matter annihilations can account for the positron spectrum at the precision level (see e.g.~\cite{Boudaud:2016jvj}). 
In this light, an astrophysical origin of the positron excess appears to be preferred.

The story of cosmic ray antiprotons shares a parallel with positrons: at the first data release by AMS-02, the antiproton spectrum came out significantly harder than expected from secondary production~\cite{tingtalk:2015}. But since then, the secondary background passed through a major revision. Cosmic ray propagation has been recalibrated to the AMS-02 boron to carbon (B/C) data~\cite{Kappl:2015bqa} (see also~\cite{Giesen:2015ufa,Evoli:2015vaa}). In addition, the experimentally established increase of the antiproton production cross section with energy has been incorporated~\cite{Kachelriess:2015wpa,Winkler:2017xor}. The updated background features significantly more high energy antiprotons. It is consistent with the hard high energy spectrum observed by AMS-02~\cite{Winkler:2017xor,Boschini:2017fxq}.

Supposing that dark matter signals dominate neither of the two antimatter fluxes at any energy
it may seem that this is the time for despair. But quite the contrary: AMS-02 has reduced experimental errors in the fluxes to the few percent level over a wide energy range~\cite{Aguilar:2014mma,Aguilar:2016vqr,Aguilar:2016kjl}. By this, it has gained sensitivity to subdominant signals which can be identified in a spectral analysis. Even if the high energy positron spectrum is dominated by an astrophysical source, the low energy part is still very useful in constraining dark matter models as well as cosmic ray propagation. In the antiproton channel AMS-02 can even realistically probe canonical thermal WIMPs -- the target of indirect dark matter searches for decades. Indeed, an antiproton excess consistent with a thermal WIMP of mass $m_{\text{DM}}\sim 80\gev$ has already been reported in the AMS-02 data~\cite{Cuoco:2016eej,Cui:2016ppb}. But the robustness of this signal needs to be investigated further.

In this work, we will attempt to systematically quantify and incorporate the dominant uncertainties in the antiproton flux. These are related to hadronic production cross sections as well as to the propagation of charged cosmic rays through the galaxy and the heliosphere. Our approach employs the combination of antiproton, positron and B/C data of AMS-02. Cosmic rays are propagated within the two-zone diffusion model~\cite{Maurin:2001sj,Donato:2001ms,Maurin:2002ua}. Positrons are consistently treated within the same framework through the pinching method~\cite{Boudaud:2016jvj}. The propagation parameters which control diffusion, convection and reacceleration are obtained by simultaneously fitting the B/C ratio~\cite{Aguilar:2016vqr} and the antiproton spectrum~\cite{Aguilar:2016kjl} of AMS-02. The size of the diffusion halo, posing a notorious difficulty for indirect dark matter detection, is efficiently constrained from the low-energy positron spectrum. Our treatment of solar modulation includes charge-sign dependent effects which are determined from the time-dependence of the antiproton flux as extracted from the PAMELA~\cite{Adriani:2012paa} and AMS-02~\cite{Aguilar:2016kjl} data. The antiproton cross sections relevant for secondary production in cosmic ray scattering are taken from the recent comprehensive analysis~\cite{Winkler:2017xor}. Nuclear fragmentation cross sections which enter the boron source term are modeled from a wide collection of accelerator data. For both, antiprotons and boron, we map the cross section uncertainties into the predicted fluxes and include them into the fit in the form of covariance matrices.

We observe an overall good agreement of the antiproton and B/C data with the background prediction. By performing a spectral analysis we derive strong constraints on hadronic dark matter annihilation. While we confirm the slight excess~\cite{Cuoco:2016eej,Cui:2016ppb} at $m_{\text{DM}} \sim 80\gev$ in the $b\bar{b}$-channel, we show that it becomes insignificant once all relevant uncertainties and the look-elsewhere effect are considered. We finally comment on possibilities to reduce uncertainties in the cosmic ray fluxes in order to further increase the sensitivity of AMS-02 to dark matter annihilations.

\section{Cosmic Ray Propagation}
Protons and most of the nuclei in cosmic rays are referred to as primaries. They correspond to galactic matter which has been energized by supernova shock acceleration also known as (first order) Fermi acceleration. When a primary species propagates through the galaxy it can scatter on the interstellar gas and create a so-called secondary cosmic ray. This production mode is very important for certain nuclei like lithium, beryllium and boron. In addition, a large fraction of the antimatter in cosmic rays is believed to be of secondary origin. Independent of their production, cosmic rays follow complicated trajectories which are controlled by the magnetic fields in the galactic halo. We shall now briefly summarize our conventions for cosmic ray propagation before turning to the solar modulation of charged particles in the heliosphere. 

\subsection{Diffusion Model}\label{sec:diffusionmodel}
On their passage through the galaxy cosmic rays scatter on magnetic field inhomogeneities. This induces a random walk which is equivalently described as spatial diffusion. Convective winds, if they exist, blow charged particles away from the galactic disc. In addition, interaction of cosmic rays with matter, light and magnetic fields leads to energy losses and annihilation, while magnetic shock waves may induce reacceleration. All relevant processes are encoded in the diffusion equation. While cosmic ray propagation codes like GALPROP~\cite{Moskalenko:1997gh,Strong:1998pw,Strong:2001gh} and DRAGON~\cite{Evoli:2008dv,Evoli:2017vim} aim at a fully numerical solution, the spatial part of the diffusion equation can also be solved analytically under slightly simplifying assumptions. In the two-zone diffusion model~\cite{Maurin:2001sj,Donato:2001ms,Maurin:2002ua}, which we employ here, diffusion is taken to occur homogeneously and isotropically in a cylinder of radius $R$ and half-height $L$ around the galactic disc. The disc of thickness $2h=0.2\kpc$ is taken to contain a constant number density of hydrogen and helium ($n_{\text{H}}=0.9\cm^{-3}$, $n_{\text{He}}=0.1\cm^{-3}$). Assuming steady state, the space-energy density $N_i$ of a stable species $i$ is related to its differential production rate (source term) $q_i$ as
\begin{align}\label{eq:diffeq}
&  -K \Delta N_i + \text{sgn$(z)$} V_c \,\partial_z N_i + \partial_E (b_{\text{halo}} N_i)  + 2 h \delta(z) \big[\partial_E (b_{\text{disc}}  N_i -K_{EE} \:\partial_E N_{i})+ \Gamma_\text{ann}\,N_{i}\big] \nonumber\\
& =2h\delta(z)q_i^{\text{disc}} + q_i^{\text{halo}}\,,
\end{align}
where $E$ denotes the total energy of $i$. The extension of the galactic disc in axial ($z$-) direction has been neglected and processes confined to the disc were multiplied by $2 h \delta(z)$ in order to keep proper normalization. We have split the source term into a disc component $q_i^{\text{disc}}$ and a halo component $q_i^{\text{halo}}$. The first term on the left-hand side accounts for spatial diffusion. Magnetohydrodynamics considerations suggest~\cite{Ptuskin:1997aa}
\begin{equation}\label{eq:standarddiffusion}
 K = K_0 \,\beta \,\left(\frac{\mathcal{R}}{\text{GV}}\right)^\delta\,,
\end{equation}
where $K_0$ is a normalization constant, $\delta$ the power law index, $\beta$ and $\mathcal{R}$ the velocity and rigidity of the cosmic ray particle. We will later also consider a modification of the diffusion term which is motivated from observed primary cosmic ray spectra (see section~\ref{sec:diffusion_break}). 
The second term in~\eqref{eq:diffeq} accounts for convection. The convective wind velocity $V_c$ controls its strength. Reacceleration by magnetic shock waves is modeled as a diffusion in momentum space which is encoded in the term~\cite{Maurin:2002ua}
\begin{align}
K_{EE}=\frac{4}{3}\frac{V_a^2}{K}\frac{p^2}{\delta(4-\delta)(4-\delta^2)}\,.
\end{align}
The Alfv\`en speed $V_a$ occurs quadratically as reacceleration corresponds to second order Fermi acceleration.\footnote{For practical purposes, the height of the reacceleration zone is taken to coincide with the disc height $h$. A difference between the two can be absorbed into a redefinition of $V_a$ which is anyway a free parameter~\cite{Maurin:2001sj}.} The term
\begin{equation}
 b_{\text{disc}}= b_{\text{coul}} + b_{\text{ion}} + b_{\text{brems}} + b_{\text{adiab}}  + b_{\text{reac}}
\end{equation}
includes energy losses in the galactic disc by Coulomb interactions, ionization and brems-strahlung, adiabatic energy losses caused by the flip of the convective wind vector at the disc as well as energy gains by reacceleration drift. We take $b_{\text{coul}}$, $b_{\text{ion}}$, $b_{\text{brems}}$ from~\cite{Strong:1998pw} and $b_{\text{adiab}}$, $b_{\text{reac}}$ from~\cite{Maurin:2002ua}. Leptonic cosmic rays in addition lose energy in the halo due to inverse Compton scattering and synchrotron emission,
\begin{equation}\label{eq:halolosses}
 b_{\text{halo}}= b_{\text{ic}} + b_{\text{synch}}= -(E^2/\tau_E)\gev^{-2}\,.
\end{equation}
As we shall only consider low energy lepton fluxes in this work, we can employ the Thomson limit and use a constant $\tau_E=10^{16}\s$ (see e.g.~\cite{Delahaye:2008ua}).
Annihilation in the galactic disc on the other hand is mainly relevant for hadronic cosmic rays. The annihilation cross sections for nuclei are taken from~\cite{Tripathi:1999nw,Tripathi:1999} and for antiprotons from~\cite{Protheroe:1981gj,Tan:1983de}.\footnote{The antiproton annihilation cross section was interpolated between the two parameterizations as in~\cite{Kappl:2011jw}.} In the case of antiprotons we also have to consider inelastic (non-annihilating) scattering with the interstellar matter. This effect is taken into account through a tertiary source term as described in~\cite{Donato:2001ms}.

We now turn to the solution of the diffusion equation for secondary cosmic rays. We will approximate secondary source terms $q^\text{sec}_{i}$ as spatially constant in the galactic disc. This amounts to assuming radially constant fluxes of the primary cosmic ray progenitors as well as constant density of the interstellar medium. While this situation is not expected to hold in reality, deviations can usually be absorbed into the propagation parameters. Local secondary fluxes are expected to be nearly unaffected as long as all species are treated within one universal framework. We will, furthermore, work in the limit $R\rightarrow \infty$ which yields identical results to a radially finite diffusion halo as long as $L\ll R$ (see e.g.~\cite{Putze:2010zn}).\footnote{Even if $L\ll R$ was not fulfilled, the difference in predicted secondary fluxes can again be compensated by a change of propagation parameters.} For secondary nuclei including antiprotons, energy losses in the halo are negligible and the spatial part of the diffusion equation can be solved analytically. The space energy density at $z=0$ is determined by the differential equation in energy \cite{Donato:2001ms,Maurin:2001sj}
\begin{equation}\label{eq:energyeq}
\left(2 h \Gamma_\text{ann} + V_c + V_c \coth{\left[ \frac{V_c L}{2 K}\right] }\right)N_i
 +  2 h \partial_E (b_\text{disc} \,N_i -K_{EE} \:\partial_E N_i ) =  2 h (q^\text{sec}_{i}+q^\text{ter}_{i})\;,
\end{equation}
where the tertiary source term $q^\text{ter}_{i}$ is only relevant for antiprotons. This equation has to be solved numerically. The interstellar flux is related to the space-energy density via $\Phi_i^\text{IS}(E)=\beta N_i /(4\pi)$. It is more common to specify fluxes in terms of the rigidity $\Phi_i^\text{IS}(\R)= \Phi_i^\text{IS}(E) dE/d\R$ or the kinetic energy per nucleon $T$.

In the case of positrons, the solution requires an additional step as energy losses occur in the disc and in the halo. We follow the ``pinching method''~\cite{Boudaud:2016jvj} and first solve the diffusion equation for secondary positrons in the high energy limit, where we include only diffusion and halo energy losses. At $z=0$ this leads to the integral equation $N_{e^+}^{\text{HE}}=\int\limits_{E}^{\infty}dE'\,q_{e^+}^\text{sec}(E')\,\eta(\lambda_D)/|b_{\text{halo}}(E)|$, where $\eta$ in terms of the diffusion length $\lambda_D$ can be taken from~\cite{Delahaye:2008ua}.
In the next step we want to include convection, reacceleration and energy losses in the disc. In order to follow the procedure for hadrons, the halo energy losses must be ``pinched'' into the disc. We substitute $b_{\text{halo}}\rightarrow 2 h \delta(z) \,b_{\text{pinched}}$ and solve the high-energy diffusion equation once again. Requiring that the solution remains unchanged by this replacement, we can fix
\begin{equation}
 b_{\text{pinched}} = \frac{1}{N_{e^+}^{\text{HE}}}\int\limits_E^\infty dE_0 \left(\frac{K(E_0)\,N_{e^+}^{\text{HE}}(E_0) }{hL}-q_{e^+}^\text{sec}(E_0)\right)\,.
\end{equation}
with $N_{e^+}^{\text{HE}}$ from above. The term $b_{\text{pinched}}$ is the ``translation'' of the halo loss term $b_{\text{halo}}$ into a disc loss term. After replacing $b_{\text{halo}}$ with $2 h \delta(z)\,b_{\text{pinched}}$ in the full diffusion equation~\eqref{eq:diffeq}, the solution for positrons proceeds completely analogous as for antiprotons and boron.

If a primary antiproton (or positron) source term is induced by dark matter annihilation, it carries a spatial dependence which is determined by the dark matter profile (see section~\ref{sec:darkmatterpbar}). The solution of the diffusion equation then requires a Bessel expansion in the radial coordinate. The procedure has been described in detail in~\cite{Barrau:2001ev,Donato:2003xg} and shall not be repeated here.\footnote{There arises a small technical difficulty as the Bessel expansion does not converge if we set $R\rightarrow\infty$. However, we verified that primary fluxes rapidly converge if we increase $R$ beyond $L$. Therefore, the practical solution is to set $R$ to a large but finite value for which we chose $R=5\,L$.}

\subsection{Solar Modulation}\label{sec:solarmodulation}

The solar system is surrounded by the heliosphere, a region of space which is permeated by the solar wind. The latter shapes the solar magnetic field whose main component is a dipole which inverts its polarity every 11 years. The two magnetic domains with inward and outward pointing field are separated by the heliospheric current sheet. Due to the solar wind outflow caused by the sun's rotation, the magnetic field lines are distended near the equator.  Since the sun's rotational axis is misaligned with the direction of the dipole, the solar magnetic field gets twisted and a wavy pattern of the current sheet emerges.

On their passage through the heliosphere, cosmic rays are affected by the solar magnetic field. The dominant effects are diffusion, drifts, convection, and adiabatic energy losses. In the widely used force field approximation \cite{Gleeson:1968zza}, solar modulation is described by a single parameter, the Fisk potential $\phi$ which only depends on time. Cosmic ray fluxes at the top of the earth's atmosphere (TOA) are related to interstellar fluxes as
\begin{equation}\label{eq:forcefield}
 \Phi^\text{TOA}(T) = \frac{2m T+T^2}{ 
2m \,(T+\frac{Z}{A}\phi)+(T+\frac{Z}{A}\phi)^2}\,\Phi^\text{IS}(T+\tfrac{Z}{A}\phi)\;. 
\end{equation}
This simple analytic solution is found if a constant radial solar wind and isotropic diffusion in the heliosphere are assumed. Unfortunately, the force field approximation cannot account for charge-sign dependent effects which have been established in cosmic ray spectra. It was argued in~\cite{Kota:1979,Jokipii:1981} that the dominant charge-breaking effect in solar modulation is connected to the heliospheric current sheet. During a negative polarity phase, negatively charged particles access the heliosphere on rather direct trajectories along the poles. Positively charged particles enter by inward drift along the current sheet. In particular, in the case of a very wavy current sheet, they spend significantly more time in the heliosphere and lose more energy. The solar magnetic field has last flipped its polarity from negative to positive between November 2012 and March 2014~\cite{Sun:2015}. After the flip, the situation reverses and negatively charged particles are more affected by solar modulation. This is evident e.g.\ in the time-dependent $e^+/e^-$ ratio measured by PAMELA between 2006 and 2015~\cite{Adriani:2016uhu}. The polarity flip is followed by a strong rise of $e^+/e^-$ at low energy.

In~\cite{Cholis:2015gna} a simple modification of the force field approximation was proposed where~\eqref{eq:forcefield} is still valid but particles are treated depending on their charge. The Fisk potential for positive ($+$) and negative ($-$) charges reads 
\begin{equation}\label{eq:forcefield2}
 \phi^{\pm}(t,\mathcal{R})= \phi_0(t) + \phi_1^{\pm}(t) \;\mathcal{F}\!\left(\tfrac{\mathcal{R}}{\mathcal{R}_0}\right)\,.
\end{equation}
The second term on the right-hand side incorporates the increased energy loss along the current sheet faced by particles whose charge sign does not match the polarity. In a positive (negative) polarity phase $\phi_1^+=0$ ($\phi_1^-=0$). We will take $\phi_0$, $\phi_1^{\pm}$ to be the parameters averaged over the time scale of the experiment.
At rigidity $\mathcal{R}\gg\mathcal{R}_0$ the particle's Larmor radius is larger than the scale of magnetic field irregularities and its motion is controlled by the average field. Here $\mathcal{R}$ stands for the rigidity before entering the heliosphere (interstellar rigidity). Down to $\mathcal{R}\sim 2\gv$ we can approximate $\mathcal{F}$ from~\cite{Cholis:2015gna} by  
\begin{equation}\label{eq:faprox}
 \mathcal{F}= \frac{\mathcal{R}_0}{\mathcal{R}}
\end{equation}
up to a normalization constant which can be absorbed into $\phi_1^{\pm}$. We set $\mathcal{R}_0=1\gv$ in the following without loss of generality. In~\cite{Cholis:2015gna} $\phi_{0},\,\phi_1^{\pm}$ were related to the strength of the solar magnetic field and the waviness of the heliospheric current sheet (tilt angle). However, the AMS-02 data were partly taken during a phase of polarity reversal. While the functional form~\eqref{eq:forcefield2} may still approximately hold in this period\footnote{The parameter $\phi_1^{\pm}$ induces a slightly stronger energy loss of negatively compared to positively charged particles or vice versa. This is a plausible ansatz for a charge-breaking effect beyond the physical motivation given above.} the connection to solar observables becomes less transparent due to the rapidly changing magnetic field configuration. Therefore, we follow a different strategy and extract $\phi_{0},\,\phi_1^{\pm}$ from cosmic ray data.

\begin{figure}[htp]
\begin{center}
  \includegraphics[width=16cm]{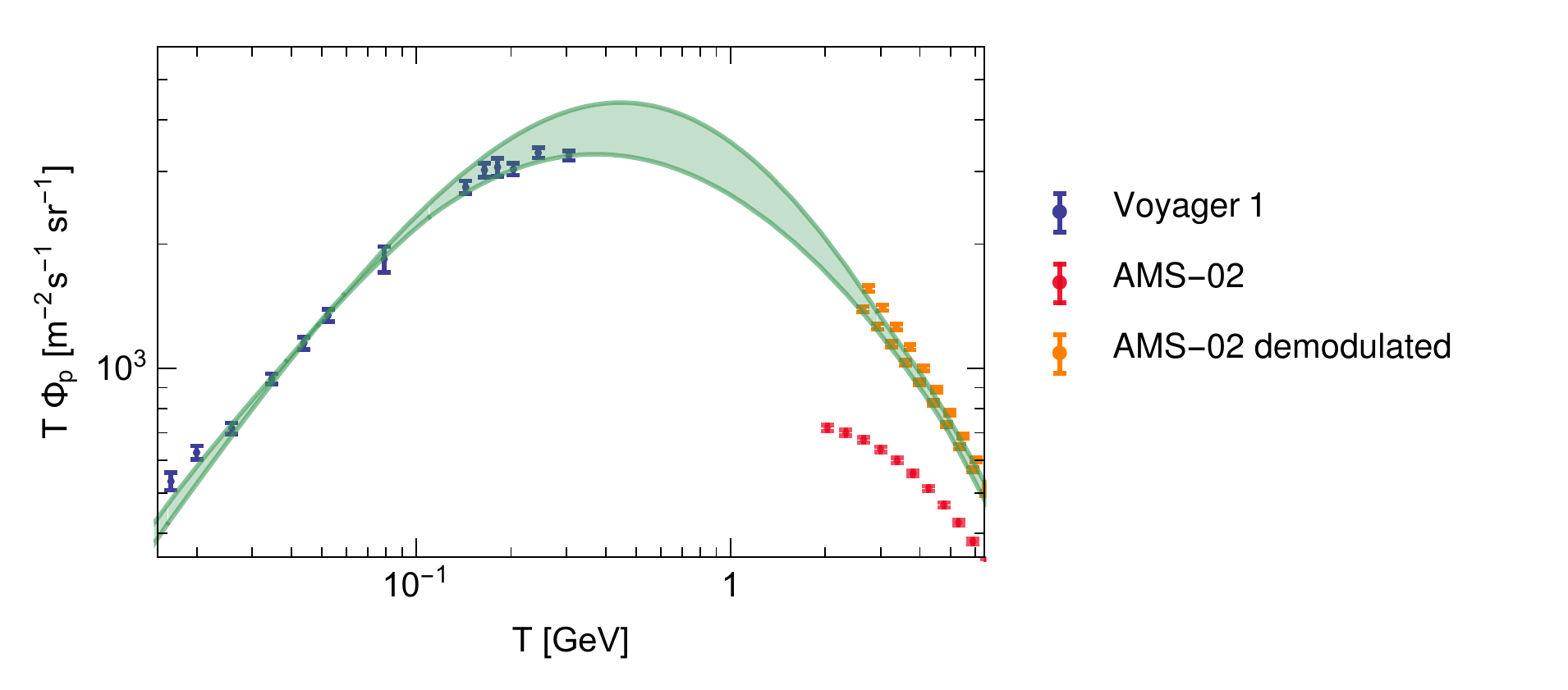}
\end{center}
\caption{Low energy proton flux measured by Voyager and AMS-02. Also shown are AMS-02 data demodulated by a Fisk potential of $0.72\gv$ (upper orange error bars) and $0.6\gv$ (lower orange error bars). The green band is the envelope of interstellar proton fluxes determined in~\cite{Vos:2015,Ghelfi:2015tvu,Corti:2015bqi}.}
\label{fig:voyager}
\end{figure}

Data from the Voyager spacecraft~\cite{Stone:2013} play an important role in pinning down the solar modulation of cosmic rays. Voyager 1 crossed the heliopause in 2012 and provided the first measurement of the interstellar proton flux. This has triggered several recent determinations of the Fisk potential for AMS-02 under the assumption of the force field approximation~\cite{Vos:2015,Ghelfi:2015tvu,Corti:2015bqi}.\footnote{In~\cite{Vos:2015} the Fisk potential $\phi^+_{\text{\tiny{AMS-02}}}=0.64\gv$ for AMS-02 is not explicitly given, but can easily be obtained from the provided interstellar flux.} These rely on different parameterizations of the interstellar flux which were fit to the AMS-02~\cite{Aguilar:2015ooa} and Voyager data simultaneously. The obtained values are in the range $\phi^{+}_{\text{\tiny{AMS-02}}}=0.60-0.72\gv$. In figure~\ref{fig:voyager} the envelope of the interstellar fluxes~\cite{Vos:2015,Ghelfi:2015tvu,Corti:2015bqi} is shown with the Voyager and AMS-02 data. To guide the eye we also depict the AMS-02 data demodulated with Fisk potentials of $\phi^{+}_{\text{\tiny{AMS-02}}}=0.60,\,0.72\gv$ which captures the range of uncertainties related to the choice of parameterization. Larger or smaller Fisk potentials appear to require unphysical inflections in the shape of the interstellar flux in order to connect the two data sets. 

The AMS-02 proton data were taken in the time interval 2011/05-2013/11, i.e. before and during the solar polarity flip. Nevertheless, there is indication that deviations from the force field approximation are small for positive charges. The dedicated solar modulation code HelMOD~\cite{Boschini:2017gic} yields virtually identical results for protons as the force field approximation with $\phi^{+}_{\text{\tiny{AMS-02}}}=0.6\gv$. In~\cite{Corti:2015bqi} it was argued that a slight evolution of the Fisk potential from $\phi^{+}_{\text{\tiny{AMS-02}}}=0.49\gv$ at $T=5\gev$ to $\phi^{+}_{\text{\tiny{AMS-02}}}=0.59\gv$ at $T=0.1\gev$ would somewhat improve the fit to Voyager and AMS-02. Even in this case, the value obtained within the force field approximation is still valid as an upper limit over the full energy range. We will therefore set $\phi^{+}_{\text{\tiny{AMS-02}}}=\phi_{0,\text{\tiny{AMS-02}}} = 0.60-0.72\gv$, $\phi^+_{1,\text{\tiny{AMS-02}}}=0$ in the following. While the upper limit on $\phi^{+}_{\text{\tiny{AMS-02}}}$ appears sufficiently robust, the lower limit will not play a role in the subsequent analysis. Compared to protons and positrons the B/C data were taken over a longer time period over which modulation may have changed. We can safely neglect this as B/C is anyway rather insensitive to the choice of the Fisk potential.\footnote{\label{footnote}An increase of $\phi^{+}_{\text{\tiny{AMS-02}}}$ would result in a stronger modulation and hence a decrease of the boron flux. At the same time, a larger Fisk potential enhances the interstellar fluxes of the boron progenitors (e.g.\ carbon) compared to the TOA fluxes measured at AMS-02. The corresponding increase of the boron source term would efficiently cancel the modulation-caused decrease of the boron flux.}

Finally, we have to account for negative charges, i.e.\ antiprotons. The corresponding AMS-02 data were taken between 2011/05 and 2015/05, i.e. before, during and after the solar polarity reversal. The strong increase of $e^+/e^-$ which sets in 2014~\cite{Adriani:2016uhu} hints at a significant charge-dependence in the modulation which may be traced back to a non-vanishing $\phi_1^-$. In order to pin down this effect, we compare the AMS-02 antiproton data~\cite{Aguilar:2016kjl} with those of PAMELA~\cite{Adriani:2012paa}. Specifically, we consider the ratio of antiproton fluxes $\bar{p}_{\text{\tiny{AMS-02}}}/\bar{p}_{\text{\tiny{PAMELA}}}$ observed at the two experiments as shown in figure~\ref{fig:pbar}.\footnote{For the direct comparison we had to rebin the AMS-02 data in order to match the PAMELA bins. This was achieved by fitting a smooth function $\Phi_{\bar{p}}^\text{smooth}$ through the AMS-02 data. For rigidity bins $\{\mathcal{R}_1,\mathcal{R}_2\}$ which had to be split at $\mathcal{R}'$, we distributed the detected events below and above $\mathcal{R}'$ according to the ratio $\int\limits_{\mathcal{R}_1}^{\mathcal{R}'} d\mathcal{R}\,\Phi_{\bar{p}}^\text{smooth}/\int\limits_{\mathcal{R}'}^{\mathcal{R}_2}d\mathcal{R}\,\Phi_{\bar{p}}^\text{smooth}$. The rebinning only affects a limited number of bins and is not expected to introduce significant systematic errors.} Within reason this ratio is insensitive to the assumed interstellar antiproton flux and, rather, depends on the difference in solar modulation between the two experiments $\Delta\phi^-= 0.2\gv + \phi^-_{1,\text{\tiny{AMS-02}}} \,\mathcal{F}(\tfrac{\mathcal{R}}{\mathcal{R}_0})$. Here we took into account that PAMELA was operating in a negative polarity period (2006-2009) and determined the difference in $\phi_0$ from the proton data~\cite{Adriani:2013as,Aguilar:2015ooa}.\footnote{We used PAMELA proton data above $T=2\gev$ for which $\phi^+_{1,\text{\tiny{PAMELA}}}$ is negligible.} For $\mathcal{F}$ we use~\eqref{eq:faprox} as interstellar rigidities fulfill $\mathcal{R}\gtrsim 2\gv$. 
The remaining free parameter can be obtained by fitting to $\bar{p}_{\text{\tiny{AMS-02}}}/\bar{p}_{\text{\tiny{PAMELA}}}$. As can be seen in figure~\ref{fig:pbar} the decrease of the AMS-02 antiproton flux compared to PAMELA towards low rigidity indeed favors a non-vanishing $\phi^-_{1,\text{\tiny{AMS-02}}}\simeq 0.7\gv$ (see section~\ref{sec:pbarandbc}). 
In order to briefly summarize our treatment of solar modulation, we employ the Fisk potential for positively and negatively charged particles
\begin{equation}\label{eq:fiskpotentials}
  \phi^+_{\text{\tiny{AMS-02}}} = \phi_{0,\text{\tiny{AMS-02}}}\,,\qquad\quad
 \phi^-_{\text{\tiny{AMS-02}}} = \phi_{0,\text{\tiny{AMS-02}}} + \phi_{1,\text{\tiny{AMS-02}}}^-\;\frac{\mathcal{R}_0}{\mathcal{R}}\,.
\end{equation}
and allow $\phi_{0,\text{\tiny{AMS-02}}}$ to float in the interval $\phi_{0,\text{\tiny{AMS-02}}}=0.6-0.72\gv$. $\phi^-_{1,\text{\tiny{AMS-02}}}$ is treated as a free parameter which will efficiently be constrained by including $\bar{p}_{\text{\tiny{AMS-02}}}/\bar{p}_{\text{\tiny{PAMELA}}}$ into our fits.

\section{Secondary Production of Charged Cosmic Rays}\label{sec:secondaryproduction}

Secondary cosmic rays descend from scattering of primary cosmic rays in the interstellar gas. A strategy to determine secondary fluxes often pursued in the literature is to start with a parameterization of primary sources and then to calculate the fluxes of all primary and secondary species simultaneously from the network of scattering and spallation reactions in the galactic disc. While this global approach certainly has its merits, it requires a huge number of inputs, e.g.\ about 2000-3000 nuclear fragmentation cross sections~\cite{Webber:2003}. In our case, where we are mainly interested in a limited number of secondary species (boron, antiprotons and positrons), it is wise to choose a more economical path.
Rather than dealing with the full network, we will determine the fluxes of the direct progenitors of boron, antiprotons and positrons from the available experimental data.
This reduces the number of relevant production cross sections to a more manageable number of 24 (48 if we include production on helium) and the relevant primary fluxes to 8. These will be parameterized in the following.

\subsection{Boron Production Cross Section}\label{sec:boronproduction}

The element boron (B) plays a significant role in cosmic ray physics. Its importance relates to the fact that boron is presumably a pure secondary which makes it an ideal target to study cosmic ray propagation effects. The two stable boron isotopes $^{11}$B and $^{10}$B mainly descend from the spallation of carbon (C), oxygen (O) and nitrogen (N), but also neon (Ne), magnesium (Mg) and silicon (Si) yield non-negligible contributions. In the spallation processes the kinetic energy per nucleon is approximately preserved, i.e.
\begin{equation}
 \left(\frac{d\sigma_{ij\rightarrow a}}{dT}\right) = \sigma_{ij\rightarrow a} \delta(T'-T)\,,
\end{equation}
where $T'$ and $T$ stand for the kinetic energy per nucleon of the incoming primary and outgoing secondary particle. This equation is also referred to as ``straight-ahead approximation''. There exist several parameterizations of fragmentation cross sections in the literature~\cite{Silberberg:1973jxa,Silberberg:1998lxa,Webber:2003,Moskalenko:2003kp}, of which the one by Webber et al.~\cite{Webber:2003} and the one implemented in GALPROP~\cite{Moskalenko:2003kp} are most commonly used. Unfortunately, the corresponding uncertainties have not been estimated in a systematic way. This is problematic as we later want to perform spectral fits to the B/C ratio in cosmic rays. Due to the small experimental errors in the flux, spallation cross sections are a comparable if not the dominant source of uncertainty. Therefore, we decided to redetermine the fragmentation cross sections for C, N, O, Ne, Mg, Si, B to B on hydrogen and helium. We shall derive the cross sections and the related uncertainties from a wide collection of experimental data. An additional motivation for this exercise are deviations from the Webber parameterization which are observed in some recent data sets.

We will follow the convention to include reactions proceeding through short-lived radioactive nuclei (e.g. $^{12}\text{C}+p\rightarrow \,^{11}\text{C}\xrightarrow{\beta^+} \,^{11}\text{B}$), referred to as ghosts, as part of the spallation cross section. The term ``short-lived'' is to be understood on astrophysical scales and stands for lifetimes $\tau < \text{kyr}$. The very long-lived isotope $^{10}$Be which decays into $^{10}$B is treated as a separate final state which will be considered in addition to $^{11}$B and $^{10}$B. We parameterize the fragmentation cross section of the nucleus $i$ to boron on a hydrogen target as
 \begin{equation}\label{eq:fragmentation}
 \sigma_{i+p\,\rightarrow\, \text{B}} = \sigma_{0,i} \,\frac{\Gamma_i^2\,(T-E_{\text{th},i})^2}{(T^2-M_i^2)^2 +\Gamma_i^2 M_i^2} + \sigma_{1,i} \left(1- \frac{E_{\text{th},i}}{T}\right)^{\xi_i} \left(1+ \frac{\Delta_i}{1+(T_\text{h}/T)^2} \right)\,.
\end{equation}
Above energy threshold $E_{\text{th},i}$ cross sections show a resonance peak whose normalization, position and width is set by the parameters $\sigma_{0,i}$, $M_i$ and $\Gamma_i$. If one subtracts the peak, there appears a steady rise which continues up to $T\sim\text{GeV}$ with its smoothness controlled by $\xi_i$. While in the older literature (e.g.~\cite{Read:1984xme}) spallation cross sections were taken to be constant above this energy, a non-trivial behavior at $T=1-5\gev$ was motivated by Webber et al.~\cite{Webber:2003} on observational grounds. As we find that fits to the experimental data indeed improve significantly when allowing for a slow change of the cross section around $T_\text{h}=2\gev$, we have added the term in the last brackets. In this way a very similar functional behavior as in the Webber parameterization can be achieved if preferred by the data. At energies $T > 5\gev$, where nuclear binding energies are irrelevant, spallation cross sections approach the asymptotic value $\sigma_{i+p\,\rightarrow\, \text{B}} = \sigma_{1,i}\, (1 + \Delta_i)$. The existence of a plateau is commonly assumed in the literature but awaits experimental proof. As total inelastic cross sections are known to increase slowly with energy beyond $T=10\gev$, one may also speculate about a proportional rise in the individual spallation cross sections~\cite{Genolini:2017dfb}. A full correlation between the cross sections is, however, ambiguous as the final state particle spectrum in inelastic collision also changes with increasing energy, where multiparticle production gains significance. Given the sparseness of high energy data we decided to follow the standard assumption of a plateau in the spallation cross sections.

\begin{figure}[htp]
\begin{center}
  \includegraphics[width=15.5cm,trim={0.2 0 4.5cm 0},clip]{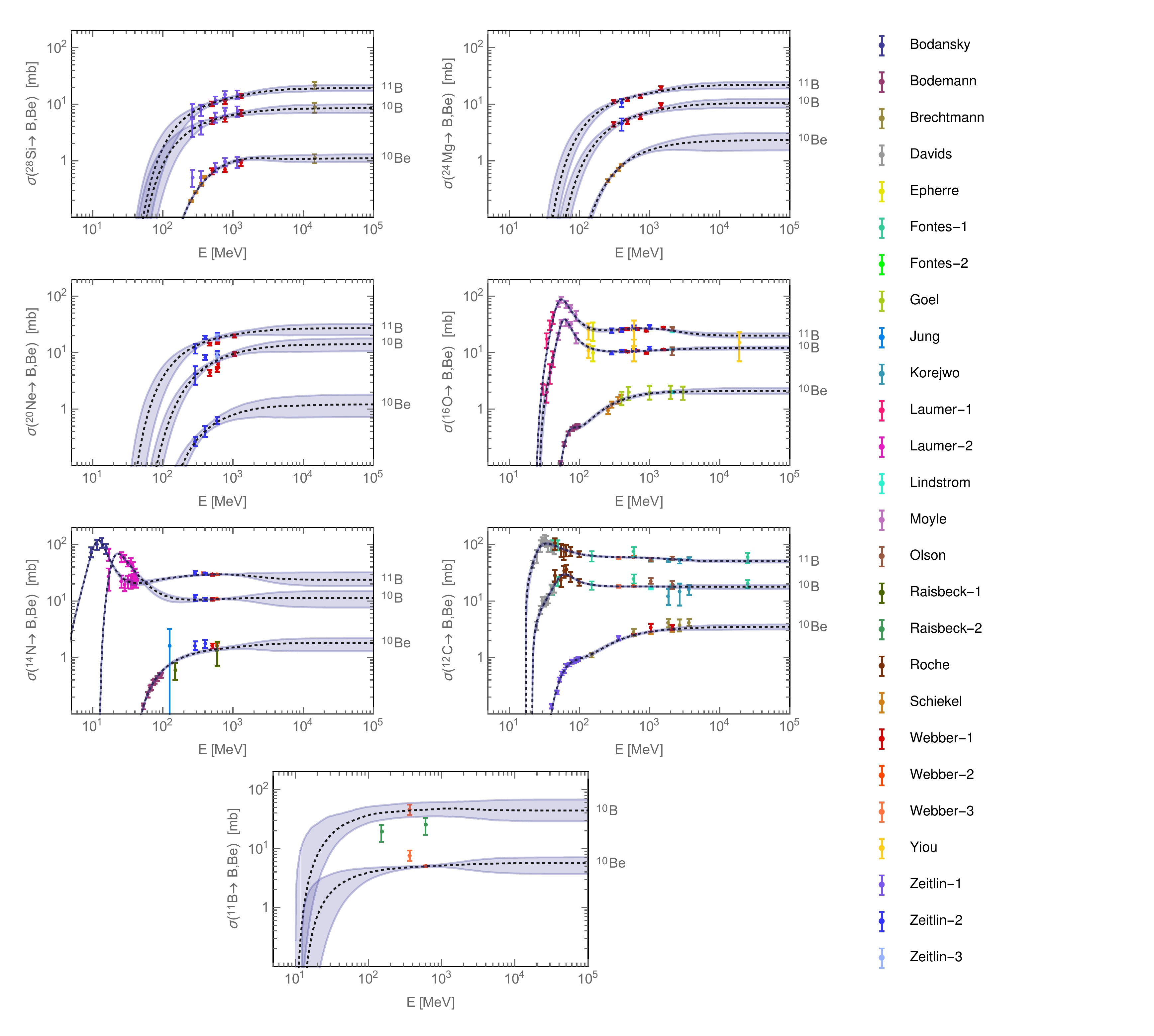}
\end{center}
\caption{Isotopic cross sections for the fragmentation of $^{28}$Si, $^{24}$Mg, $^{20}$Ne, $^{16}$O, $^{14}$N, $^{12}$C, $^{11}$B to $^{10,11}$B and $^{10}$Be. Also shown are our fits and the corresponding uncertainty bands. Experimental data are taken from Bodansky~\cite{Bodansky:1975}, Bodemann~\cite{Bodemann:1993}, Brechtmann~\cite{Brechtmann:1988rs}, Davids~\cite{Davids:1970wh}, Epherre~\cite{Epherre:1969qxr}, Fontes-1~\cite{Fontes:1977qq}, Fontes-2~\cite{Fontes:1971rn}, Goel~\cite{Goel:1969}, Jung~\cite{Jung:1970xr}, Korejwo~\cite{Korejwo:2000pf,Korejwo:2002ts}, Laumer-1~\cite{Laumer:1974zza}, Laumer-2~\cite{Laumer:1973zz}, Lindstrom~\cite{Lindstrom:1975mi}, Moyle~\cite{Moyle:1979zz}, Olson~\cite{Olson:1983jz}, Raisbeck-1~\cite{Raisbeck:1974zz}, Raisbeck-2~\cite{Raisbeck:1971ie}, Roche~\cite{Roche:1976zz}, Schiekel~\cite{Schiekel:1996}, Webber-1~\cite{Webber:1990kb}, Webber-2~\cite{Webber:1990kc}, Webber-3~\cite{Webber:1998}, Yiou~\cite{Yiou:1969}, Zeitlin-1~\cite{Zeitlin:2007sm}, Zeitlin-2~\cite{Zeitlin:2011qg}, Zeitlin-3~\cite{Zeitlin:2001ye}. In some cases the original data were processed in order to arrive at isotopic cross sections (see text).}
\label{fig:spallation}
\end{figure}

Sufficient experimental data to perform our fits only exist for the most abundant isotope of the respective element (e.g.\ $^{12}$C in the case of carbon). This leaves us with 20 considered isotopic fragmentation cross sections for which the collected data are shown in figure~\ref{fig:spallation}. A number of comments are in order: in some references~\cite{Webber:1990kb,Zeitlin:2007sm,Zeitlin:2011qg,Zeitlin:2001ye} only the charge-changing cross sections have been measured. These were translated into isotopic cross sections by using the hydrogen mass fractions given in~\cite{Webber:1990kc}. If the latter were not specified (i.e.\ for Si, Mg and Ne) we used the Webber code as extracted from DRAGON in order to predict how the cross section is divided among the isotopes. In the case of silicon, a previous step was required as the charge-changing cross section was only measured down to Z=6 (carbon). Luckily, data down to Z=4,5 exist for a polyethylene target~\cite{Gupta:2013}. We determined the charge changing cross section to Z=4,5 on hydrogen by assuming that the ratio of cross section to Z=6 compared to Z=4,5 is the same for hydrogen and polyethylene.\footnote{This is justified as a very similar behavior of the charge changing cross sections down to Z=6 is observed for hydrogen and polyethylene.} The isotopic cross sections were then obtained as described above. We refrained from assigning a systematic error to this procedure as silicon only contributes to the boron production at the few percent level. 
Turning to the low energy part of the fragmentation cross sections, sufficient data to fit the resonance peak above threshold do not exist for Ne, Mg and Si. This is acceptable for our purposes as we will later deal with cosmic ray boron at $T>0.4\gev$ which is practically unaffected by the threshold behavior of the cross sections. Nevertheless, as a tiny effect may still arise due to the reacceleration of very low energy cosmic rays, we attempt to capture at least roughly the threshold behavior for the dominant boron progenitors C, N and O. In the case of nitrogen a very complicated structure with various peaks emerges which we slightly smooth out by combining energy bins in sets of two~\cite{Bodansky:1975}.\footnote{We conservatively assume $20\%$ errors as no uncertainties were given in this reference.} The low-energy data on the cross section for $^{12}\text{C}\,\rightarrow\,^{10}\text{B}$ were obtained from the isobaric cross section to A=10~\cite{Davids:1970wh,Roche:1976zz} by subtracting the (tiny) beryllium contribution~\cite{Bodemann:1993}.

We determined the best fit parameters for each individual cross section. The parameters $\sigma_{0,i}$, $M_i$, $\Gamma_i$ (considered only for C, N, O) were kept fixed at their best fit values as variations hardly affect the boron flux in the relevant energy range (see above). The probability distribution of $\sigma_{1,i}$, $\xi_i$, $\Delta_i$  was determined from a $\Delta\chi^2$-metric with three degrees of freedom (d.o.f.).\footnote{For some cross sections the parameter $\Delta_i$ is not well constrained due to the lack of high energy data. In order to avoid unphysical values of $\Delta_i$ we imposed the (conservative) constraint $|\Delta_i|<0.5$.} Parameter values and uncertainties (for $\sigma_{1,i}$, $\xi_i$, $\Delta_i$) are given in table~\ref{tab:fragcross}. In figure~\ref{fig:spallation} we show the median cross sections and the uncertainty bands ($\pm 1\sigma$) following from the derived probability distribution on the cross section parameters. Correlations between uncertainties at different energies cannot be made visible in the figure, but are taken into account in the following.

\begin{table}[htp]
\begin{center}
\begin{tabular}{|c|ccccccc|}
\hline
\rowcolor{light-gray}  &   &  &   &  &&&\\[-3mm]
\rowcolor{light-gray}  Channel & $ \!\!E_{\text{th},i}\,[\text{MeV}]\!\!\!$ & $\!\sigma_{0,i}\,[\text{mb}]\!$ & $\!\!M_i\:[\text{MeV}]\!\!$ &  $\!\Gamma_i\,[\text{MeV}]\!$ & $\!\sigma_{1,i}\,[\text{mb}]\!$ & $\xi_i$ &  $\Delta_i$  \\[2mm]
\hline 
  &   & & & & & &\\[-3mm]
 $\!\!^{28}\text{Si}\rightarrow$ $^{11}\text{B}\!$  &26.7   & & && $15.1^{+2.4}_{-1.6}$ & $7.1^{+3.4}_{-2.2}$ & $\!0.30^{+0.14}_{-0.21}$\!\\[2mm]
 $\!\!^{28}\text{Si}\rightarrow$ $^{10}\text{B}\!$  &29.0   & & && $7.5^{+1.7}_{-0.9}$& $6.4^{+4.2}_{-2.3}$& $\!0.12^{+0.26}_{-0.29}$\!\\[2mm]
 $\!\!^{28}\text{Si}\rightarrow$ $^{10}\text{Be}\!$ &40.9   & & & & $1.71^{+0.15}_{-0.12}$& $12.1^{+0.6}_{-0.5}$& $\!-0.36^{+0.14}_{-0.10}$\!\\[2mm]
 $\!\!^{24}\text{Mg}\rightarrow$ $^{11}\text{B}\!$  &21.0   & & & & $17.6^{+2.0}_{-1.8}$ & $7.4^{+2.1}_{-1.9}$ & $\!0.29^{+0.15}_{-0.25}$\!\\[2mm]
 $\!\!^{24}\text{Mg}\rightarrow$ $^{10}\text{B}\!$  &22.5   & & & & $8.7^{+1.3}_{-1.1}$  & $10.4^{+2.7}_{-2.5}$ & $\!0.24^{+0.18}_{-0.29}$\!\\[2mm]
 $\!\!^{24}\text{Mg}\rightarrow$ $^{10}\text{Be}\!$ &36.0   & & & & $2.27^{+0.45}_{-0.23}$ & $10.9^{+1.5}_{-0.9}$& $\!0.01^{+0.33}_{-0.34}$\!\\[2mm]
 $\!\!^{20}\text{Ne}\rightarrow$ $^{11}\text{B}\!$  &23.6   & & & & $23.3^{+5.3}_{-2.0}$ & $6.7^{+3.8}_{-1.9}$ & $\!0.17^{+0.22}_{-0.37}$\!\\[2mm]
 $\!\!^{20}\text{Ne}\rightarrow$ $^{10}\text{B}\!$  &23.6   & & & & $13.1^{+3.0}_{-2.0}$ & $14.3^{+4.5}_{-3.4}$ & $\!0.13^{+0.25}_{-0.36}$ \!\\[2mm]
 $\!\!^{20}\text{Ne}\rightarrow$ $^{10}\text{Be}\!$ &37.2   & & & & $1.19^{+0.63}_{-0.30}$& $11.0^{+4.3}_{-3.0}$& $\!0.0^{+0.32}_{-0.34}$\!\\[2mm]
 $\!\!^{16}\text{O}\rightarrow$ $^{11}\text{B}\!$   &23.6   & 240 & 45.9 & 34.2 & $31.4^{+1.7}_{-1.1}$ & $4.0^{+1.1}_{-0.6}$ & $\!-0.36^{+0.08}_{-0.07}$\!\\[2mm]
 $\!\!^{16}\text{O}\rightarrow$ $^{10}\text{B}\!$  & 26.9   & 104 & 55.0 & 29.0 & $10.9^{+0.3}_{-0.2}$& $1.3^{+0.4}_{-0.2}$ & $\!0.10\pm 0.08$\!\\[2mm]
 $\!\!^{16}\text{O}\rightarrow$ $^{10}\text{Be}\!$ & 36.6   & 1.3 & 62.3 & 45.2 & $2.28^{+0.17}_{-0.16}$ & $4.9\pm0.3$ & $\!-0.07^{+0.17}_{-0.15}$\!\\[2mm]
 $\!\!^{14}\text{N}\rightarrow$ $^{11}\text{B}\!$  & 3.1    & 193 & 10.6 & 7.3  & $31.6\pm 0.8$& $9.6^{+1.8}_{-1.6}$ & $\!-0.24^{+0.28}_{-0.19}$\!\\[2mm]
 $\!\!^{14}\text{N}\rightarrow$ $^{10}\text{B}\!$  & 12.5   & 360 & 16.9 & 14.1 & $11.6^{+0.8}_{-0.5}$ & $3.7^{+3.3}_{-1.5}$ & $\!-0.02^{+0.36}_{-0.34}$\!\\[2mm]
 $\!\!^{14}\text{N}\rightarrow$ $^{10}\text{Be}\!$  & 34.3  & & & & $1.6\pm 0.09$ & $2.36^{+0.09}_{-0.08}$ & $\!0.15^{+0.24}_{-0.34}$\!\\[2mm]
 $\!\!^{12}\text{C}\rightarrow$ $^{11}\text{B}\!$   & 17.3  & 330& 18.9 & 23.5& $60.2^{+1.1}_{-0.9}$ & $0.47^{+0.15}_{-0.11}$& $\!-0.16^{+0.04}_{-0.05}$\!\\[2mm]
 $\!\!^{12}\text{C}\rightarrow$ $^{10}\text{B}\!$  &  21.4  & 34.1& 46.9 & 46.9 & $18.4^{+0.4}_{-0.3}$& $0.75^{+0.07}_{-0.06}$ & $\!-0.03^{+0.08}_{-0.10}$\!\\[2mm]
 $\!\!^{12}\text{C}\rightarrow$ $^{10}\text{Be}\!$  & 29.5  & 0.24 & 14.9 & 242 & $3.5\pm 0.2$& $7.0^{+0.5}_{-0.4}$ & $\!0.0\pm 0.13$ \!\\[2mm]
 $\!\!^{11}\text{B}\rightarrow$ $^{10}\text{B}\!$  &  10.1  &      &      &     & $49.0^{+14.4}_{-12.7}$ & $2.7^{+1.6}_{-1.9}$ & $\!-0.04^{+0.36}_{-0.32}$\!\\[2mm]
 $\!\!^{11}\text{B}\rightarrow$ $^{10}\text{Be}\!$  & 12.3  & & & & $5.2^{+0.4}_{-0.2}$ & $2.1^{+2.8}_{-1.5}$ & $\!0.08^{+0.28}_{-0.39}$\!\\[2mm]
\hline
\end{tabular}
\end{center}
\caption{Fit parameters entering the fragmentation cross section parameterization~\eqref{eq:fragmentation}.}
\label{tab:fragcross}
\end{table}

In table~\ref{tab:webbercomp}, we compare the fragmentation cross sections (summed over the two boron isotopes) from our fits with those from the Webber parameterization at two different energies. It can be seen that substantial differences exist for Si, Mg, Ne, for which we obtain systematically higher cross sections than Webber. For these three elements we employed recent data sets which did not enter in~\cite{Webber:2003}. The effect on cosmic ray boron production, to which Si, Mg, Ne contribute about $10\%$, is not huge but may amount to a few percent. For C, N, O, our cross sections are in overall good agreement with those by Webber. Only for carbon, there appears a $\sim10\%$ discrepancy at high energy. The Webber parameterization shows a strong decrease of the carbon spallation cross section between $T=1\gev$ and $T=10\gev$ which is not preferred by our fit. In~\cite{Webber:2003} it is argued that this decrease is backed up by the data of Korejwo et al.~\cite{Korejwo:2000pf,Korejwo:2002ts}. Given that Korjewo et al. measured $\sigma({^{12}}C\rightarrow B)=65,\:65.3,\:72.8\mb$ at $T=1.87,\:2.69,\:3.66\gev$ this argument is not plausible at all. The only data point~\cite{Fontes:1977qq} at higher energy ($T=25\gev$) also speaks against the strong decrease albeit with large uncertainty. Unfortunately, existing data are insufficient to fully settle this issue and upcoming experimental efforts are eagerly awaited. 

\begin{table}[htp]
\begin{center}
\begin{tabular}{|l|cc|cc|}
\hline
\rowcolor{light-gray}  &   &  &   &  \\[-3mm]
\rowcolor{light-gray}  Channel & $\sigma(1\gev)$ & $\sigma_{\text{Webber}}(1\gev)$ & $\sigma(10\gev)$ & $\sigma_{\text{Webber}}(10\gev)$  \\[2mm]
\hline 
  &   & & &\\[-3mm]
 $^{28}\text{Si}\rightarrow$ $\text{B}$    & $19.7^{+0.8}_{-0.9}$& 13.2 & $27.2^{+2.6}_{-2.7}$ & 14.2 \\[2mm]
 $^{24}\text{Mg}\rightarrow$ $\text{B}$    & $23.0^{+0.9}_{-1.2}$& 20.4 & $31.8^{+3.3}_{-3.4}$ & 20.4 \\[2mm]
 $^{20}\text{Ne}\rightarrow$ $\text{B}$    & $30.2^{+2.1}_{-1.7}$& 21.9 & $40.4^{+5.9}_{-6.6}$ & 21.1 \\[2mm]
 $^{16}\text{O}\rightarrow$ $\text{B}$     & $37.6^{+0.5}_{-0.4}$& 36.9 & $32.3^{+1.9}_{-1.8}$ & 34.7 \\[2mm]
 $^{14}\text{N}\rightarrow$ $\text{B}$     & $40.4^{+1.2}_{-1.1}$& 41.2 & $35.6^{+8.6}_{-6.1}$ & 37.1 \\[2mm]
 $^{12}\text{C}\rightarrow$ $\text{B}$     & $76.0^{+0.8}_{-0.9}$& 75.4 & $68.6^{+2.5}_{-2.6}$ & 61.0 \\[2mm]
 $^{11}\text{B}\rightarrow$ $^{10}\text{B}$& $47.3^{+14.3}_{-12.2}$& 40.5 & $44.4^{+22.8}_{-15.1}$ & 39.0 \\[2mm]
 \hline
\end{tabular}
\end{center}
\caption{Comparison of the fragmentation cross sections into boron ($^{11}\text{B}\,$+$\,^{10}$B) determined in this work with the parameterization by Webber.}
\label{tab:webbercomp}
\end{table}

Besides the spallation of nuclei on hydrogen, also the spallation on helium contributes significantly to the boron flux. In~\cite{Ferrando:1988tw} it was suggested to account for scattering on helium by multiplying the protonic fragmentation cross sections by an energy-dependent enhancement factor of the form
\begin{equation}\label{eq:heliumenhance}
 e^{\mu |(Z_i-Z_f) - f_i \delta|^\nu }\,,
\end{equation}
where $Z_i$ and $Z_f$ denote the atomic number of the initial and final state nucleus respectively. The exponent $\nu$ as well as the functions $\delta,\;\mu$ are extracted from iron fragmentation on a helium target at three different beam energies~\cite{Ferrando:1988tw}. The coefficients $f_i$ depend on the initial nucleus and are determined by interpolating between carbon, oxygen and iron fragmentation on helium.

While we employ the functional form~\eqref{eq:heliumenhance}, we refrain from using the parameters given in~\cite{Ferrando:1988tw}. This is because the exponent $\nu=1.43$ ~\cite{Ferrando:1988tw} leads to a very steep increase of cross sections with $Z_i$ which is inconsistent with data on aluminum spallation by helium~\cite{Webber:1990kb}. Furthermore, $\mu$ and $\delta$ determined in~\cite{Ferrando:1988tw} lead to unphysical results for silicon and magnesium fragmentation at low energies. This can be seen from silicon carbon scattering~\cite{Zeitlin:2007sm} which is expected to show a similar low-energy behavior as silicon helium scattering. Our approach is, therefore, to refit the enhancement factor using carbon, nitrogen, oxygen and aluminum instead of iron data. We find
\begin{equation}
 \mu=0.29-\frac{0.056\,T}{\text{GeV}}\,,\qquad \delta=1.93-\frac{0.45\,T}{\text{GeV}}\,,\qquad \nu=0.57\,,
\end{equation}
where we assumed $\mu$ and $\delta$ to be linear functions of energy for $T<1.6\gev$ as in~\cite{Ferrando:1988tw}. Above this energy $\mu$ and $\delta$ are taken to be constant.\footnote{There are no data above $T=1.6\gev$, but the enhancement factor should become constant around this energy~\cite{Ferrando:1988tw}.} The parameters $f_{\text{C,N,O}}$ are fitted to the data sets~\cite{Ferrando:1988tw,Webber:1990kb}, while $f_{\text{Ne,Mg,Si}}$ are obtained by linear interpolation between the values for oxygen and aluminum
\begin{align}
 f_\text{C}&= -0.35\pm 0.45\,,\quad f_{\text{N}}=-0.8\pm 0.64\,,\quad f_{\text{O}}=0.04\pm 0.35\,,\nonumber\\
 f_{\text{Ne}}&=0.24\pm 0.37\,,\;\quad f_{\text{Mg}}=0.45\pm 0.38\,,\quad\:
 f_{\text{Si}}=0.65\pm 0.4\,.
\end{align}
Before we finish the discussion on spallation cross sections, we should emphasize that new precision data are urgently required. They are needed to eliminate systematic errors in the parameterizations of all relevant spallation cross sections (on hydrogen and helium) and to establish the high energy behavior. The data we employed in this work were taken over a period of more than four decades in time, using hugely varying experimental techniques. In some cases claimed uncertainties appear suspiciously small. We believe that our analysis contributes in identifying some of the most urgent measurements to be performed (e.g. carbon spallation at energies of a few GeV), but further work is certainly needed.

\subsection{Antiproton Production Cross Section}\label{sec:antiprotoncrosssection}

Secondary antiprotons in cosmic rays mainly originate from proton proton collisions, but also processes involving helium yield a seizable contribution. Since antiprotons are generated with a smooth phase space distribution, the full differential cross section must be modeled at all relevant energies. Experimental data again play a crucial role in this as the cross section is dominated by soft QCD processes and cannot be calculated from first principles. A complication arises from the fact that about half of the antiprotons in cosmic rays stem from the decay of antineutrons which escape detection in laboratory experiments. This contribution must be modeled on the basis of symmetry arguments. In addition, antiprotons partly stem from the decay of strange hyperons which are metastable on detector scales. As in most experiments this contribution is rejected through cuts, it must be determined from the phase space distribution of the parent hyperons. A careful analysis of all relevant processes has been performed in~\cite{Kappl:2014hha,Winkler:2017xor} which we will briefly review for completeness. 

The (Lorentz) invariant differential cross section $f_{pp\rightarrow \bar{p}}\equiv f_{\bar{p}}$ can be expressed as
\begin{equation}\label{eq:contributions}
  f_{\bar{p}}\equiv E\frac{d^3\sigma_{\bar{p}}}{dp^3} = f_{\bar{p}}^0 \:( 2 + \iso + 2 \,\hyp)\,,
\end{equation}
where the energy and three-momentum of the final state antiproton are denoted by $E$ and $p$ respectively. The index $0$ indicates the prompt part of the antiproton production, while $\hyp$ stands for the hyperon induced contribution. Both are multiplied by the factor of two in order to account for antineutrons. A possible asymmetry between antineutron and antiproton production due to isospin effects is included through $\iso$. The prompt cross section
\begin{equation}\label{eq:tsallis}
 f_{\bar{p}}^0 = R \:\sigma_{pp\,\text{in}} \, c_5 \,(1-x_R)^{c_6} \,\big[1 + X (\mt - m_p)\big]^{-\frac{1}{X c_7}}\,,
\end{equation}
is modeled in terms of the transverse momentum $\pt$ and the radial scaling variable $x_R= E^*/E^*_{\text{max}}$ with $E^*$ denoting the antiproton energy in the center-of-mass frame and $E^*_{\text{max}}=(s-8 m_p^2)/(2\sqrt{s})$ the maximal energy. The function $R$ accounts for near threshold production of antiprotons. It is taken to be unity at $T>10\gev$ and 
\begin{equation}\label{eq:nearthreshold}
 R= \left[ 1 + c_9 \left( 10 - \tfrac{\sqrt{s}}{\text{GeV}}\right)^5\right] \exp\left[ c_{10} \left( 10 - \tfrac{\sqrt{s}}{\text{GeV}}\right)^2 \left(x_R- x_{R,\text{min}}\right)^2\right]\quad \text{for}\;\; T\leq 10\gev\,.
\end{equation}
There occur two additional terms in $f_{\bar{p}}^0$ which violate radial scaling through their explicit energy-dependence: the inelastic cross section
\begin{equation}\label{eq:inelasticpar}
 \sigma_{pp,\,\text{in}} = c_{11} + c_{12}\, \log{\sqrt{s}} + c_{13}\, \log^2{\sqrt{s}}\,,
\end{equation}
grows slowly with $\sqrt{s}$. In addition, the term
\begin{equation}\label{eq:X}
 X = c_8\, \log^2\left[\frac{\sqrt{s}}{4\,m_p}\right]\,,
\end{equation}
induces a flattening in the $\pt$-distributions at large energy. It originates from multiple scattering of protons which goes along with the production of hard $\pt$-jets. Coming back to~\eqref{eq:contributions}, the hyperon contribution can be expressed as $\hyp=(0.81\pm 0.04)\,(\bar{\Lambda}/\bar{p})$ with
\begin{equation}\label{eq:lambdapara}
\bar{\Lambda}/\bar{p}= c_1 +  \frac{c_2}{1+(c_3/s)^{c_4}}\,,
\end{equation}
where the second term on the right-hand side accounts for the increased strange hadron production which sets in at $\sqrt{s}\sim 100\gev$. Finally, the asymmetry between antineutron and antiproton production is written as
\begin{equation}\label{eq:isofit}
\iso =  \frac{c_{14}}{1+(s/c_{15})^{c_{16}}}\,.
\end{equation}
A non-vanishing $\iso$ may be present at low energy, but the experimental data are not fully conclusive. In any case, the asymmetry disappears at high energy. The parameterization contains in total 16 parameters $c_{1}\dots c_{16}$ which were fitted to a large set of experimental data in~\cite{Winkler:2017xor}. In table~\ref{tab:pbarcross} we provide median values and uncertainties for the $c_i$.\footnote{In some cases, the median values differ marginally from the best fit values provided in~\cite{Winkler:2017xor}.}

\begin{figure}[htp]
\begin{center}   
 \includegraphics[width=14cm]{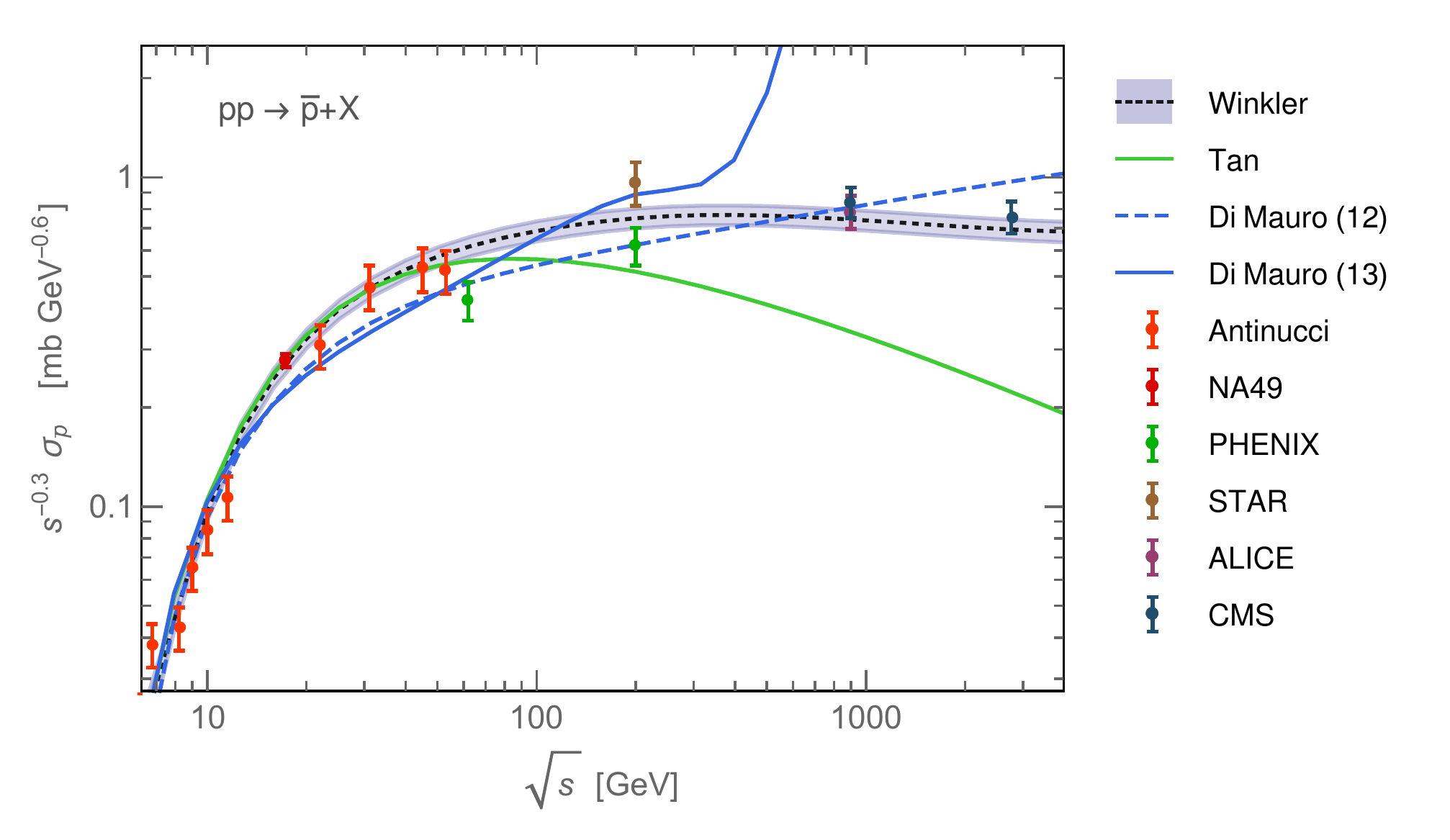}
\end{center}
\caption{Antiproton production cross section employed in this work~\cite{Winkler:2017xor} compared to the cross section derived from the parameterizations of Tan et al.~\cite{Tan:1982nc} and di Mauro et al.~\cite{diMauro:2014zea}. The contribution from antineutron decay is not included in this figure. Experimental data were taken from~\cite{Antinucci:1972ib,Abelev:2008ab,Anticic:2009wd,Adare:2011vy,Aamodt:2011zj,Chatrchyan:2012qb} and processed as described in~\cite{Winkler:2017xor}.}
\label{fig:comparisonpbar}
\end{figure}

\begin{table}[htp]
\begin{center}
\begin{tabular}{|cccc|}
\hline
\rowcolor{light-gray}  &   &  &    \\[-4mm]
\rowcolor{light-gray}  $c_1$ & $\qquad\qquad c_2\qquad\qquad$ & $\qquad\qquad c_3\qquad\qquad$ & $c_4$  \\[1mm]
\hline 
 &   &  &    \\[-3mm]
  $0.31\pm 0.04$ & $0.30\pm 0.06$ & $(153^{+65}_{-57})^2\gev^2$ & $1.0\pm 0.3$\\[1mm]
\hline  
\rowcolor{light-gray}  &   &  &    \\[-4mm]
\rowcolor{light-gray}  $c_5$ & $c_6$ & $c_7$ & $c_8$  \\[1mm]
\hline
 &   &  &    \\[-3mm]  
   $0.0467\pm 0.0038$ & $7.77\pm 0.10$ & $(0.168\pm 0.001)\gev$ & $(0.0380\pm 0.0006)\gev^{-1}$  \\[1mm]
 \hline
 \rowcolor{light-gray}  &   &  &    \\[-4mm]
\rowcolor{light-gray}  $c_9$ & $c_{10}$ & $c_{11}$ & $c_{12}$  \\[1mm]
\hline 
 &   &  &    \\[-3mm]
  $0.0010\pm 0.0004$ & $0.7\pm 0.04$ & $(30.9\pm 0.4)\mb$ & $(-1.74\pm 0.17)\mb$\\[1mm]
\hline  
\rowcolor{light-gray}  &   &  &    \\[-4mm]
\rowcolor{light-gray}  $c_{13}$ & $c_{14}$ & $c_{15}$ & $c_{16}$  \\[1mm]
\hline
 &   &  &    \\[-3mm]  
   $(0.71\pm 0.02)\mb$ & $0.20^{+0.30}_{-0.18}$ & $(31^{+47}_{-25})^2\gev^2$ & $1.0\pm 0.3$  \\[1mm]
   \hline
\end{tabular}
\end{center}
\caption{Parameters entering the antiproton production cross section~\eqref{eq:contributions} and related uncertainties.}
\label{tab:pbarcross}
\end{table}
In figure~\ref{fig:comparisonpbar}, we compare the integrated antiproton production cross section derived from~\eqref{eq:contributions} with the cross section obtained from the parameterizations of Tan et al.~\cite{Tan:1982nc} and di Mauro et al.~\cite{diMauro:2014zea}. As can be seen, the cross section predicted by Tan et al. falls short at high energy as it does not account for the violation of radial scaling. Di Mauro et al. provide two parameterizations (in equations (12) and (13) of~\cite{diMauro:2014zea}) which were fit to data at $\sqrt{s}\leq 200\gev$. They differ substantially in the high energy regime due to different extrapolations.

The cross sections involving helium were predicted in~\cite{Winkler:2017xor} from an empirical model which was introduced in~\cite{Baatar:2012fua} for proton carbon scattering and first applied to helium in~\cite{Kappl:2014hha}. The invariant antiproton production cross section is expressed in terms of $f_{\bar{p}}$ as
\begin{equation}\label{eq:helium}
f_{ij\rightarrow \bar{p}}   = \frac{\sigma_{ij,\text{in}}}{\sigma_{pp,\text{in}}}\,\Big(\langle \nu_i \rangle  F_\text{pro}(x_f)
+\langle \nu_j \rangle  F_\text{tar}(x_f)\Big)\,
f_{\bar{p}}\,,
\end{equation}
where $i=p,\text{He}$ and $j=p,\text{He}$ stand for the projectile and the target particle respectively. The ratios of inelastic cross sections are taken to be 
$\sigma_{pp,\text{in}}:\sigma_{p\text{He},\text{in}}:\sigma_{\text{He}\text{He},\text{in}}=1:3.2:7.7$~\cite{Kappl:2014hha}. The projectile and target overlap functions are defined in terms of the Feynman scaling variable $x_f=\pl^*/(2\sqrt{s})$~\cite{Baatar:2012fua}, where $\pl^*$ denotes the longitudinal antiproton momentum in the center-of-mass frame. The average number of interacting nucleons in the projectile and target $\langle \nu_{i,j} \rangle$ can be expressed in terms of the inelastic cross sections $\langle \nu_{i} \rangle = A_i\,\sigma_{pp,\text{in}}/\sigma_{i p,\text{in}}$, where $A_i$ denotes the mass number of the nucleus $i$.

\begin{figure}[htp]
\begin{center}   
 \includegraphics[width=7.5cm]{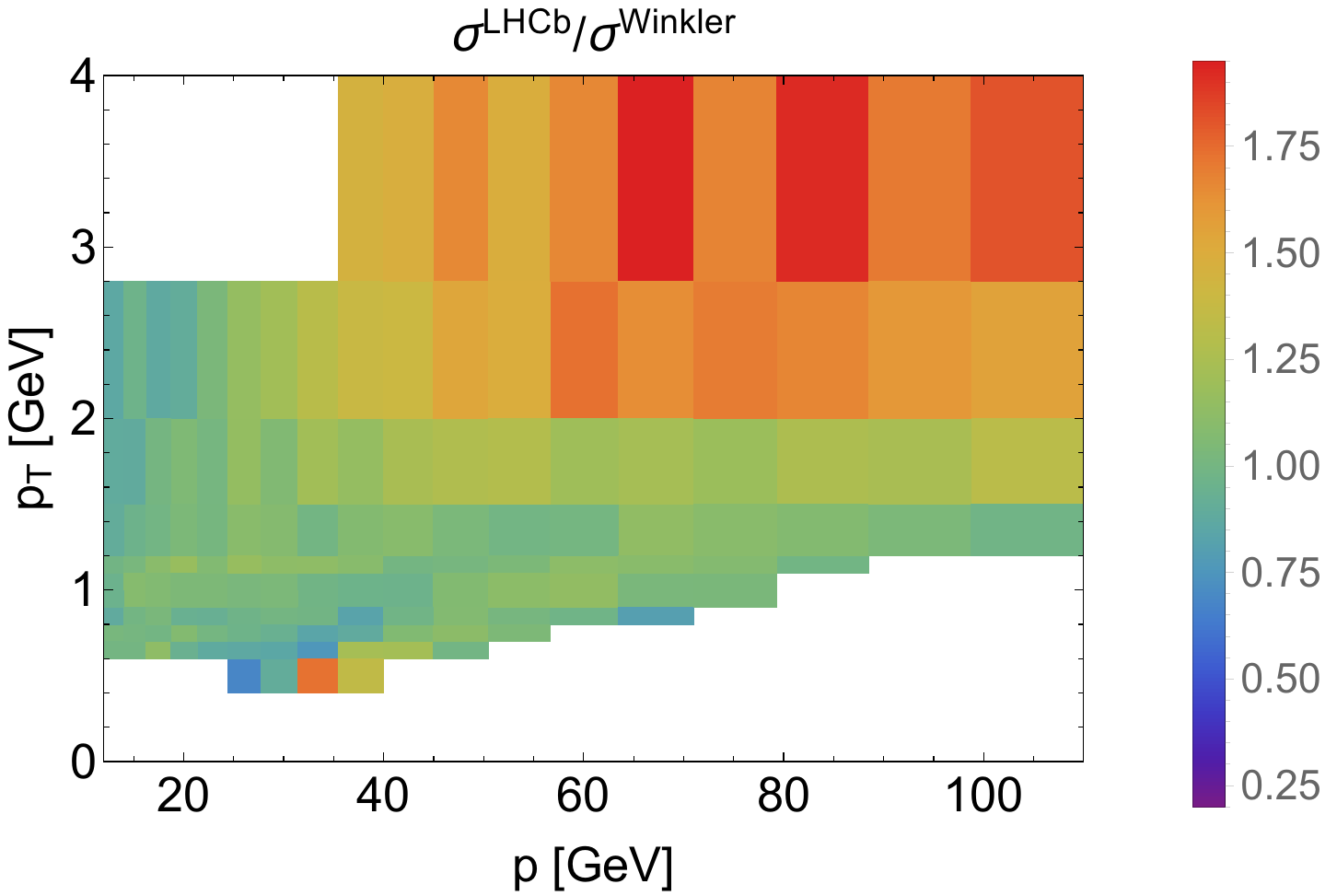}
 \hspace{1mm}
 \includegraphics[width=7.5cm]{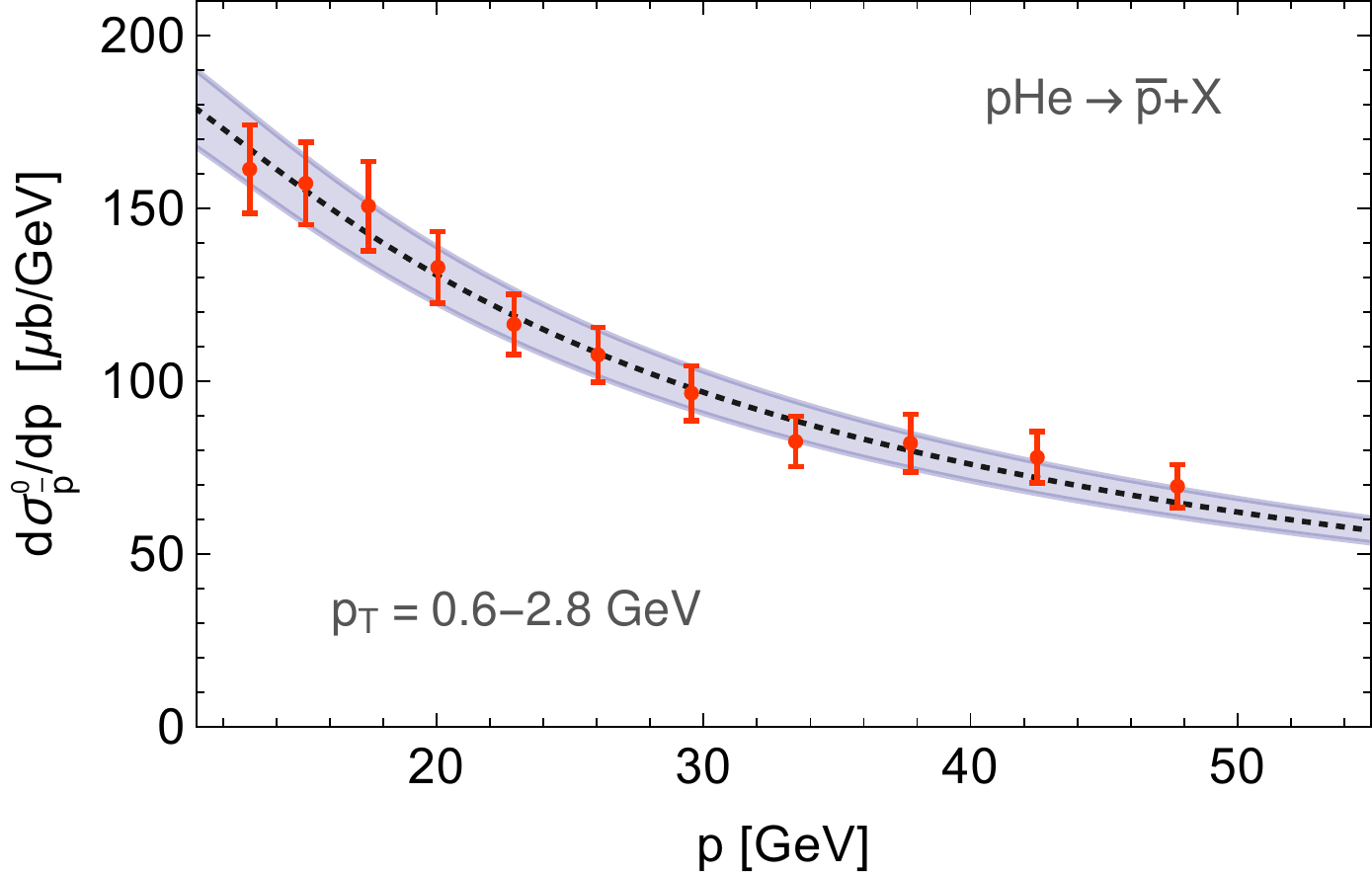}
\end{center}
\caption{Prompt antiproton production cross section in proton helium scattering. In the left panel the full differential cross section predicted in~\cite{Winkler:2017xor} is compared to the LHCb data. The right panel refers to the cross section integrated over $\pt=0.6-2.8\gev$. Uncertainties on the prediction (blue band) are derived from the uncertainties in the parameters $c_{1}-c_{16}$~\cite{Winkler:2017xor}.}
\label{fig:comparisonlhcb}
\end{figure}

The antiproton production in proton helium scattering has recently been measured for the first time with the LHCb-SMOG detector~\cite{LHCb:2017tqz} which provides an important test for the parameterization~\cite{Winkler:2017xor}. Incoming protons with energy $6.5\tev$ were scattered on a helium gas target at rest. The data refer to the prompt antiproton production which can be obtained by replacing $f_{\bar{p}}$ with $f_{\bar{p}}^0$ in~\eqref{eq:helium}. In addition, the target component (the term including $F_\text{tar}$) has to be multiplied by $(1+0.5\,\iso)$ to account for the difference between a proton and a mixed proton-neutron target.\footnote{Note that the factor $\iso$ does not appear in~\eqref{eq:helium}. While the relative number of produced antiprotons to antineutrons differs between a proton and neutron target if $\iso \neq 0$, the total number of antinucleons does not.} This factor is, however, almost negligible at the considered energy.
A detailed comparison of the prediction~\cite{Winkler:2017xor} with the LHCb data is provided in figure~\ref{fig:comparisonlhcb}. It can be seen that very good agreement is obtained for $\pt\leq 2\gev$. At higher $\pt$ the measured cross section somewhat exceeds the prediction. But as $\pt > 2\gev$ contributes $\lesssim 1\%$ to the total cross section this difference has negligible impact on the cosmic ray antiproton flux. Since the $\pt$-integrated cross section is in remarkably good agreement with the data (see figure~\ref{fig:comparisonlhcb}) we will employ the parameterization~\eqref{eq:helium} without modification.

\subsection{Positron Production Cross Section}

Secondary positrons, similar to antiprotons, mainly stem from proton proton scattering as well as from processes involving helium. In most cases positrons descend as final states from the decay chains of pions and kaons. There exist several parameterizations of the inclusive cross section for positron production in the literature. These rely either on analytic fits to experimentally measured meson spectra~\cite{Badhwar:1977,Tan:1984ha,Blum:2017iol} or on Monte Carlo simulation~\cite{Kamae:2006bf}. In our analysis, we will not attempt to fit the positron flux over the full energy range. Rather we will use the minimal secondary positron flux to constrain the size of the diffusion halo $L$. Different from the case of boron and antiprotons, we do not require an uncertainty band for positron production, but merely a robust lower limit. Therefore, we do not attempt to evaluate the production cross section ourselves and, instead, employ the parameterization of Kamae et al.~\cite{Kamae:2006bf}.\footnote{A number of typos have been corrected with the kind help of the authors.} It was pointed out in~\cite{Delahaye:2008ua} that this parameterization yields positron fluxes up to a factor of two smaller compared to~\cite{Badhwar:1977,Tan:1984ha} which will result in a conservative bound on $L$.

\subsection{Progenitors of Secondary Cosmic Rays}\label{sec:progenitors}

A robust parameterization of primary fluxes\footnote{We use a loose terminology here and call the progenitors of boron, antiprotons and positrons primaries. Strictly speaking the progenitor fluxes contain a secondary admixture themselves.} is another important ingredient in predicting the source terms of secondary cosmic rays. While antiprotons and positrons mainly stem from collisions involving proton and helium, the production of boron results from spallation of carbon, oxygen, nitrogen and subdominantly silicon, magnesium and neon. We model the interstellar fluxes as a function of rigidity $\R$ as
\begin{align}\label{eq:primariesFit}
\Phi_i^{\text{IS}}(\R)= \frac{\R}{\sqrt{\R^2+\R_{l,i}^2}} \,\alpha_i\,\left(\frac{\R}{\text{GV}}\right)^{-\gamma_i}\left(1+\left(\frac{\R}{\R_b}\right)^{\Delta\gamma/s}\right)^{s}\,.
\end{align}
The first factor on the right-hand side allows to fit the low energy part of the spectra, while $\alpha_i$ and $\gamma_i$ set the normalization and the power law index of the flux. The last term accounts for the observed spectral hardening at rigidity $\R_b$. As there is indication that primary cosmic rays share the position and form of the spectral break~\cite{Tomassetti:2015doz,Genolini:2017dfb}, we choose the parameters $\R_b$, $\Delta\gamma$ and $s$ to be universal among the considered species. The origin of the spectral break will be discussed in section~\ref{sec:diffusion_break}. The isotopic composition of fluxes, which is assumed to be rigidity-independent, is estimated from the cosmic ray data base. We neglect a deuteron contamination of the proton flux and set $^3\text{He}:\,\! ^4\text{He}= 0.15:0.85$, $^{12}\text{C}:\,\! ^{13}\text{C}= 0.93:0.07$, $^{14}\text{N}:\,\! ^{15}\text{N}= 0.55:0.45$, $^{16}\text{O}:\,\! ^{17}\text{O} : \,\!^{18}\text{O}= 0.96:0.02:0.02$. The Ne, Mg, Si fluxes, which contribute only subdominantly to boron production, are identified with the leading isotopes $^{20}\text{Ne}$, $^{24}\text{Mg}$ and $^{28}\text{Si}$ for simplicity.

\begin{figure}
	\centering
	\includegraphics[height=8cm]{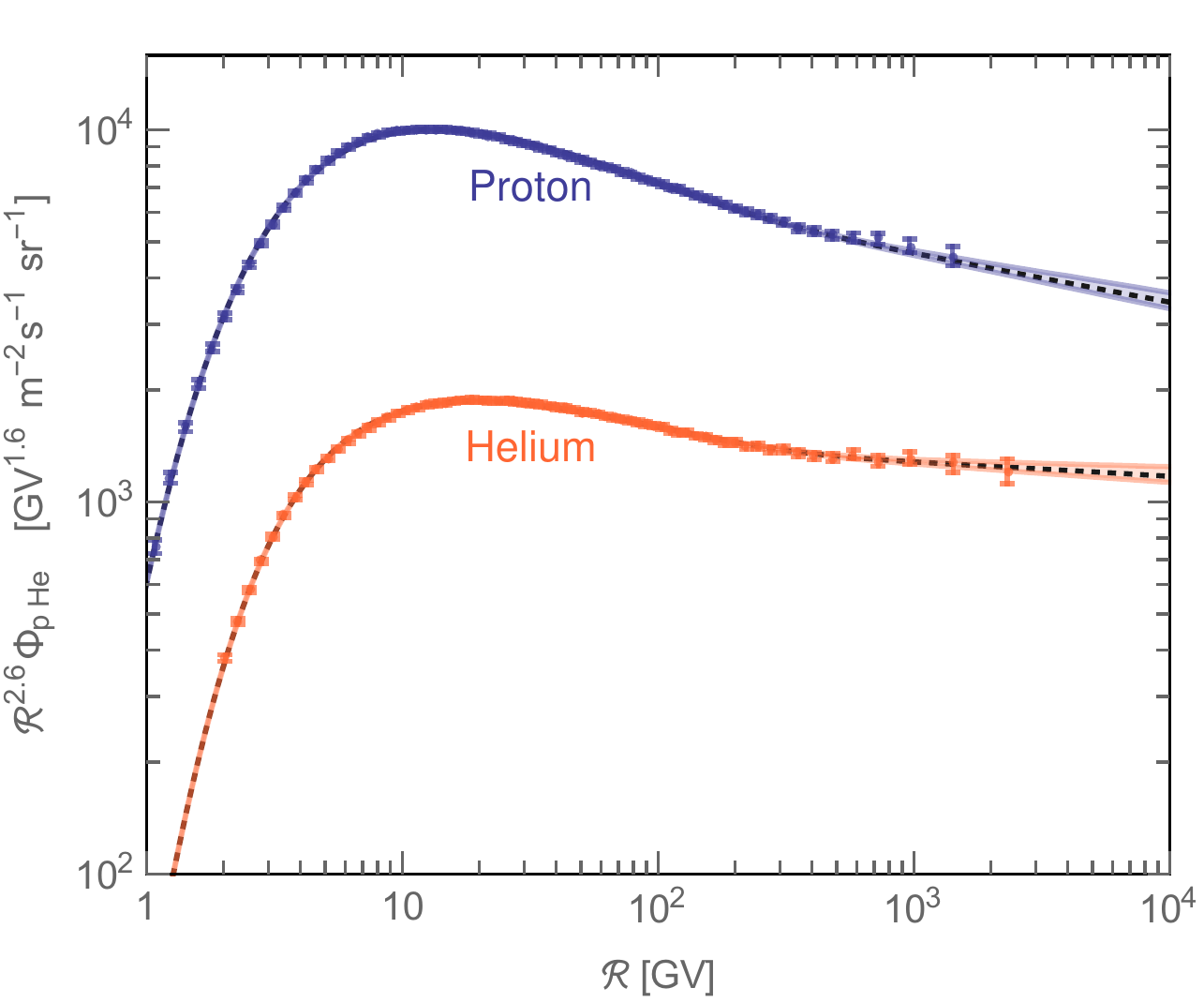}
	\caption{Proton and helium fluxes measured by AMS-02 and our fits with the corresponding uncertainty bands.}
	\label{fig:primaries}
\end{figure}

\begin{figure}
	\centering
  \includegraphics[height=9cm]{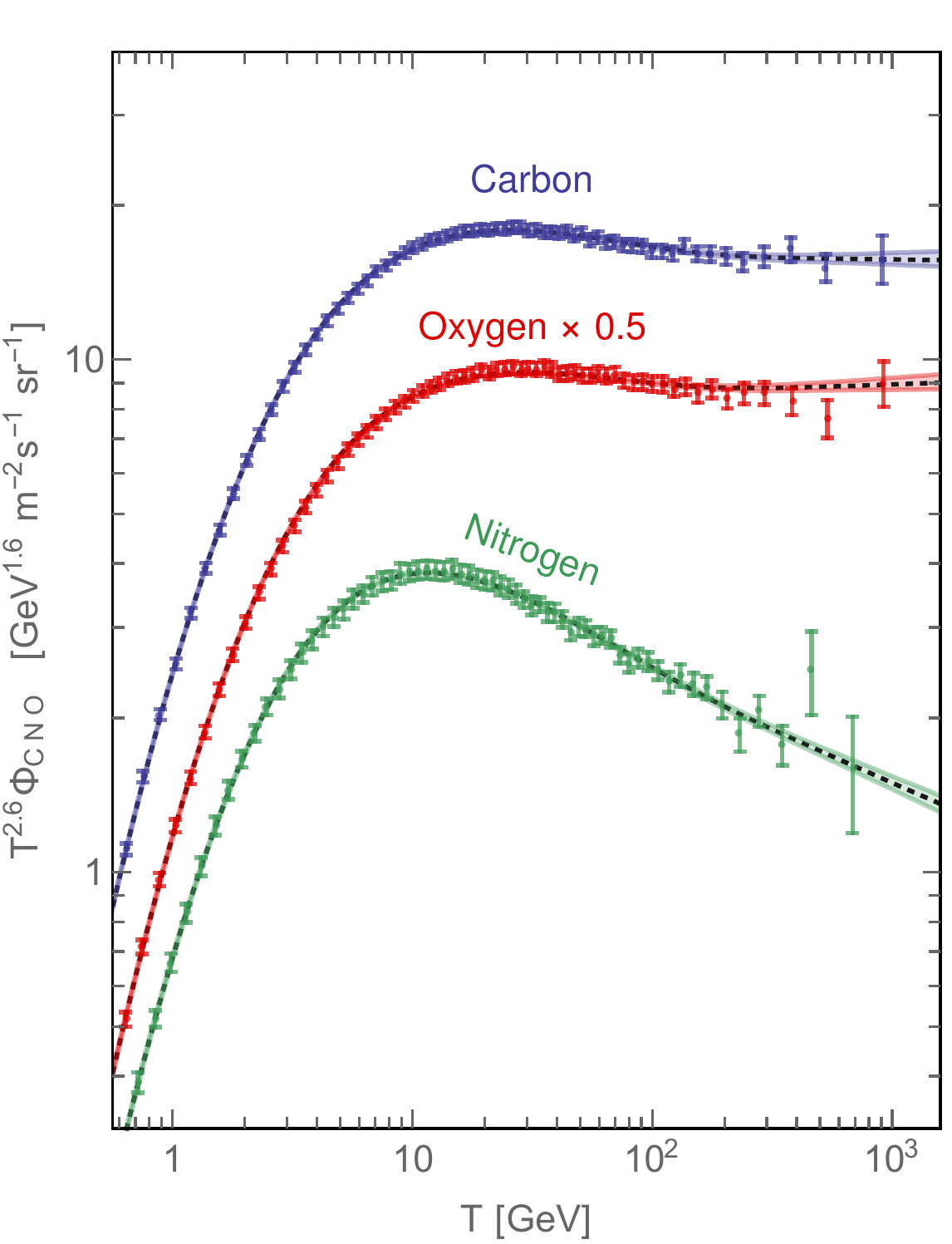}
  \hspace{4mm}
	\includegraphics[height=9cm]{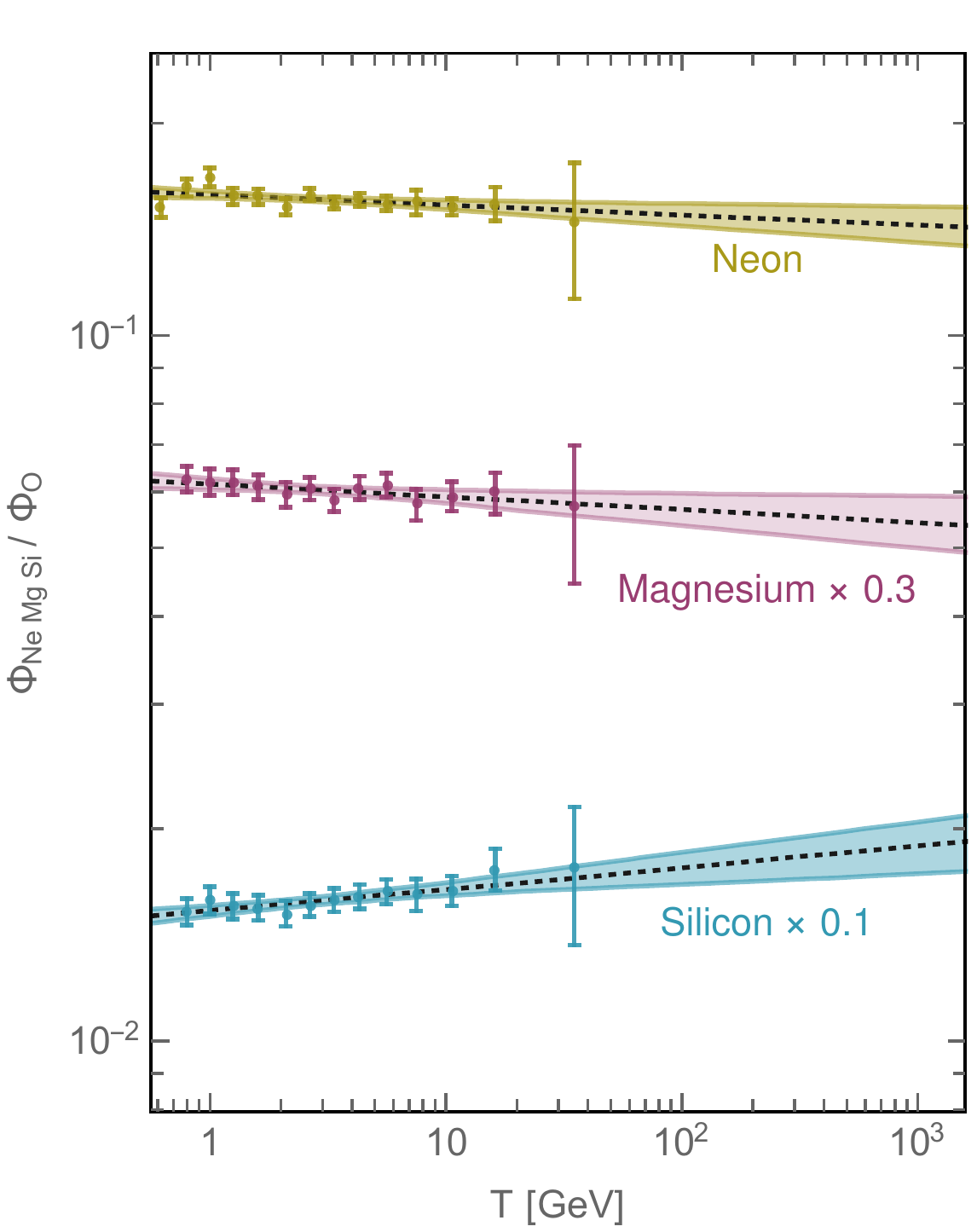}
	\caption{Fluxes of C, N, O measured by AMS-02 (left panel) and Ne/O, Mg/O, Si/O measured by HEAO (right panel). Our fits and the corresponding uncertainty bands are also shown.}
	\label{fig:nemgsi}
\end{figure}

The parameterization~\eqref{eq:primariesFit} was fit to the published (H, He) or preliminary (C, N, O) data of AMS-02~\cite{Aguilar:2015ooa,Aguilar:2015ctt,talkXSCR}, see figures~\ref{fig:primaries} and~\ref{fig:nemgsi}. In the case of C, N, O, fluxes are given in terms of the kinetic energy per nucleon and the parameterization was translated accounting for the isotopic composition. Uncertainty bands were determined in a slightly simplified two step procedure: first, we derived the probability distribution of the break parameters. Then we kept the break parameters fixed at their best fit values and determined the probability distributions for the individual $\alpha_i$, $\gamma_i$, $\R_{l,i}$, employing a $\Delta\chi^2$-metric with 3 d.o.f. for each species. In table~\ref{tab:primariesFit} we present the resulting median parameters and uncertainties. The large power-law index $\gamma_N$ is explained as the nitrogen flux carries a strong secondary component. For the parameters characterizing the spectral break, we obtain $\R_b = 275^{+23}_{-22}$ GeV, $\Delta\gamma= 0.157^{+0.020}_{-0.012}$ and $s = 0.074^{+0.008}_{-0.007}$. As the AMS-02 data refer to TOA fluxes, the interstellar fluxes depend on assumptions regarding solar modulation. In order to derive the values in table~\ref{tab:primariesFit}, we assumed the Fisk potential $\phi^+_{2,\text{\tiny{AMS-02}}}=0.72\gv$. Interstellar fluxes assuming any other Fisk potential for AMS-02 can still be parameterized using the values of table~\ref{tab:primariesFit}. One simply has to modulate the so-obtained fluxes with the difference between $0.72\gv$ and the true Fisk potential of AMS-02.

For Ne, Mg, Si, AMS-02 data are not yet available. As ratio data are less affected by systematic errors if one compares different experiments, we fit Ne/O, Mg/O, Si/O as extracted from HEAO~\cite{Engelmann:1990} with the function $\lambda_i (T/\text{GeV})^{\zeta_i}$. Fits are visualized in figure~\ref{fig:nemgsi}, parameter values and uncertainties are given in table~\ref{tab:primariesFit}. The absolute Ne, Mg and Si fluxes are then obtained by multiplying the ratios with the oxygen flux derived from AMS-02.

\begin{table}[htp]
\begin{center}
\begin{tabular}{|cccc|}
\hline
\rowcolor{light-gray}&&&\\[-3mm]
\rowcolor{light-gray} species&\(\alpha_i\, [\text{m}^{-2}\text{sr}^{-1}\text{s}^{-1}\text{GV}^{-1}]\)&\(\R_{l,i}\,[\text{GV}]\)&\(\gamma_i\)\\[1mm]
\hline 
&&&\\[-3mm]
$p$ & $(2.79\pm 0.01)\cdot 10^4$  & $2.74\pm 0.04$ & $2.889\pm 0.001$\\[2mm]
He & $(4.01\pm 0.02)\cdot 10^3$ & $2.97\pm 0.05$ & $2.795\pm 0.001$ \\[2mm]
C & $123\pm 1$ & $3.91\pm 0.09$ & $2.765\pm 0.002$\\[2mm]
O & $119\pm 1$ & $4.19\pm 0.11$ & $2.743^{+0.002}_{-0.003}$ \\[2mm]
N & $57\pm 1$ & $5.68\pm 0.16$ & $2.968^{+0.004}_{-0.003}$\\[2mm]
\hline
\rowcolor{light-gray}&&&\\[-3mm]
\rowcolor{light-gray} ratio& $\lambda_i$ & $\zeta_i$ &\\[1mm]
\hline 
&&&\\[-3mm]
Ne/O & $0.158 \pm 0.002 $  & $-0.01\pm 0.01$ & \\[2mm]
Mg/O & $0.205 \pm 0.004 $  & $-0.02\pm 0.01$ & \\[2mm]
Si/O & $0.153 \pm 0.003 $  & $0.03 \pm 0.01$ & \\[2mm]
\hline
\end{tabular}
\caption{Parameters entering the boron progenitor fluxes~\eqref{eq:primariesFit} (upper part) and parameters determining the ratios Ne/O, Mg/O, Si/O (lower part).}
\label{tab:primariesFit}
\end{center}
\end{table}

\subsection{Secondary Source Terms}

We can now use the derived cross sections and progenitor cosmic ray fluxes to determine the secondary source terms of boron, antiprotons and positrons. Source terms are defined as differential production rates per volume, time and energy,
\begin{equation}\label{eq:secsource}
 q_a^\text{sec} = \sum\limits_{i,j} 4\pi \int dT' \left(\frac{d\sigma_{ij\rightarrow a}}{dT}\right) \rho_{j} \Phi_i(T')\,,
\end{equation}
where $T'$ and $T$ are the kinetic energy per nucleon (or simply the kinetic energy in case of a lepton) of the incoming primary $i$ and the produced secondary particle $a$. It is sufficient to consider hydrogen and helium as targets $j$ in the galactic disc. We set $n_{\text{H}}=0.9\cm^{-3}$ and $n_{\text{He}}=0.9\cm^{-3}$ (see section~\ref{sec:diffusionmodel}). The incoming fluxes $\Phi_i$ and the differential cross sections $(d\sigma_{ij\rightarrow a}/dT)$ are taken from the previous sections. For the case of antiprotons one may equivalently use the cross section tables published in~\cite{Winkler:2017xor}. In the case of boron, we determine the source terms of the two isotopes $^{11}$B and $^{10}$B separately, assuming the isotopic composition of primaries specified in the previous section. In the first step, we used the cross sections derived in section~\ref{sec:boronproduction} for all isotopes of the same element (e.g. we assumed that the fragmentation cross section of $^{13}$C is identical to the one of $^{12}$C). In order to account for errors, we then applied a correction factor which was estimated from the Webber parameterization~\cite{Webber:2003}.\footnote{We employed~\cite{Webber:2003} and calculated $q_{\text{B}}^{\text{sec}}$ one time with isotope-dependent and one time with isotope-independent cross sections (using the cross sections of the leading isotopes). The correction factor corresponds to the ratio of the two.} We note, however, that the correction only amounts to $\sim 1\%$ and can as well be neglected. Finally, we have to include boron production though the radioactive decay of $^{10}$Be. For this purpose we simply add the source term of $^{10}$Be to the source term of $^{10}$B. This amounts to neglecting propagation effects on the abundance of $^{10}$Be which would result from its long lifetime. The corresponding error on the total boron flux can be estimated to be $\lesssim 2\%$~\cite{Maurin:2002ua}. We account for this by including an additional $2\%$ normalization uncertainty in the boron flux on top of the uncertainties related to cross sections and primary fluxes. 

\begin{figure}[htp]
\begin{center}
  \includegraphics[width=11cm]{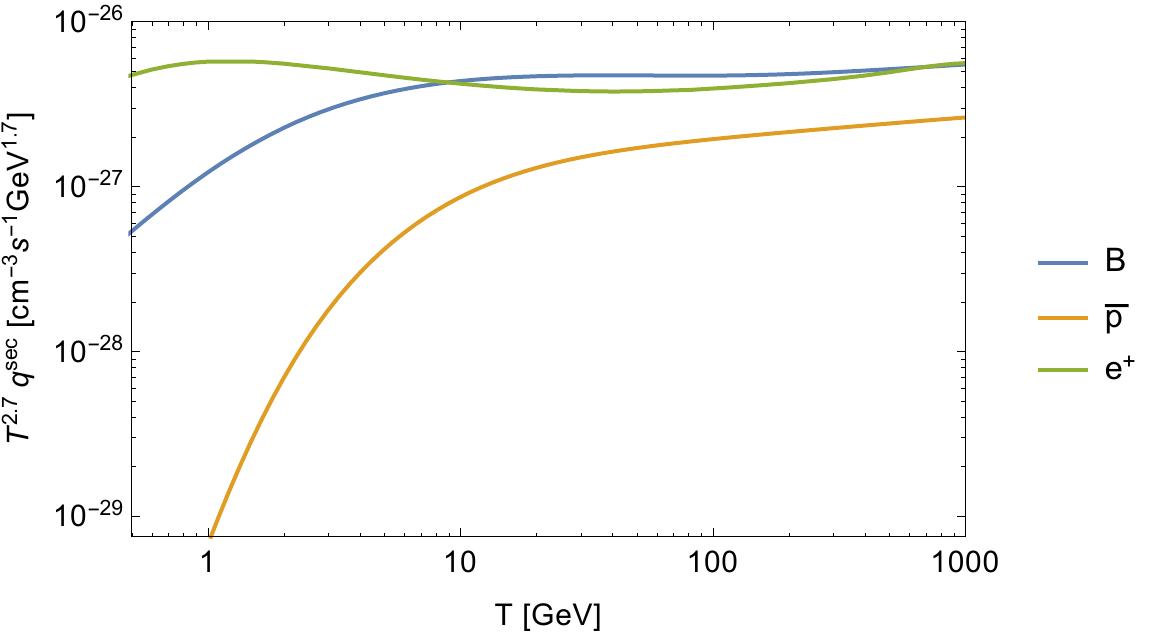}
\end{center}
\caption{Secondary source terms of boron, antiprotons and positrons.}
\label{fig:secondarysource}
\end{figure}

In figure~\ref{fig:secondarysource} we depict the secondary source terms of boron\footnote{The $^{10}$B source term contains a contribution from nucleon stripping of $^{11}$B. The latter can be calculated once the propagation parameters are fixed and the $^{11}$B-flux is known. In figure~\ref{fig:secondarysource} we used the best fit configuration of section~\ref{sec:diffusion_break}.}, antiprotons and positrons. It can be seen that $q_{\bar{p}}^{\text{sec}}$ decreases more rapidly towards low energies compared to $q_{\text{B}}^{\text{sec}}$ and $q_{e^+}^{\text{sec}}$ due to the higher threshold for antiproton production. At high energy, all source terms show an approximate power law behavior which is set by the progenitor fluxes. However, a slight increase of $q_{\bar{p}}^{\text{sec}}$ and $q_{e^+}^{\text{sec}}$ relative to $q_{\text{B}}^{\text{sec}}$ appears due to the violation of scaling which affects the antiproton and positron production cross sections (see section~\ref{sec:antiprotoncrosssection}).

\section{Primary Antiprotons from Dark Matter Annihilation}\label{sec:darkmatterpbar}

While the secondary production of antimatter is well-established, one may also speculate about a primary component due to dark matter annihilation in the galactic halo. In the case of positrons, astrophysical contributions beyond secondary production (e.g.\ from pulsars) may well add to the flux. This complicates the dark matter analysis as one would have to deal with two unknown components simultaneously. Therefore, we decided to focus on the antiproton signal from dark matter annihilation in this work. We will later employ positrons to constrain cosmic ray propagation parameters, but this will only require knowledge of the secondary positron flux.

The primary antiproton source term induced by dark matter annihilation reads
\begin{equation}\label{eq:primary}
  q^\text{prim}_{\bar{p}}  =\frac{\rho_{\text{DM}}^2}{m_{\text{DM}}^2}\, 
\frac{\langle \sigma 
v\rangle}{2}\,\frac{dN}{dT}\,, 
\end{equation}
where $dN/dT$ denotes the antiproton energy spectrum per annihilation process. We will consider $b\bar{b}$ and $WW$ as final states and extract $dN/dT$ for the two channels from~\cite{Cirelli:2010xx}. While this choice may seem selective, we note that antiproton spectra in hadronic channels exhibit a similar shape. If a dark matter signal was present in the experimental data, we would typically observe an excess in $\bar{b}b$ or $WW$ even if the true annihilation channel is not captured. The annihilation cross section $\langle \sigma v\rangle$ (averaged over the velocity distribution) determines the normalization of the primary signal. If dark matter is identified with a thermal WIMP, $\langle \sigma v\rangle$ is directly related to the dark matter density in the universe. For the canonical case of a velocity-independent $\sigma v$, the observed density corresponds to~\cite{Drees:2009bi}
\begin{equation}\label{eq:relic}
  \langle \sigma v\rangle \simeq 
10^{-26}\cm^3\s^{-1}\times\frac{1}{\sqrt{g_*(T_F)}}\:\frac{m_{\text{DM}}}{T_F}\,, 
\end{equation} 
where $T_F\simeq m_{\text{DM}}/20$ denotes the freeze-out temperature and $g_*$ the effective number of degrees of freedom which can be taken from~\cite{Laine:2006cp}.

The dark matter density profile $\rho_{\text{DM}}$ determines the normalization of the primary source term. While N-body simulations of cold dark matter suggest a Navarro-Frenk-White (NFW) profile~\cite{Navarro:1995iw}, the backreaction of baryons on the dark matter halo is a field of active research. The local dark matter density $\rho_0$ constitutes another source of uncertainty. As the propagation of antiprotons washes out local features in the dark matter profile, the relevant quantity is the dark matter density averaged over a larger ($\sim$kpc) scale. The latter is best assessed by global measurements which are converging at $\rho_0\sim 0.4\gev\cm^{-3}$ albeit with sizable uncertainties~\cite{Catena:2009mf,Salucci:2010qr,Pato:2015dua,McMillan:2016}.

In order to capture the uncertainties in $\rho_{\text{DM}}$, we consider three representative dark matter halos which follow a generalized NFW (gNFW) profile
\begin{equation}
  \rho_{\text{DM}} = \rho_0\, \left(\frac{R_0}{r}\right)^\gamma \left(\frac{R_0 +r_s}{r+r_s}\right)^{3-\gamma}\,,
\end{equation}
where $r$ denotes the distance from the galactic center, $R_0$ the distance between galactic center and sun and $r_s$ the scale radius. The parameter $\gamma$ determines the contraction of the profile. 
If not specified explicitly, we assume a standard NFW profile ($\gamma=1$) and $\rho_0=0.38\gev\cm^{-3}$~\cite{McMillan:2016}. We will, however, also present results for a more conservative cored profile ($\gamma=0$) with lower $\rho_0=0.3\gev\cm^{-3}$ as well as a more aggressive cuspy profile ($\gamma=1.3$) with $\rho_0=0.45\gev\cm^{-3}$. In order to arrive at self-consistent halo models we took $R_0=8.2\kpc$ and $r_s$ for the NFW profile from~\cite{McMillan:2016} and determined the scale radii for the other two cases by keeping the amount of dark matter fixed within the radius $r=50\kpc$. The parameters for the three profiles are summarized in table~\ref{tab:profiles}.

\begin{table}[htp]
\begin{center}
\begin{tabular}{|l|cccc|}
\hline
\rowcolor{light-gray}  &   &  &&   \\[-3mm]
\rowcolor{light-gray}  Profile & $\rho_0\;[\text{GeV}\cm^{-3}]$ & $r_s\;[\text{kpc}]$ & $\,R_0\;[\text{kpc}]\,$ & $\;\;\;\gamma\;\;\;$  \\[2mm]
\hline 
  &   & && \\[-3mm]
NFW & 0.38 & 18.6 & 8.2& 1  \\[2mm]
gNFW$_{\gamma=0}$ & 0.30 & 12.3 & 8.2 & 0  \\[2mm]
gNFW$_{\gamma=1.3}$ & 0.45 & 17.2 & 8.2 & 1.3  \\[2mm]
 \hline
\end{tabular}
\end{center}
\caption{Dark matter profiles considered in this work.}
\label{tab:profiles}
\end{table}

We comment that a number of hydrodynamic simulations of Milky Way type galaxies which include baryonic effects have been performed. Most of them hint at a dark matter profile with slope $\gamma \sim 1$ or slightly larger down to the innermost part ($r\lesssim 1\kpc$), where a flattening may occur~\cite{Guedes:2011ux,Roca-Fabrega:2015gma,Zhu:2015jwa,Schaller:2015mua} (see however~\cite{Mollitor:2014ara}). For the diffusion halos considered here, this suggests primary antiproton fluxes similar or larger than for the standard NFW profile. The cored profile with $\gamma=0$ is somewhat disfavored, but shall, nevertheless, be included for the sake of a conservative approach.

\section{Combined Analysis of Charged Cosmic Ray Data}

The AMS-02 experiment has provided high precision data on the antiproton flux and the B/C ratio. In this section we aim at investigating the consistency of the AMS-02 data with a secondary origin of the two species and at determining the favored propagation configuration. Experimental errors have shrunk to a level where they no longer dominate over uncertainties in the predicted fluxes. Therefore, we will carefully include uncertainties in the production of boron and antiprotons in our analysis. Since the secondary fluxes of boron and antiprotons are insensitive to a particular combination of propagation parameters, we will employ positrons in the last step, to lift this degeneracy.

\subsection{B/C and the Diffusion Break}\label{sec:diffusion_break}

The ratio of B/C in cosmic rays plays an important role in pinning down the propagation parameters. In the first step, we use B/C to investigate the origin of the spectral hardening observed in primary cosmic ray fluxes at rigidity $\R \gtrsim 200\gv$ (see section~\ref{sec:progenitors}).\footnote{See~\cite{Serpico:2015caa} for a summary of possible physics scenarios behind the spectral hardening of primary fluxes.} A plausible possibility is that this feature is already imprinted onto the primary source terms. In this case it might be related to non-linear effects in diffusive shock
acceleration~\cite{Ptuskin:2012qu} or to different types of sources contributing to the observed spectra~\cite{Ohira:2015ega}. Alternately, one may attribute the hardening to a propagation effect. The rigidity scaling of the diffusion term is set by the power spectrum of turbulences in the magnetic plasma. A break in the diffusion coefficient could arise due to the transition from diffusion on cosmic ray self-generated turbulence at low rigidity to diffusion on external turbulence at high rigidity~\cite{Blasi:2012yr}.
Similarly, an effective break results if the inner and outer part of the diffusion halo are dominated by turbulences of different type~\cite{Tomassetti:2012ga}. The following modification of the diffusion term has been suggested~\cite{Genolini:2017dfb}
\begin{equation}\label{eq:breakdiffusion}
 K = \frac{K_0 \,\beta \,\left(\frac{\mathcal{R}}{\text{GV}}\right)^\delta}{\left(1+\left(\frac{\R}{\R_b}\right)^{\Delta\delta/s}\right)^{s}}\,.
\end{equation}
The origin of the spectral hardening affects the spectra of secondary cosmic rays. If it is attributed to primary sources, boron would simply inherit the break from its progenitors at virtually the same rigidity (as fragmentation preserves $T$). In the regime where diffusion dominates the propagation, B/C would thus resemble a flat power law. If, however, the hardening relates to diffusion, the boron flux would be affected twice: by the progenitor fluxes and by its own propagation. Hence, a spectral break would be observable in B/C. In~\cite{Genolini:2017dfb} the high energy part of B/C was used to distinguish between the two hypotheses within a simplified propagation model. We consider it worth repeating this analysis over the full energy, since we can also benefit from the fragmentation cross sections and primary fluxes derived in this work. Before we proceed, we should mention a caveat: the shape of primary spectra has also been explained in terms of nearby sources~\cite{Thoudam:2011aa,Bernard:2012pia,Tomassetti:2015xem,Kachelriess:2017yzq}. In this case local primary spectra would be markedly different from spatially averaged spectra in the galactic disc and our analysis would not hold. On the other hand, it has been argued in~\cite{Blasi:2011fm,Salati:2016owk} that a significant local fluctuation in the primary fluxes is unlikely within the established propagation models.

We determine the boron flux, as outlined in section~\ref{sec:diffusionmodel}, for the standard diffusion term~\eqref{eq:standarddiffusion} and for the modified diffusion term~\eqref{eq:breakdiffusion}. For the latter, we fix $\mathcal{R}_b=275\gv$, $\Delta\delta=0.157$ and $s=0.074$ as required by the observed primary spectra. We checked that we can neglect uncertainties on these three parameters in our fits without changing results noticeably.\footnote{We will, however, include uncertainties in the break parameters of primary fluxes.} For both diffusion terms, the parameters controlling the interstellar boron flux are, thus, $K_0$, $\delta$, $L$, $V_c$, $V_a$ which must be determined by a fit to the AMS-02 data. Due to a degeneracy, there are indeed only four combinations of propagation parameters which enter, namely $K_0/L$, $\delta$, $V_c$, $V_a/\sqrt{L}$. In order to arrive at B/C at the top of the atmosphere, we account for solar modulation as described in section~\ref{sec:solarmodulation}. B/C remains virtually invariant within the considered range of Fisk potentials $\phi_{0,\text{\tiny{AMS-02}}}=0.6-0.72\gv$ (see footnote~\ref{footnote}). For our fits to converge we fix $\phi_{0,\text{\tiny{AMS-02}}}=0.72\gv$ without affecting the results.

We include the uncertainties related to the boron source term in the form of a covariance matrix $\Sigma^\text{B/C,source}$ which we determine as described in appendix~\ref{sec:appendix}. On top of $\Sigma_{ij}^\text{B/C,source}$, the experimental errors of AMS-02 are added. The full covariance matrix reads $
\Sigma_{ij}^\text{B/C} = \Sigma_{ij}^\text{B/C,source} + \left(\sigma_i^{\text{AMS}} \right)^2\delta_{ij}
$
with $\sigma_i^\text{AMS}$ denoting the quadratic sum of statistical and systematic errors in the $i$th bin. In the absence of detailed information provided by the AMS-02 collaboration, we took systematic errors to be uncorrelated.\footnote{Even if this approximation is oversimplistic, it is not expected to impact our fits dramatically as uncertainties contained in $\Sigma_{ij}^\text{source}$ exceed systematic errors of AMS-02.} The $\chi^2$ test statistic is defined as
\begin{equation}\label{eq:chibc}
 \chi^2_{\text{B/C}} = \sum\limits_{i,j=1}^{67} \Big((B/C)_i - (B/C)_{i}^{\text{AMS}}\Big)  \left({\Sigma^\text{B/C}}^{\:-1}\right)_{ij} \Big((B/C)_j - (B/C)_{j}^{\text{AMS}}\Big)\,,
\end{equation}
where $(B/C)_{i}^{\text{AMS}}$ denotes the measured ratio in bin $i$.

\begin{figure}[htp]
\begin{center}
  \includegraphics[height=7.7cm]{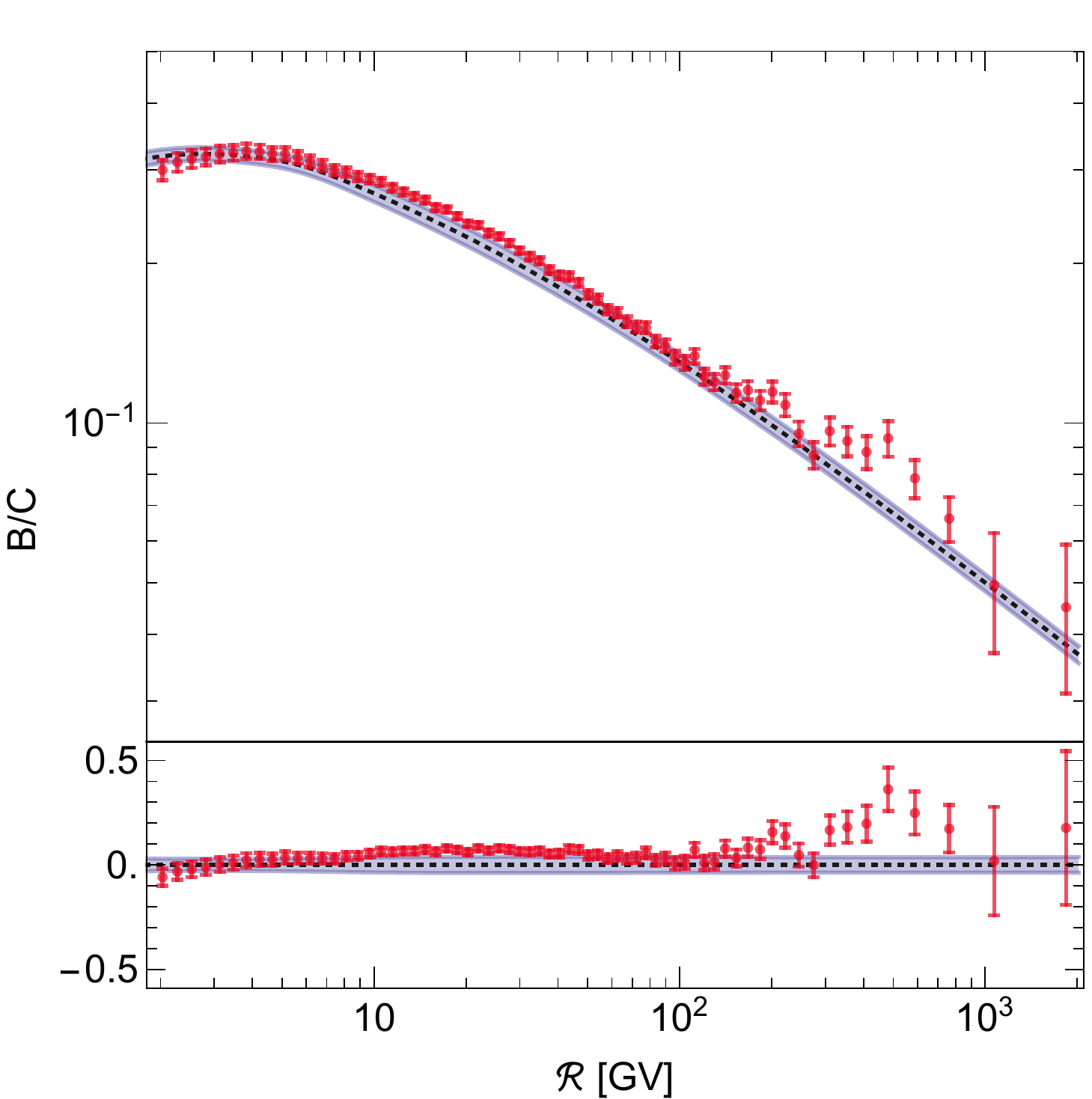}\hspace{3mm}
  \includegraphics[height=7.7cm]{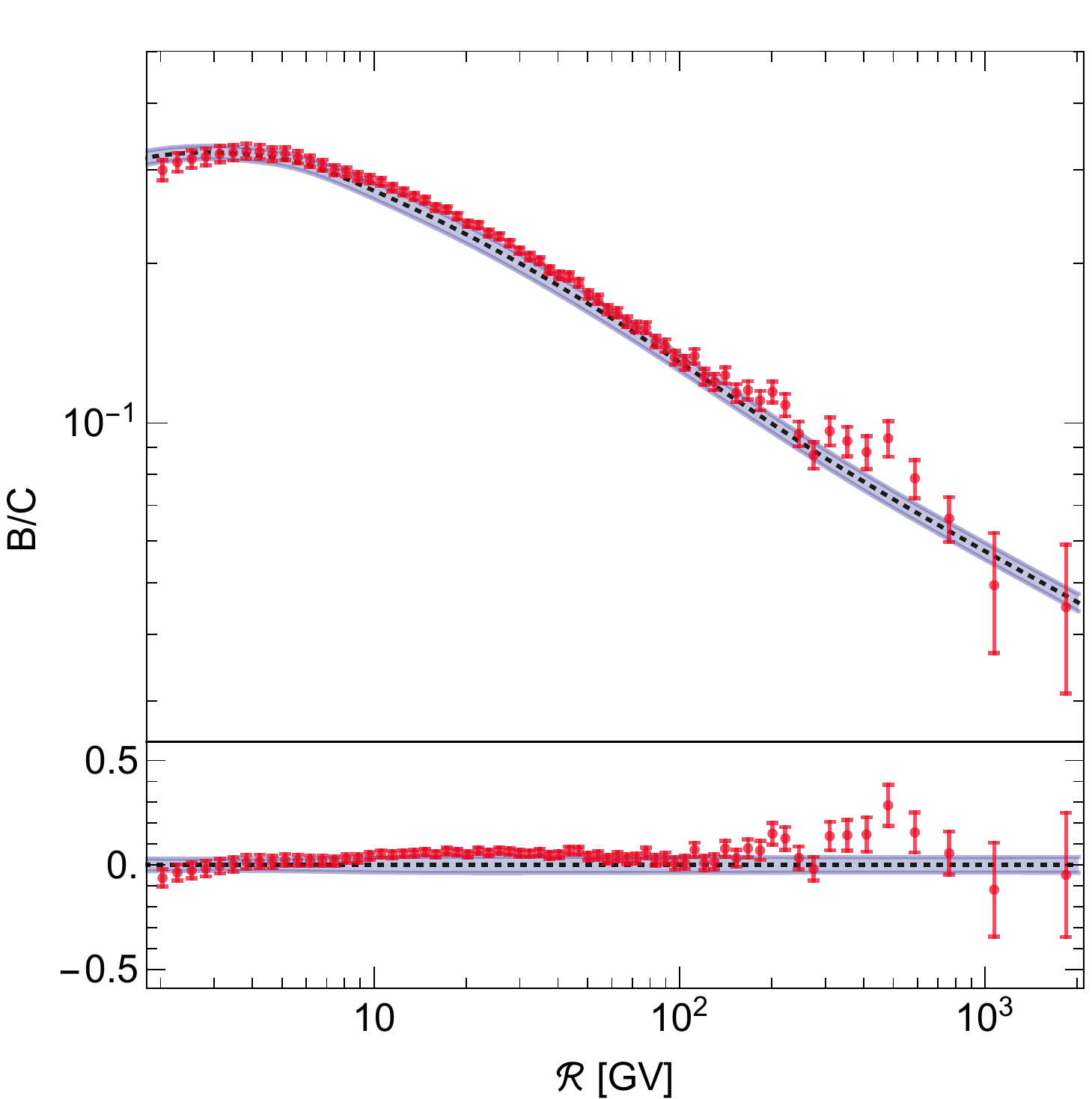}\\
\end{center}
\caption{Fit to the AMS-02 B/C spectrum assuming standard diffusion (left panel) and assuming a spectral break in the diffusion coefficient (right panel). Residuals are shown in the lower subpanels.}
\label{fig:bc}
\end{figure}

In table~\ref{tab:fitsecondary} we provide the best fit propagation parameters separately for the diffusion terms~\eqref{eq:standarddiffusion} and~\eqref{eq:breakdiffusion}. The corresponding B/C ratios compared to the AMS-02 data are shown in figure~\ref{fig:bc}. The error bands indicate the diagonal part of the boron production uncertainties contained in $\Sigma_{ij}^\text{B/C,source}$. We note, however, that $\Sigma_{ij}^\text{B/C,source}$ carries a strong degree of correlation, in particular the high energy part.\footnote{When determining $\chi^2_{\text{B/C}}$ we, of course, took into account the full covariance matrix.} Both fits yield an acceptable $\chi^2_{\text{B/C}}$, but we observe that a substantially better fit is obtained with the modified diffusion term including the break ($\Delta\chi^2_{\text{B/C}} =16.2$). As the break in the modified diffusion term was fixed by primary fluxes, the improvement comes without the cost of additional free parameters. We now want to quantify the preference for a diffusion break within frequentist statistics. For this purpose we generated a large number of mock data sets under the hypothesis of standard diffusion.\footnote{We performed a Cholesky decomposition of $\Sigma_{ij}^\text{B/C}$ to generate mock data with correlated uncertainties.} For each mock data set we determined the minimal $\chi^2_{\text{B/C}}$ for standard diffusion and for modified diffusion allowing $K_0$ and $\delta$ to float. 
The chance for an accidental improvement as large as $\Delta\chi^2_{\text{B/C}} \geq 16.2$ with modified diffusion turns out to be 1/250000.
Formally, this corresponds to a $4.5\,\sigma$ exclusion of standard diffusion against the alternative hypothesis of modified diffusion. Our results are in good agreement with~\cite{Genolini:2017dfb} and indicate strong preference for a break in the diffusion coefficient. In the following, we will hence fix the diffusion term to the form of~\eqref{eq:breakdiffusion}. Upcoming cosmic ray data on other secondary nuclei will provide important tests for this assumption. The strong spectral break observed in the lithium flux~\cite{talkXSCR} may already be seen as another hint for a diffusion break.

Turning to the other propagation parameters, we observe that B/C does not favor significant convection or reacceleration effects. Seizable reacceleration velocities $V_a\gtrsim 20\ \text{km/s}$ tend to produce a bump in B/C at rigidities of a few GV which is not observed in the AMS-02 data. As can be seen, there is a shape in the residuals (see figure~\ref{fig:bc}) which is not unexpected in the presence of correlated errors. If anything, it indicates some difficulty in reproducing a sufficiently concave B/C spectrum at low rigidity. While convective winds lead to reduction of the flux at $\mathcal{R}\lesssim 100\gv$ -- seemingly going in the right direction -- this decrease is too smooth over rigidity to improve the fit. It would be interesting to explore if convection can affect the boron flux more favorably in the presence of a non-vanishing spatial gradient in the convective wind. The slope of the diffusion coefficient was kept as a free parameter in our analysis. However, special mention should be made of the fit values $\delta=0.507$ and $\delta+\Delta\delta=0.35$ below and above the spectral break. These values are remarkably close to 0.5 and 0.33 corresponding to a Kraichnan~\cite{Kraichnan:1965zz} and Kolmogorov~\cite{Kolmogorov:1941} spectrum of turbulence. If the break in the diffusion coefficient results indeed from the interplay between turbulences of two types~\cite{Blasi:2011fm}, one may wonder if they can be related to the Kraichnan and Kolmogorov theories respectively.

\subsection{Antiproton and B/C Fit}\label{sec:pbarandbc}

\begin{figure}[htp]
\begin{center}
  \includegraphics[height=7.7cm]{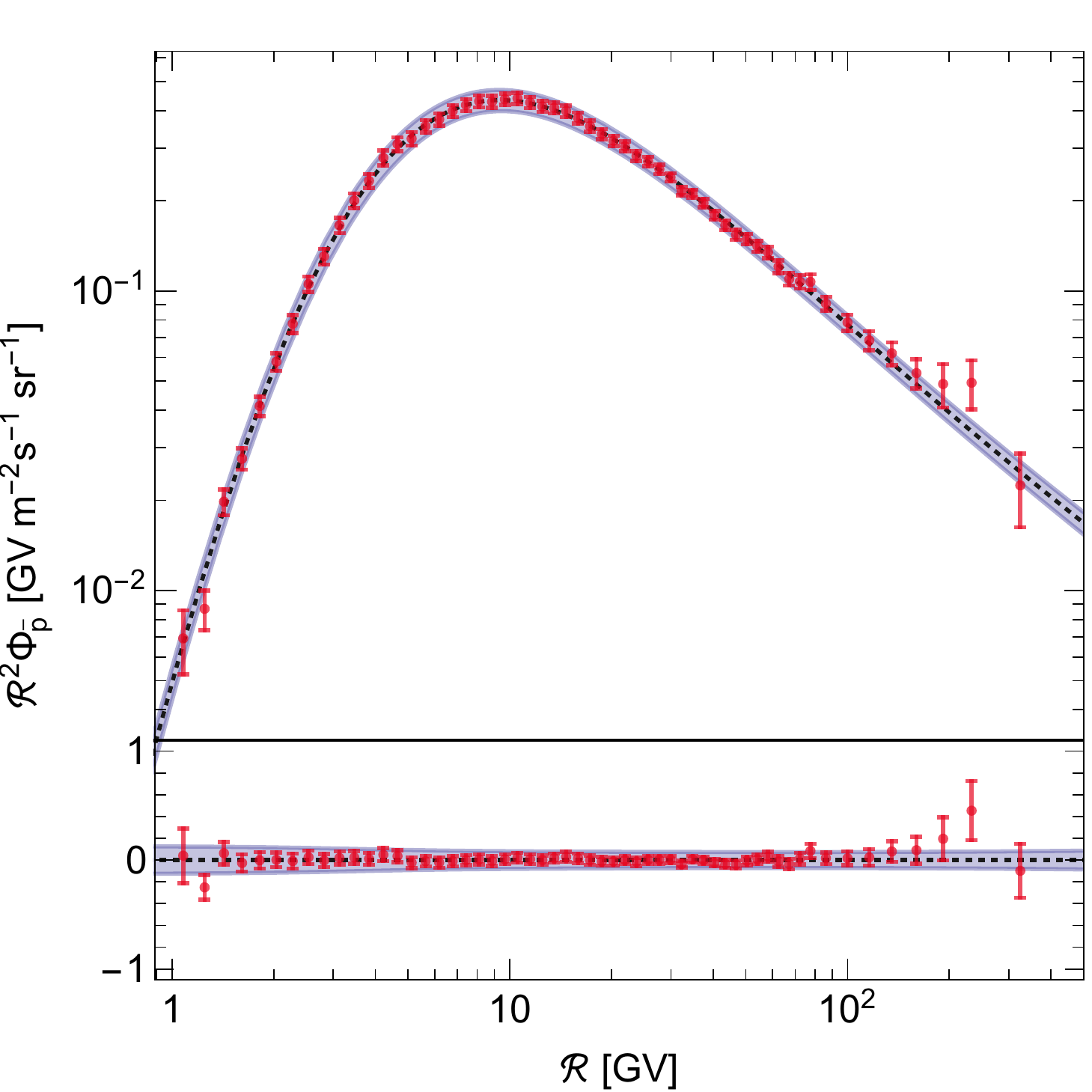}\hspace{3mm}
  \includegraphics[height=7.7cm]{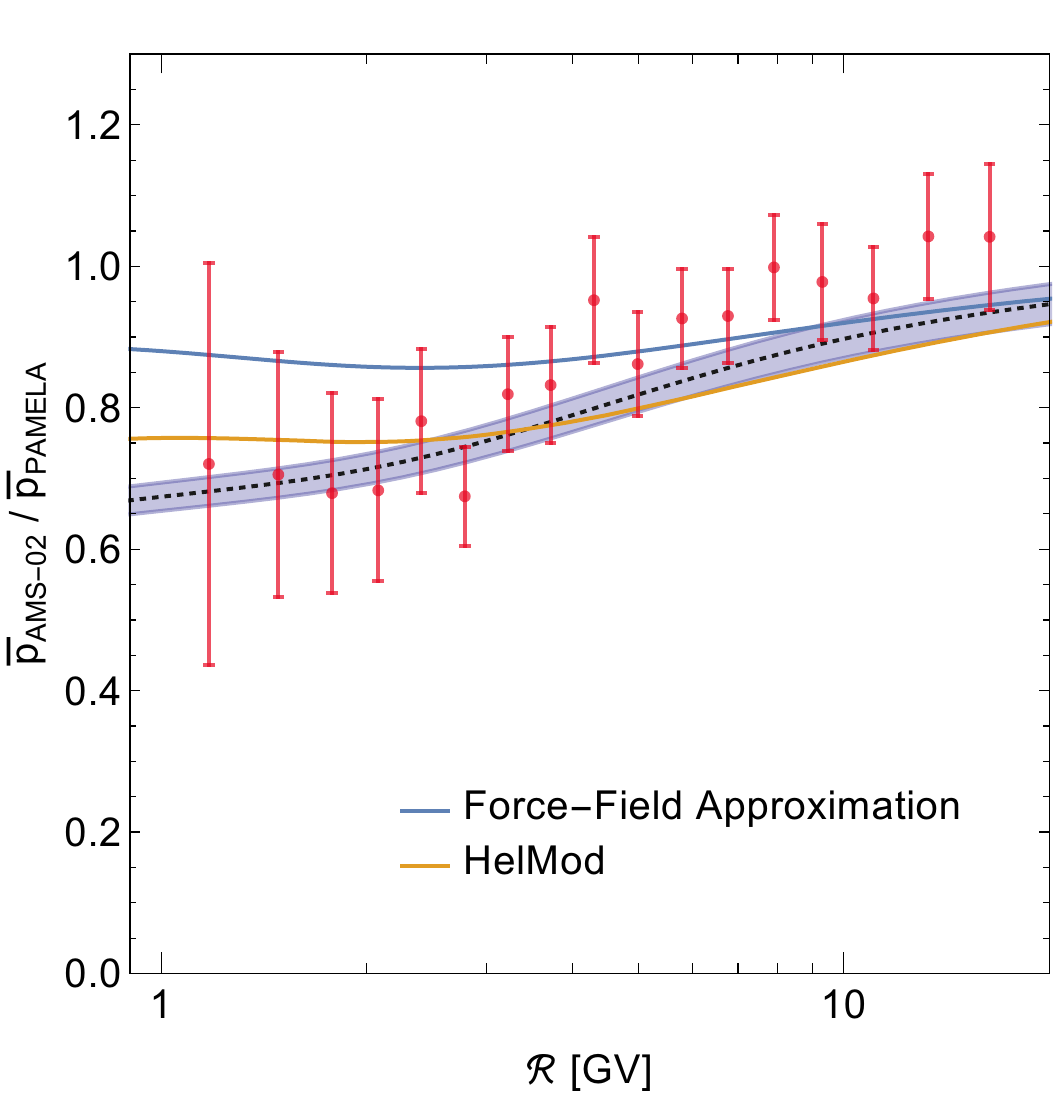}
\end{center}
\caption{Fit to the AMS-02 antiproton spectrum (left panel) and the ratio of antiproton spectra observed at AMS-02 and PAMELA (right panel). The ratio depends on the assumptions of solar modulation and is also shown for the standard force field approximation and for the HelMod model.}
\label{fig:pbar}
\end{figure}

At the time when the first AMS-02 antiproton data were released~\cite{tingtalk:2015}, the spectrum was considered surprisingly hard. This triggered speculations about a possible primary component due to dark matter annihilations~\cite{Hamaguchi:2015wga,Lin:2015taa,Chen:2015kla}. But subsequent analyses revealed that the secondary antiproton flux had been underestimated and might indeed account for the shape of the observed spectrum. An important role is played by the increase of antiproton production cross sections due to scaling violation~\cite{Kachelriess:2015wpa,Winkler:2017xor}. We now want to extend on~\cite{Winkler:2017xor} and investigate whether the AMS-02 antiproton spectrum is consistent with secondary production at the precision level. We rigorously include uncertainties in the antiproton source term in the form of the covariance matrix $\Sigma^\text{$\bar{p}$,source}$. The latter is obtained in complete analogy to $\Sigma^\text{B/C,source}$ by mapping uncertainties in cross section and primary flux parameters into uncertainties in the antiproton flux (see appendix~\ref{sec:appendix}). The full covariance matrix $\Sigma^\text{$\bar{p}$}$ includes the experimental errors of AMS-02 and enters the $\chi^2$-test as in~\eqref{eq:chibc}. We account for solar modulation via the Fisk potential $\phi^-_{\text{\tiny{AMS-02}}}$ defined in~\eqref{eq:fiskpotentials}.
The two force field parameters $\phi_{0,\text{\tiny{AMS-02}}}$ and $\phi_{1,\text{\tiny{AMS-02}}}^-$ are allowed to float, but we impose $\phi_{0,\text{\tiny{AMS-02}}}=0.6-0.72\gv$ for consistency with Voyager. The propagation parameters and the solar modulation parameters are determined by a simultaneous fit to the AMS-02 antiproton data~\cite{Aguilar:2016kjl} and to the ratio of antiproton spectra observed at AMS-02 and PAMELA~\cite{Adriani:2012paa} (see section~\ref{sec:solarmodulation}).\footnote{As uncertainties in the ratio are strongly dominated by PAMELA we can treat $\bar{p}_{\text{\tiny{AMS-02}}}/\bar{p}_{\text{\tiny{PAMELA}}}$ as independent.} 
We included an additional 2\% correlated normalization uncertainty in our fit to $\bar{p}_{\text{\tiny{AMS-02}}}/\bar{p}_{\text{\tiny{PAMELA}}}$ which accounts for a small systematic offset observed between PAMELA and AMS-02 at high energy. The best fit parameters can be found in table~\ref{tab:fitsecondary}. The corresponding spectra and uncertainty bands are visualized in figure~\ref{fig:pbar}.

The quality of the fit to the antiproton spectrum is remarkably good, we obtain $\chi_{\bar{p}}^2=21.3$ for 57 rigidity bins.\footnote{While this value of $\chi^2$ may look suspiciously small, we remind the reader that the absolute $\chi_{\bar{p}}^2$ does not have a rigorous statistical interpretation until the systematic errors of AMS-02 have been taken into account.} Residuals have no particular shape and are controlled by statistical fluctuations. The shape of $\bar{p}_{\text{\tiny{AMS-02}}}/\bar{p}_{\text{\tiny{PAMELA}}}$ is also well reproduced up to a slight offset. The latter is likely caused by systematics between PAMELA and AMS-02. The charge asymmetric term $\phi_{1,\text{\tiny{AMS-02}}}^-$ in the Fisk potential is crucial in accounting for the observed decrease in the ratio towards low rigidity. This is in contrast to the standard force field approximation ($\phi_{1,\text{\tiny{AMS-02}}}^-=0$), which predicts a nearly constant $\bar{p}_{\text{\tiny{AMS-02}}}/\bar{p}_{\text{\tiny{PAMELA}}}$. As a cross check for our treatment of solar modulation we also calculated the ratio of fluxes with the HelMod code and found reasonable agreement with our fit (see figure~\ref{fig:pbar}). 

While one is tempted to interpret the good fit to be in favor of a secondary origin of antiprotons, the comparison with B/C is still to be made. Within our assumptions on cosmic ray propagation the antiproton flux and B/C should be explainable by an identical set of propagation parameters. If one compares the best fit propagation parameters of table~\ref{tab:fitsecondary} it is striking that antiproton data -- in contrast to B/C -- favor a large reacceleration velocity. While such high $V_a$ would have considerable impact on the antiproton spectrum, the effect on the boron flux would be more dramatic.  Due to the lower threshold energy for boron compared to antiproton production, there is more low energy boron available which can be reshuffled to high energies through reacceleration (see figure~\ref{fig:secondarysource}).
Large $V_a$ leads to a bump in B/C -- not seen in the AMS-02 data. In order to investigate the compatibility further, we perform a simultaneous fit to the B/C and antiproton spectra of AMS-02. Again we include $\bar{p}_{\text{\tiny{AMS-02}}}/\bar{p}_{\text{\tiny{PAMELA}}}$ to constrain solar modulation.

\begin{figure}[htp]
\begin{center}
  \includegraphics[height=7.6cm]{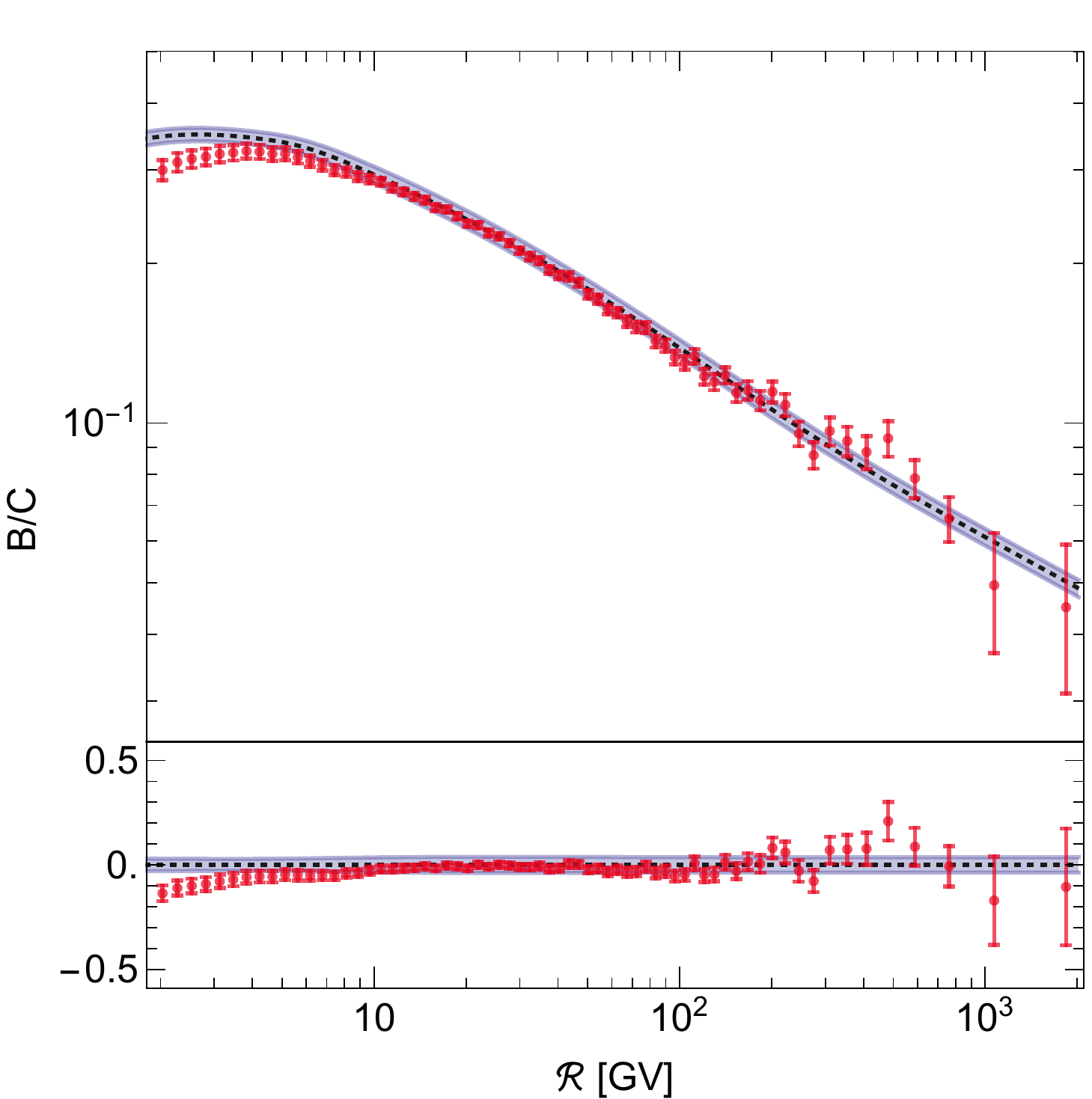}\hspace{3mm}
  \includegraphics[height=7.6cm]{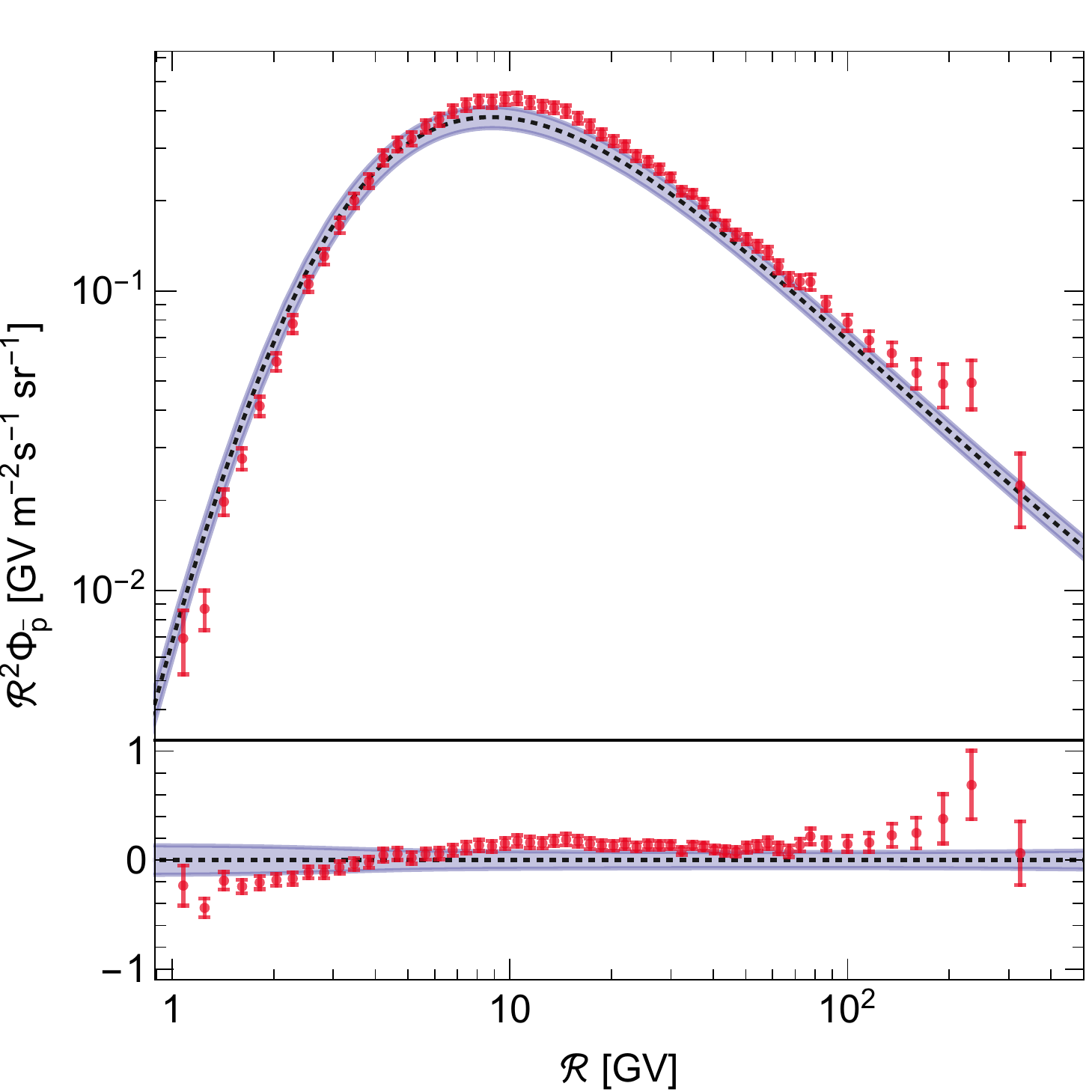}\\[2mm]
  \includegraphics[width=7.8cm]{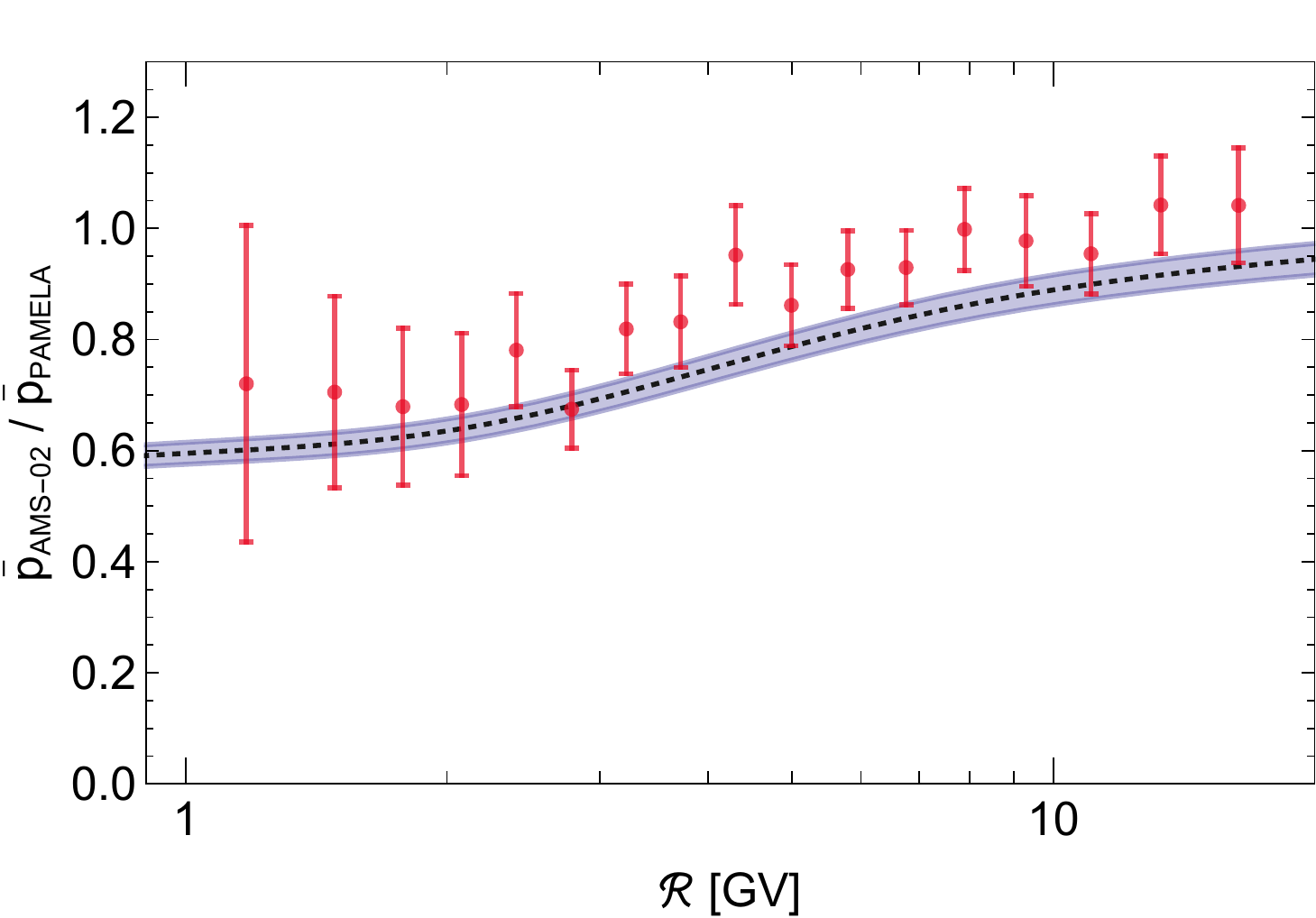}
\end{center}
\caption{Best fit spectra of the combined B/C + $\bar{p}$ fit.}
\label{fig:bcandpbar}
\end{figure}

The favored parameters of the joined fit are shown in the last column of table~\ref{tab:fitsecondary}, the corresponding fluxes and uncertainties are depicted in figure~\ref{fig:bcandpbar}. Remarkably, B/C and antiprotons can be fit simultaneously with $\chi^2/\text{d.o.f.}<1$. This implies that both spectra are, indeed, consistent with pure secondary production. The fit is considerably better than one may conclude by eye due to correlations in the uncertainties in $\Sigma_{ij}^\text{source}$. Nevertheless, we observe a clear rise in $\chi_{\bar{p}}^2$ compared to the fit without B/C. In the high energy regime, there appears a slight offset between predicted antiproton flux and data which is, however, within the margin of cross section uncertainties. The increase in $\chi_{\bar{p}}$ is indeed mainly driven by the low energy spectrum. The combined fit picks a reacceleration velocity significantly lower than the fit with antiprotons alone. This manifests itself in the antiproton flux exceeding the data at $\R < 4 \gv$ and falling short above. The residuals exhibit a modest peak at rigidity $\R\sim 10 \gv$. The 12 AMS-02 bins at $\R=7-20\gv$ increase $\chi^2_{\bar{p}}$ by $\sim 10$ compared to the pure antiproton fit. This increase is not larger since cross section uncertainties can partly explain a peak: proton proton scattering resides in a scaling regime at $\sqrt{s}\simeq 10 - 50\gev$. However, at $\sqrt{s}< 10\gev$ near-threshold effects contained in the function $R$~\eqref{eq:nearthreshold} play a role. Lowering of the parameter $c_9$ within uncertainties would manifest itself in a decrease of the antiproton spectrum at $R\lesssim 10\gv$. Simultaneously, the asymmetry between antineutron and antiproton production (cf.~\eqref{eq:isofit}) may increase the antiproton flux at $\R\lesssim 30\gv$ compared to the median parameter choice. Among other possibilities, the interplay between these two effects could lead to a smooth bump in the residuals at $\R\sim 10\gv$. Despite the overall consistency of B/C and antiprotons with standard secondary production, there is still room for modifications. If an alternative hypothesis is able to capture the shape of residuals, it may still be statistically preferred at a significant level. In section~\ref{sec:darkmatter} we will explore whether a significant improvement of the fit arises in the presence of a primary antiproton component from dark matter annihilation.

\begin{table}[htp]
\begin{center}
\begin{tabular}{|l|cccc|}
\hline
\rowcolor{light-gray}  &   &  &   & \\[-3mm]
\rowcolor{light-gray} Best Fit & \parbox[c]{2.1cm}{\centering{B/C\\[-1mm]\scriptsize{(w/o break)}}}& \parbox[c]{2.3cm}{\centering{B/C\\[-1mm]\scriptsize{(w/ break)}}} &
\parbox[c]{2.3cm}{\centering{\pbar\\[-0.5mm]\scriptsize{(w/ break)}}} & \parbox[c]{2.3cm}{\centering{B/C + \pbar\\[-1mm]\scriptsize{(w/ break)}}} \\[3mm]
\hline
  &   & & &\\[-3mm]
$K_0\;[\frac{\text{kpc}^2}{\text{Gyr}}]$ & $39.6\cdot L_{4.1}$ & $34.3\cdot L_{4.1}$ & $39.5\cdot L_{4.1}$ & $32.5\cdot L_{4.1}$ \\[2mm]
$\delta$ & 0.479 & 0.507& 0.446 & 0.506 \\[2mm]
$V_a\;[\frac{\text{km}}{\text{s}}]$ & 0 & 0 & $59.7\cdot\sqrt{L_{4.1}}$ & $15.6\cdot\sqrt{L_{4.1}}$ \\[2mm]
$V_c\;[\frac{\text{km}}{\text{s}}]$ & 0 & 1.3 & 0 & 0 \\[2mm]
$\Delta\delta$ & \multirow{3}{1cm}{\rotatebox{65}{no break $\;$}} & 0.157& 0.157 & 0.157 \\[2mm]
$\mathcal{R}_b\;[\text{GV}]$& & 275 & 275 & 275\\[2mm]
$s$ &   & 0.074 & 0.074 & 0.074 \\[2mm]
$\phi_0\;[\text{GV}]$ & 0.72  & 0.72 & 0.72 & 0.72 \\[2mm]
$\phi_1\;[\text{GV}]$ &       &  & 0.66 & 0.84 \\[2mm]
\hline
  &   & & &\\[-3mm]
$\chi^2_{\text{B/C}}$ (67 bins)  & 64.2 & 48.0  & & 55.1  \\[2mm]
$\chi^2_{\bar{p}}$ (57 bins)  &   &  &21.3  & 47.9   \\[2mm]
$\chi^2_{\text{AMS/PAM}}$ (17 bins)& & & 10.9 & 12.6 \\[2mm]
\hline
\end{tabular}
\end{center}
\caption{Best fit propagation and solar modulation parameters corresponding to the B/C, the $\bar{p}$ and the combined B/C + $\bar{p}$ fit. For the B/C fit the cases without and with diffusion break are considered. The goodness of fit is indicated by the $\chi^2$-values. The acronym $L_{4.1}$ stands for $L / 4.1\kpc$.}
\label{tab:fitsecondary}
\end{table}

\subsection{Positron Constraints on the Diffusion Halo}\label{sec:positron}

The cosmic ray positron flux experiences a spectral hardening at $\R \gtrsim 10 \gv$ which has been established by the PAMELA collaboration~\cite{Adriani:2008zr}. Within standard assumptions, this shape cannot be explained by secondary production. However, it was pointed out that the excess can be reconciled with secondary positrons if one employs a mechanism to avoid energy losses in the galactic halo~\cite{Katz:2009yd,Blum:2013zsa,Lipari:2016vqk,Blum:2017iwq}. Furthermore, there exist attempts to describe positrons as secondaries within modified diffusion models~\cite{Cowsik:2013woa,Kappl:2016qug} or through invoking acceleration of secondaries in supernova remnants~\cite{Blasi:2009hv,Mertsch:2009ph,Mertsch:2014poa}. Alternatively, a primary contribution to the flux may resolve the positron puzzle. While a dark matter interpretation is difficult to reconcile with complementary indirect detection probes~\cite{Abdo:2010dk,Ackermann:2011wa,Galli:2009zc,Slatyer:2009yq,Huetsi:2009ex}, pulsars might account for the excess without conflicting other observations~\cite{Aharonian:1995zz,Hooper:2008kg} (see however~\cite{Abeysekara:2017old}).
In this section we will assume standard propagation of positrons but will otherwise stay agnostic about the origin of the positron anomaly. Independent of which contributions are added to the positron flux, the secondary background alone must not overshoot the data of AMS-02. 

Although our focus is on the antiproton flux from dark matter annihilation, positrons will still play an important role in the dark matter analysis. The primary antiproton flux strongly depends on the size of the diffusion zone $L$. This is easily understood as primary antiprotons originate from everywhere in the dark matter halo, but only those inside $L$ may ever reach the earth. The secondary spectra of antiprotons and B/C only constrain the combination $L/K_0$, not the absolute size $L$. Positrons can lift this degeneracy. Different from hadronic cosmic rays, positrons experience energy losses in the diffusion halo which limits the distance from which positrons reach the earth. As a consequence, they are not very sensitive to the boundaries of the diffusion halo, but rather to the diffusion coefficient $K_0$. The fact that hadronic and positron secondary fluxes depend on different combinations of $K_0$ and $L$ can be used to efficiently constrain $L$~\cite{Lavalle:2014kca,Boudaud:2016jvj}.

\begin{figure}[htp]
\begin{center}
  \includegraphics[width=13cm]{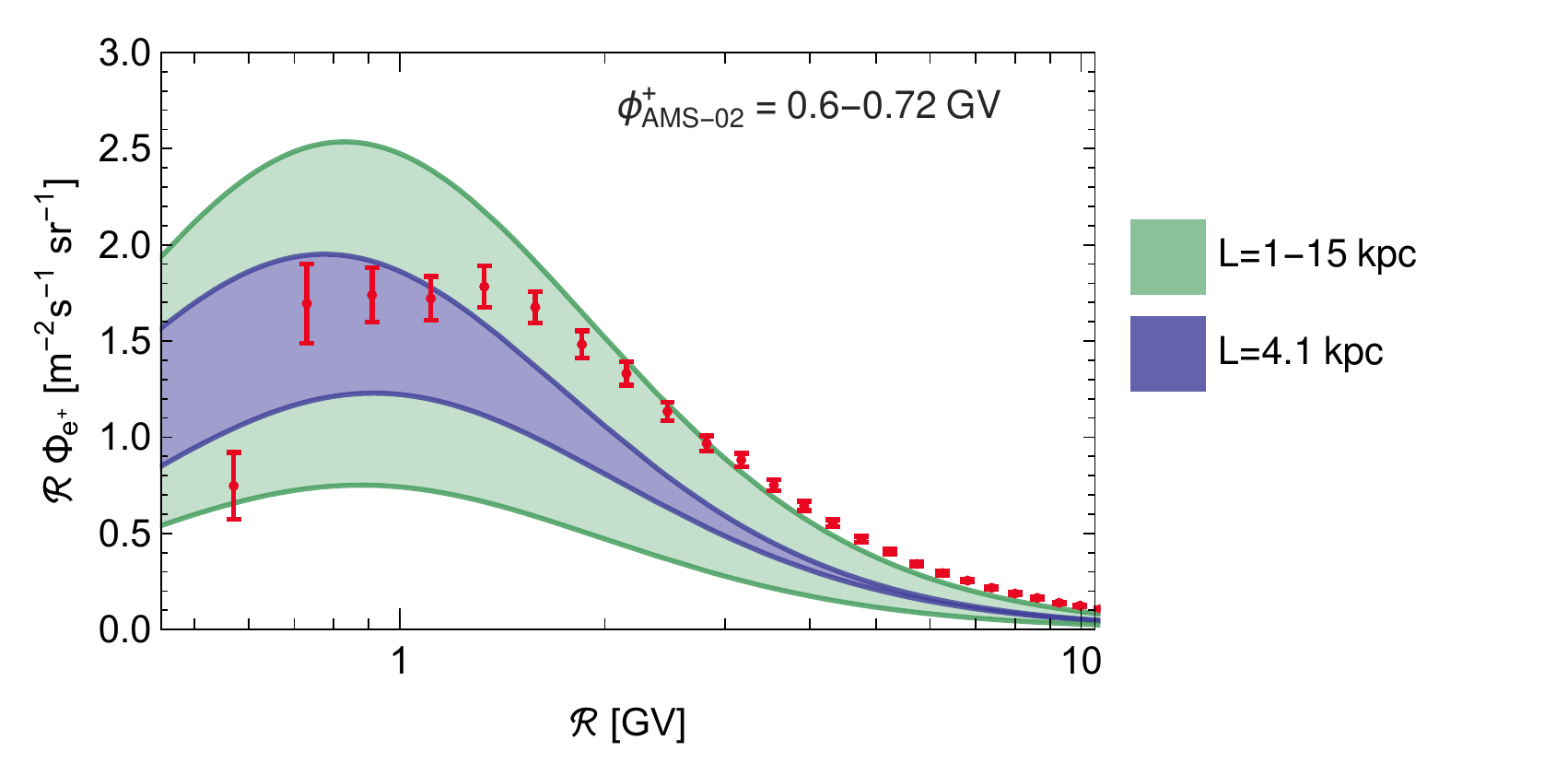}
\end{center}
\caption{Positron flux for the propagation parameters from the B/C + $\bar{p}$ fit (see table~\ref{tab:fitsecondary}). The size of the diffusion halo has been set to the values stated in the plot legend. The width of the bands includes the uncertainty in solar modulation corresponding to $\phi^+_{\text{\tiny{AMS-02}}}=0.6-0.72\gv$.}
\label{fig:positron}
\end{figure}

We now wish to determine the minimal allowed $L$ for the propagation configuration which best fits the antiproton and B/C data. For this purpose we fix $K_0/L$, $\delta$, $V_c$ and $V_a/\sqrt{L}$ to the values shown in the last column of table~\ref{tab:fitsecondary} and calculate the corresponding positron spectra as described in section~\ref{sec:diffusionmodel}. The positron flux modulated with $\phi^+_{\text{\tiny{AMS-02}}}=0.6-0.72\gv$ is shown in figure~\ref{fig:positron}. Our limit is derived for $\phi^+_{\text{\tiny{AMS-02}}}=0.72\gv$ which yields the smallest (most conservative) positron flux within the considered range. We decrease L until the secondary positron flux exceeds the 95\% CL upper limit in one of the bins. As it is always the first bin which sets the strongest constraint, we do not need to assign a statistical penalty for the choice of bin. The lower limit we obtain is $L=4.1\kpc$. Our constraint is significantly weaker than $L>8.5\kpc$ derived in~\cite{Boudaud:2016jvj}. The difference finds its explanation in the propagation configurations considered in~\cite{Boudaud:2016jvj}. These stem from an older B/C analysis~\cite{Maurin:2001sj} and feature large $V_A$. Strong reacceleration increases the low energy positron flux and tightens the constraint on $L$. For the reacceleration velocity favored by our B/C + $\bar{p}$ fit, the weaker constraint applies. In the next section we will use the lower limit $L=4.1\kpc$ to derive constraints on dark matter annihilation. As the best fit propagation parameters are hardly affected by including primary antiprotons in the fit, it is justified to keep the lower limit on $L$ fixed in the following.

Caveats in our constraint on $L$ exist if halo energy losses for positrons are substantially stronger than considered in section~\ref{sec:diffusionmodel} or if solar modulation deviates strongly from the force field approximation. On the other hand such deviations are hardly observed in the AMS-02 proton spectrum (see section~\ref{sec:solarmodulation}). In addition, we remind the reader that the positron source term was derived with the cross section parameterization of Kamae et al.~\cite{Kamae:2006bf} which yields the lowest (most conservative) positron flux among the alternatives. In this light, we consider $L>4.1\kpc$ as a sufficiently robust lower limit. Further indications against a diffusion halo $L\lesssim 4\kpc$ arise from the diffuse gamma ray background~\cite{Ackermann:2012pya} and from radio data~\cite{Bringmann:2011py,DiBernardo:2012zu,Orlando:2013ysa}.

\section{Dark Matter Search}\label{sec:darkmatter}

We confirmed in section~\ref{sec:pbarandbc} that the observed antiproton and B/C spectra are consistent with secondary production. But at the precision level, some structures appeared in the residuals of the fit. A modest peak at $\R\sim 10\gv$ matches with the excess pointed out in~\cite{Cuoco:2016eej,Cui:2016ppb}. There, it was tentatively interpreted as a matter signal. We now want to investigate the robustness of the signal with respect to uncertainties in the secondary backgrounds. 
We consider $b\bar{b}$ and $WW$ as exemplary dark matter annihilation channels. The corresponding primary antiproton fluxes are derived from the source terms~\eqref{eq:primary}. We choose the standard NFW profile (see table~\ref{tab:profiles}) and consider the dark matter mass ranges $m_{\text{DM}}=7-3000\gev$ and $82-3000\gev$ for $b\bar{b}$ and $WW$ respectively. The annihilation cross section is taken to be a free parameter, we only require $\langle \sigma 
v\rangle>0$. The size of the diffusion halo, which controls the normalization of the primary flux, is set to the minimum $L=4.1\kpc$ derived in section~\ref{sec:positron}. A combined fit to the antiproton and B/C spectra is performed -- this time including the primary antiprotons. The primary flux component must be strongly subdominant not to spoil the combined fit. Therefore, it is justified to neglect uncertainties in the primary flux related to annihilation spectra. Uncertainties in the secondary fluxes are, however, fully taken into account.

\begin{table}[htp]
\begin{center}
\begin{tabular}{|l|cc|}
\hline
\rowcolor{light-gray}   &&\\[-3mm]
\rowcolor{light-gray} Channel  & $\bar{\text{b}}$b & WW \\[2mm]
\hline
&&\\[-3mm]
 $m_{\text{DM}}$  & $78.7\gev$ & $85.2\gev$ \\[2mm]
$\langle \sigma v \rangle\;[\frac{\text{cm}^3}{\text{s}}]$  & $0.91\cdot 10^{-26}$ & $1.0\cdot 10^{-26}$  \\[2mm]
\hline
&&\\[-3mm]
$K_0\;[\frac{\text{kpc}^2}{\text{Gyr}}]$  & 34.0 & 33.7\\[2mm]
$L\,[\text{kpc}]$  & 4.1 & 4.1\\[2mm]
$\delta$ & 0.499 & 0.500\\[2mm]
$V_a\;[\frac{\text{km}}{\text{s}}]$ & 15.0 & 15.1 \\[2mm]
$V_c\;[\frac{\text{km}}{\text{s}}]$ & 0 & 0 \\[2mm]
$\Delta\delta$ & 0.157 & 0.157 \\[2mm]
$\mathcal{R}_b\;[\text{GV}]$&  275 & 275\\[2mm]
$s$ & 0.074 & 0.074 \\[2mm]
$\phi_0\;[\text{GV}]$ & 0.72 & 0.72 \\[2mm]
$\phi_1\;[\text{GV}]$ &  0.95 & 0.96 \\[2mm]
\hline
&&\\[-3mm]
$\chi^2_{\text{B/C}}$& 53.2 & 53.6\\[2mm]
$\chi^2_{\bar{p}}$ & 43.2  & 45.0  \\[2mm]
$\chi^2_{\text{AMS/PAM}}\;\;$ & 14.5  & 14.6 \\[2mm]
$\Delta\chi^2$ & 4.7  & 2.4 \\[2mm]
$p_\text{local}$ & 0.015 ($2.2\,\sigma$)& 0.061 ($1.6\,\sigma$)\\[2mm]
$p_\text{global}$ & 0.14 ($1.1\,\sigma$)& 0.25 ($0.7\,\sigma$)\\[2mm]
\hline
\end{tabular}
\end{center}
\caption{Propagation, solar modulation and dark matter parameters yielding the best fit to the B/C and antiproton data. Dark matter annihilations into $b\bar{b}$ and $WW$ are considered. Also shown is the goodness of fit and the significance at which the pure secondary hypothesis is ``excluded'' against a dark matter interpretation.} 
\label{tab:fitdm}
\end{table}

\begin{figure}[htp]
\begin{center}
  \includegraphics[height=7.6cm]{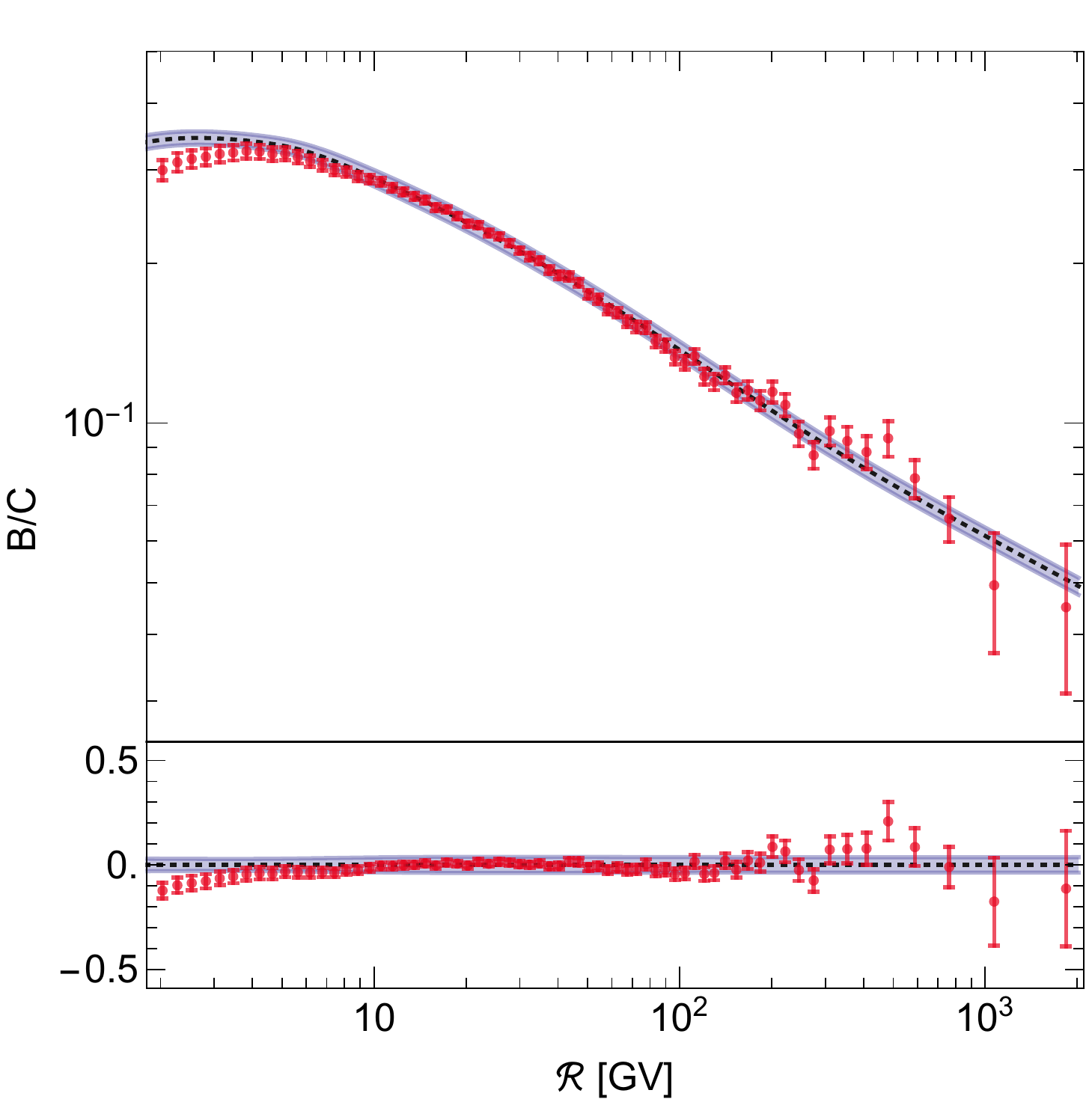}\hspace{3mm}
  \includegraphics[height=7.6cm]{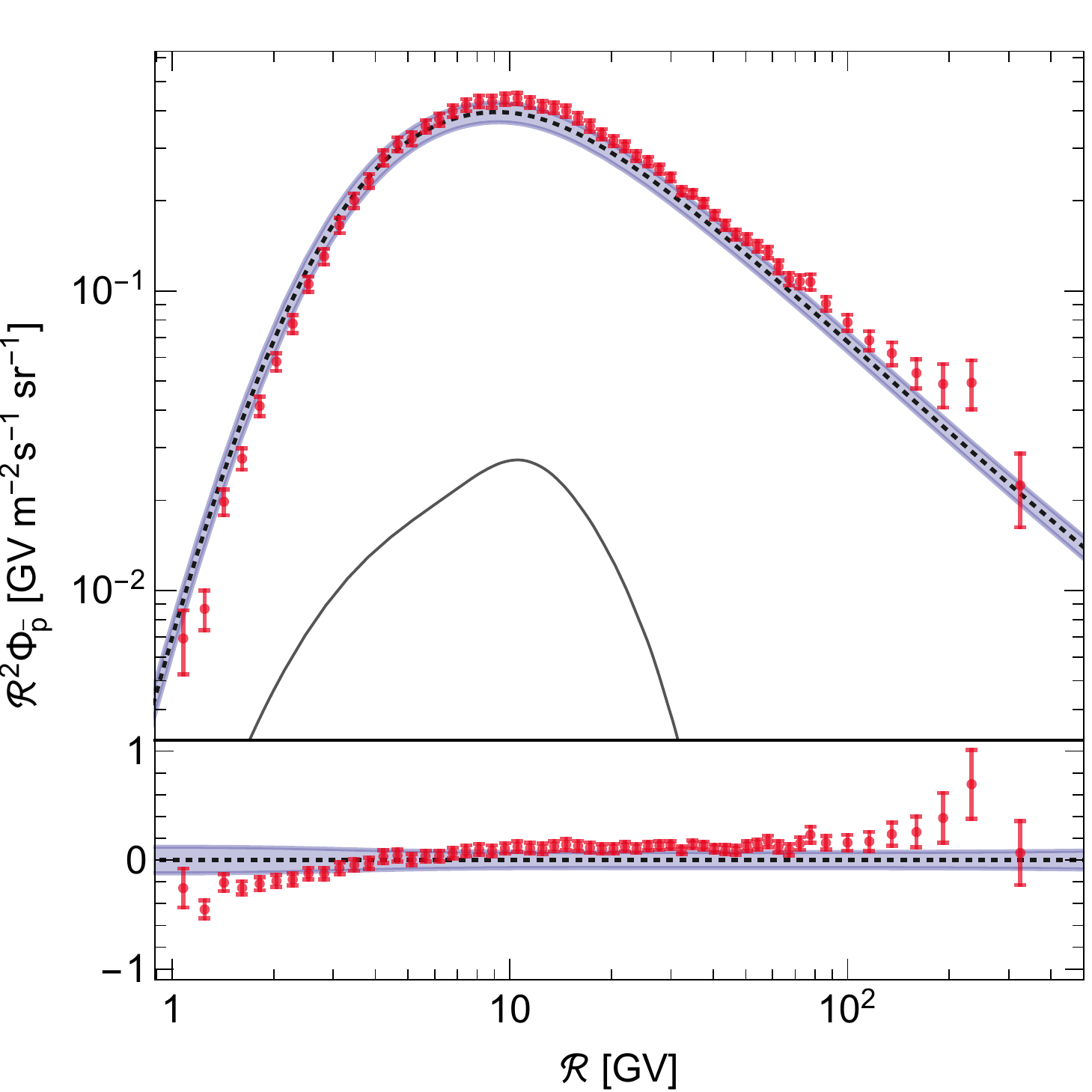}\\[2mm]
  \includegraphics[width=7.8cm]{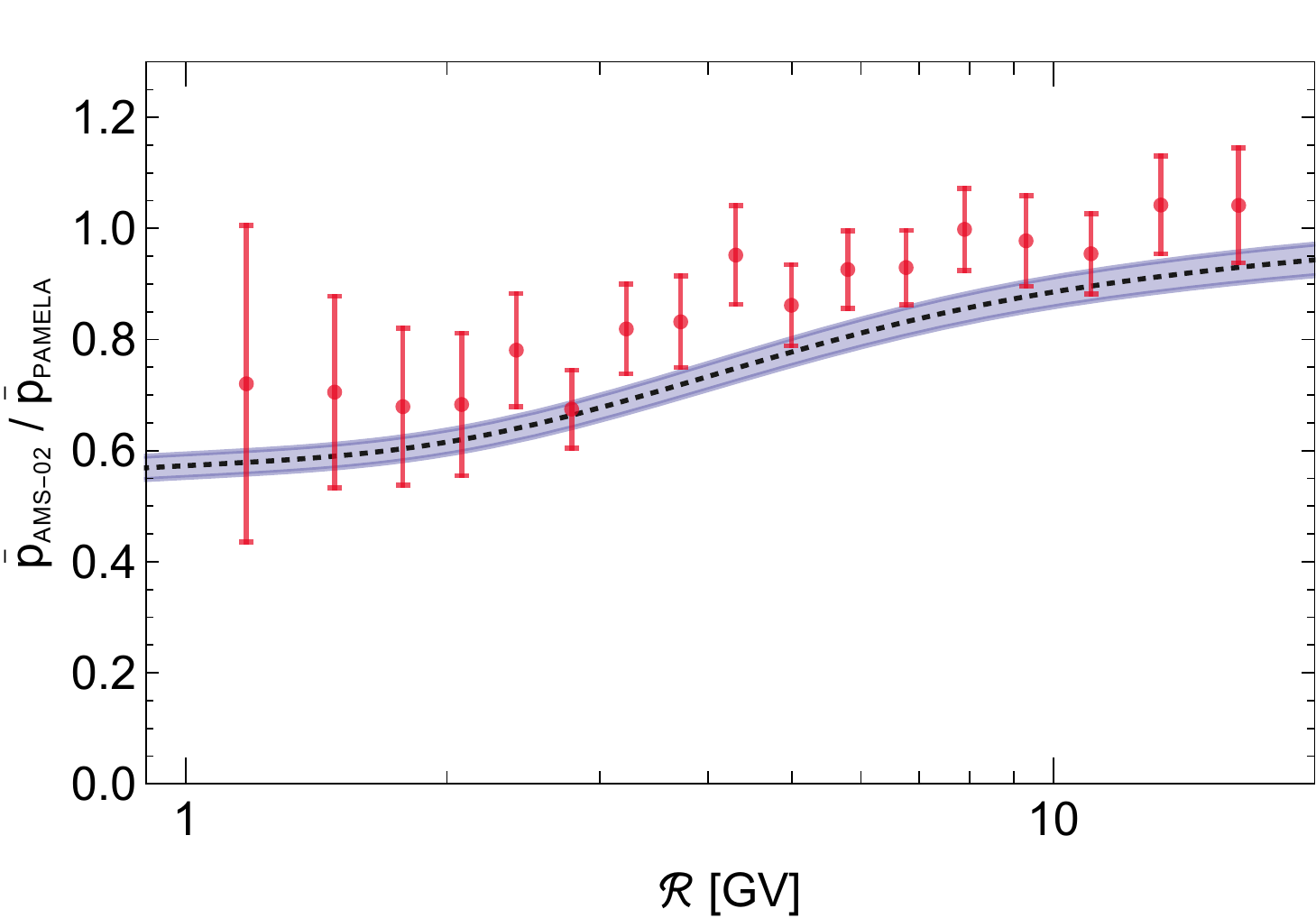}
\end{center}
\caption{Best fit spectra of the combined B/C + $\bar{p}$ fit including a primary antiproton component from dark matter annihilation into $b\bar{b}$. The best fit primary antiproton flux is indicated in the upper right panel (solid black line).}
\label{fig:bcandpbaranddm}
\end{figure}

Dark matter annihilations lead to a slight improvement of the fit which is more pronounced in the $b\bar{b}$ channel. The best fit parameters can be found in table~\ref{tab:fitdm}. Propagation parameters do not change considerably compared to the fit without dark matter. A slightly stronger solar modulation of
antiprotons is preferred to mitigate the increase of the low energy flux caused by the primary component. The favored dark matter mass resides at $m_{\text{DM}}\sim80\gev$ for both channels (slightly above threshold for $WW$). As can be seen in figure~\ref{fig:bcandpbaranddm}, the corresponding primary flux (scaled by $\R^2$) peaks at $\R\sim 10\gv$ and (partly) absorbs the residuals at this rigidity. The preferred annihilation cross section $\langle \sigma 
v\rangle\sim 10^{-26}\cm^3/\text{s}$ matches the expectation of a thermal WIMP up to a factor of two. The normalization is sensitive to the considered dark matter profile. It may, furthermore, be augmented by leptonic channels which leave the antiproton flux unaffected. The observed best fit properties are, hence, consistent with a thermal WIMP interpretation. 

In the next step, we determine the significance of the excess corresponding to the observed $\Delta\chi^2$. At fixed $m_{\text{DM}}$, $\Delta\chi^2$ under the background hypothesis is expected to follow a $0.5 \chi_{0\,\text{d.o.f}}^2+0.5 \chi_{1\,\text{d.o.f}}^2$ distribution as the alternative (background + signal) has one additional positive parameter $\langle\sigma v \rangle$. The local significance is hence $2.2\,\sigma$ for $b\bar{b}$ and $1.6\,\sigma$ for $WW$. The global significance is affected by the look-elsewhere effect: if the excess is a statistical fluctuation it may have occurred at any mass. Therefore, we generated a large sample of mock experimental data under the background hypothesis and determined the largest excess due to fluctuations in the considered range of $m_{\text{DM}}$.\footnote{Naively, one may think that global probability distribution of $\Delta\chi^2$ is simply a $\chi^2$-distribution with two degrees of freedom. This assumption is wrong as Wilks' theorem does not apply to cases, where a parameter (in this case $m_{\text{DM}}$) is only defined under the alternative hypothesis~\cite{Davies:1987zz,Gross:2010qma}. Therefore, Monte Carlo simulation cannot be avoided.} In $14\%$ ($25\%$) of the mock data we find an excess at least as large as the one observed in the $b\bar{b}$ ($WW$) channel. Formally, this corresponds to a global significance of $1.1\,\sigma$ ($0.7\,\sigma$) for $b\bar{b}$ ($WW$). Clearly, the case for dark matter in the AMS-02 data is not very strong. 

We can directly compare our results to the previous analyses. While we have confirmed the presence of a modest antiproton excess at the same rigidity as in~\cite{Cuoco:2016eej,Cui:2016ppb}, the significance of the excess is substantially lower in our analysis. The decrease in significance is likely driven by the inclusion of cross section uncertainties. We outlined in section~\ref{sec:pbarandbc} how variations in cross section parameters can lead to a ``bumpier'' secondary spectrum around $\R\simeq 10\gv$. This possibility effectively enters our fit via the covariance matrix $\Sigma^\text{$\bar{p}$,source}$. An additional reduction in significance occurs as -- different from~\cite{Cuoco:2016eej} -- we included the low energy spectrum which tends to exceed the data. But differences compared to~\cite{Cuoco:2016eej,Cui:2016ppb} also exist in the considered species and in the underlying propagation model.

In order to either eliminate or establish the excess, a further decrease of uncertainties is desirable. New measurements of antiproton and boron production cross sections at low energy would be very helpful in this regard. Even annihilation cross sections on the interstellar matter will have to be revisited in order to reach percent level accuracy. Even if the excess persisted after raising the precision, alternative explanations to dark matter would have to be explored -- unless one would find complementary evidence in cleaner channels like antideuterium or antihelium~\cite{Korsmeier:2017xzj}. An increase of the reacceleration velocity could, for example, completely flatten out the residuals in the antiproton spectrum at $\sqrt{s}\sim 10\gev$. While this possibility was disfavored by B/C, it could be revived through a modification of the diffusion model at very low energy. In~\cite{Jones:2000qd,Ptuskin:2005ax} e.g.\ an enhancement of the diffusion term at $\R\lesssim \text{few GeV}$ has been motivated which would reduce the low energy fluxes of antiprotons and boron. Given the present (in)significance of the excess, we shall for now refrain from investigating this further. 

\begin{figure}[htp]
\begin{center}
  \includegraphics[width=14cm]{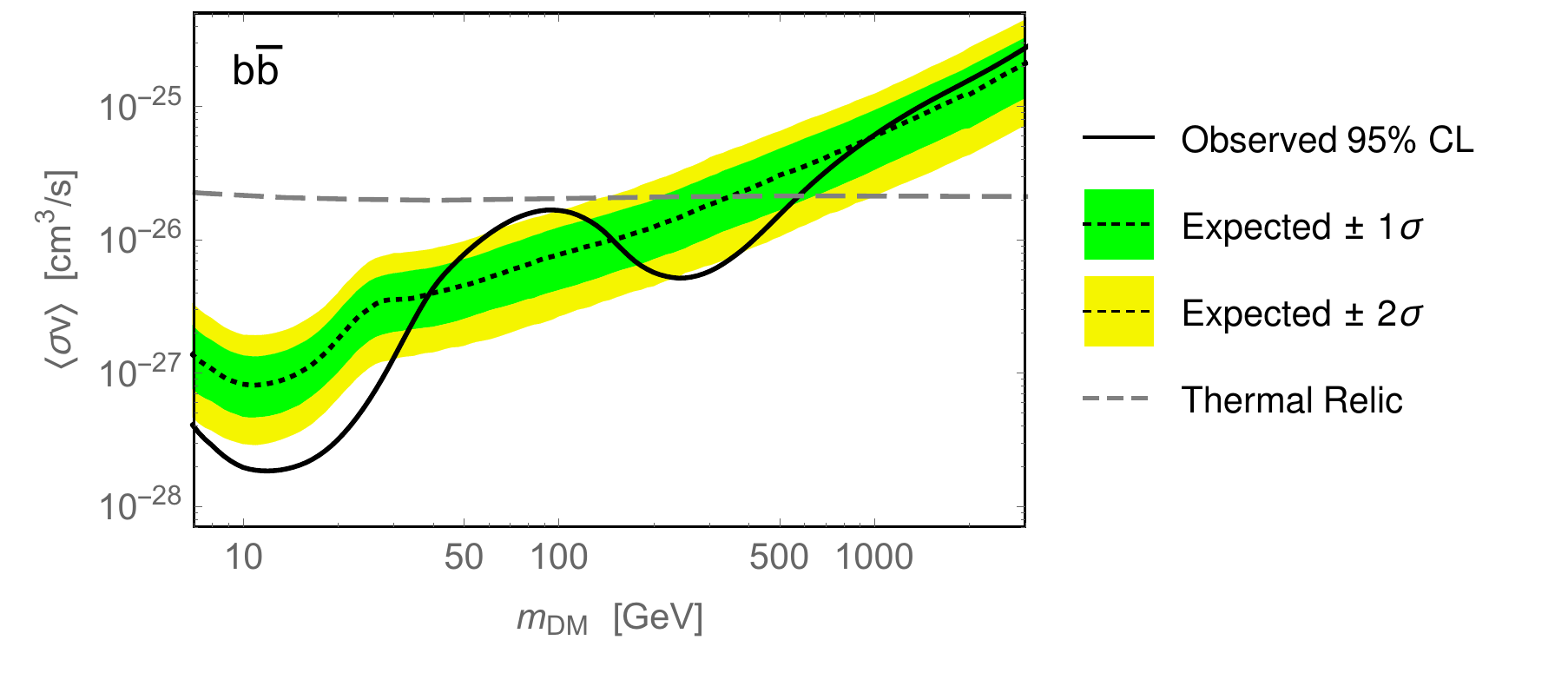}\\[3mm]
   \includegraphics[width=14cm]{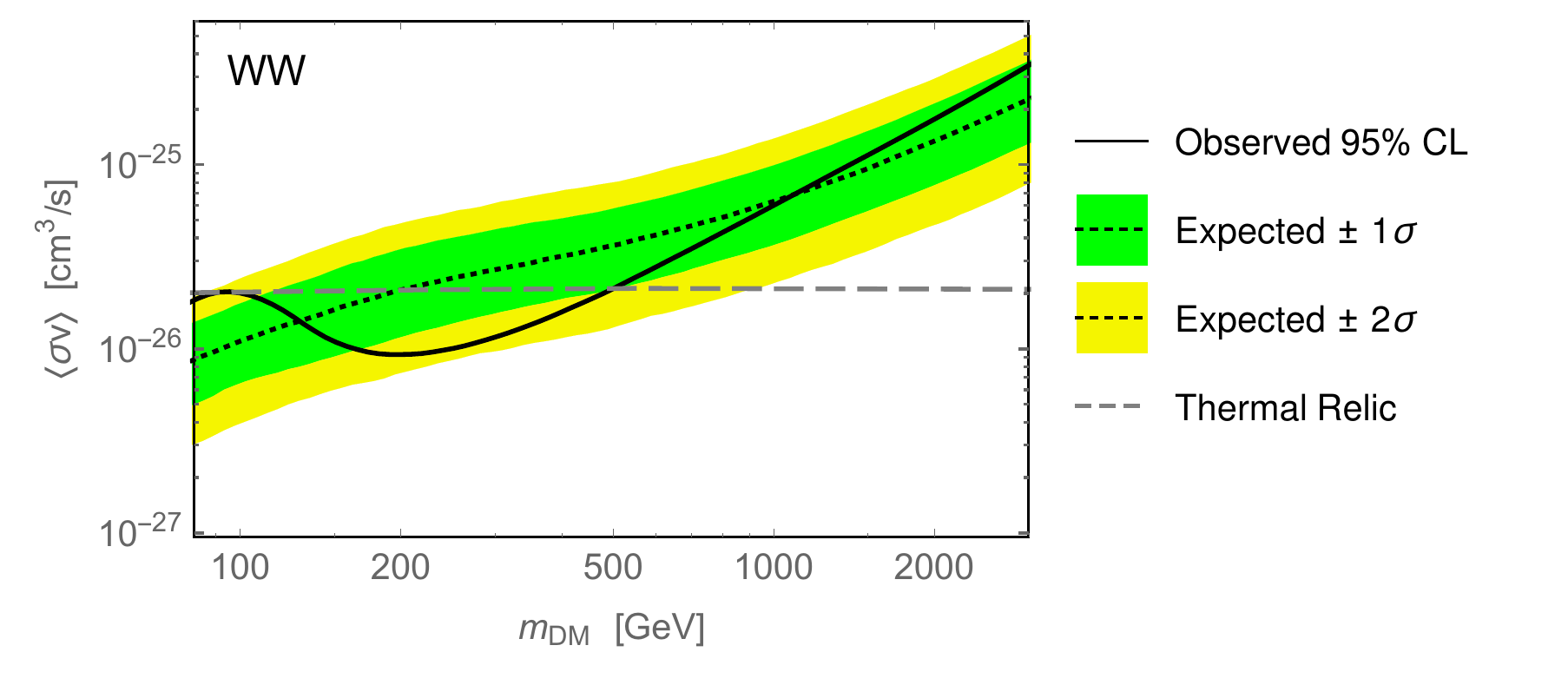}
\end{center}
\caption{Constraints on dark matter annihilation into $b\bar{b}$ and $WW$ derived from the antiproton and B/C data of AMS-02. Expected limits are also shown.}
\label{fig:dmlimits}
\end{figure}

Rather, we will now derive limits on the dark matter annihilation cross section. We again consider $b\bar{b}$ and $WW$ final states and assume the NFW dark matter profile. The diffusion halo size is set to the lower limit $L=4.1\kpc$. This is the most conservative choice, as any larger halo would result in stronger constraints on dark matter. For each $m_{\text{DM}}$ we allow the propagation parameters to float and determine the best fit annihilation cross section and corresponding $\chi^2_\text{best}(m_{\text{DM}})$ imposing $\langle \sigma v \rangle > 0$. The 95\% CL upper limit on $\langle \sigma v \rangle$ is then obtained by requiring $\chi^2(m_{\text{DM}},\langle \sigma v\rangle)-\chi^2_\text{best}(m_{\text{DM}})=2.71$.\footnote{In principle, the propagation parameters should be refitted for every combination of $m_{\text{DM}},\,\langle \sigma v\rangle$. We checked for a few examples that the refitting has negligible impact on the constraints. Therefore, we avoided this time-consuming procedure.} In addition to the actual limit, we derive the expected limit under the background hypothesis. The latter is extracted from a large sample of generated mock data by taking the median limit within the sample.\footnote{Variations in the propagation parameters were neglected when deriving the expected limit.} Observed and expected 95\% CL upper limits on the annihilation cross section are shown in figure~\ref{fig:dmlimits} together with the $1$ and $2\,\sigma$ uncertainty bands on the expected limit.

\begin{figure}[htp]
\begin{center}
  \includegraphics[width=15.5cm]{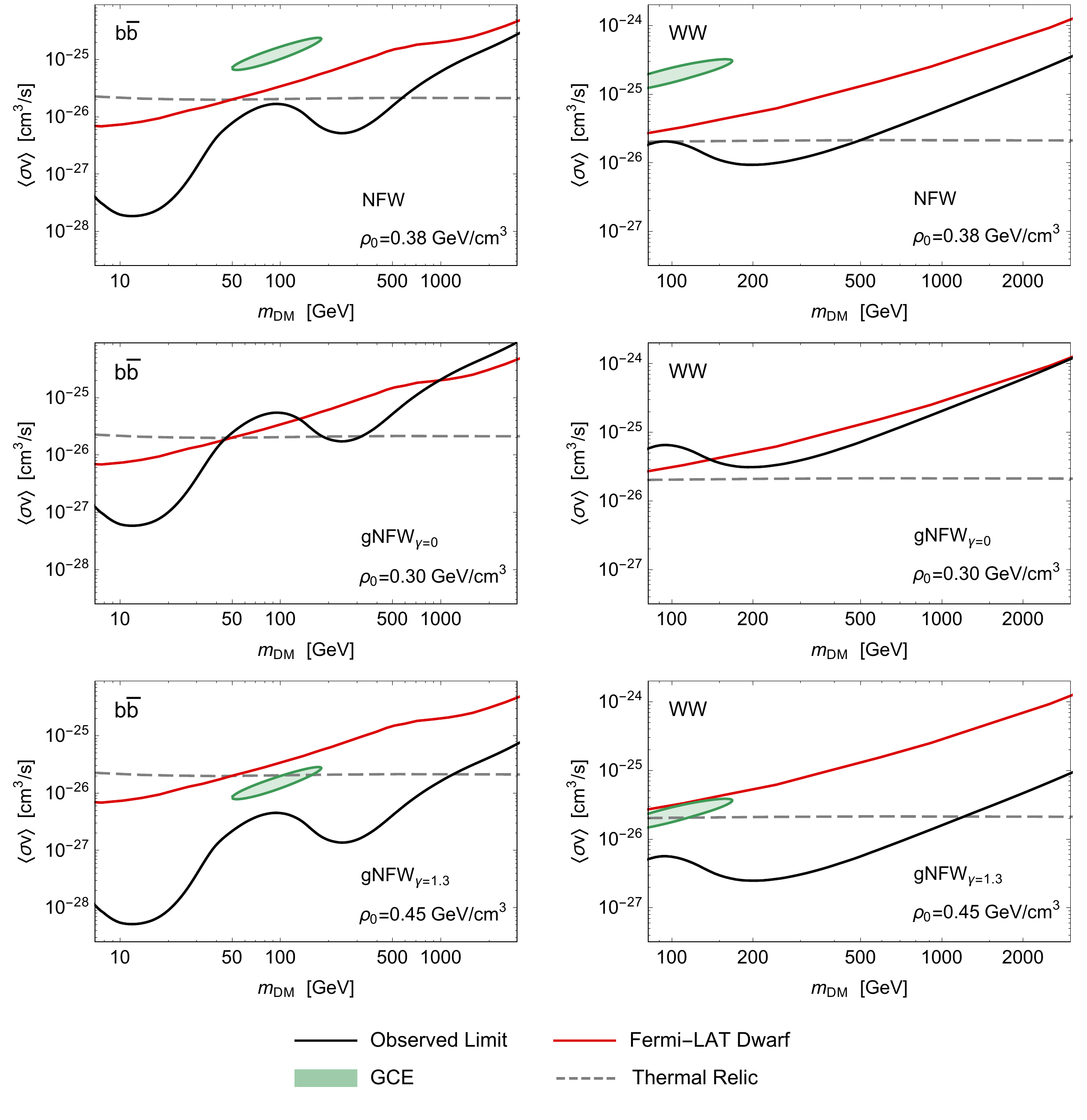}
\end{center}
\caption{Limits on dark matter annihilation as in figure~\ref{fig:dmlimits} for the canonical NFW profile, a cored and a contracted gNFW profile (see table~\ref{tab:profiles}). Also shown are the gamma ray constraints from dwarf galaxies as well as the confidence regions for the dark matter interpretation of the galactic center excess (see text).}
\label{fig:dmlimits2}
\end{figure}

The discussed excess at $m_{\text{DM}}\sim 80\gev$ is clearly visible by the weakening of limits around this energy. However, the downward fluctuation in the $b\bar{b}$-constraint around $m_{\text{DM}}\sim 10\gev$ is actually even more pronounced. At high mass $m_{\text{DM}}\gtrsim\text{TeV}$ observed and expected limits are very close. Dark matter explanations which had been introduced to account for a hardening of the high energy antiproton spectrum are not required. The parameter space of thermal WIMPs is severely constrained. For the considered NFW profile, thermal WIMPs with mass $m_{\text{DM}}<570\gev$ are excluded if they annihilate into bottom quarks.\footnote{The exclusion holds for a velocity-independent annihilation cross section.} Limits in the $WW$ channel are only slightly weaker. Similar exclusions are also expected for other hadronic dark matter annihilation channels, while leptonic channels provide a loophole to the antiproton constraints. In figure~\ref{fig:dmlimits2} we illustrate the dependence of limits on the properties of the dark matter halo. For the cored profile with $\rho_0 = 0.3\gev\cm^{-3}$, exclusion still holds for $m_{\text{DM}}<45\gev$ and $m_{\text{DM}}=185-320\gev$ in the $b\bar{b}$ channel. Constraints on $WW$ shift slightly above the thermal cross section in the whole mass range. This choice of profile might be a bit too conservative, but already for small deviation from the canonical NFW profile WIMP masses $m_{\text{DM}}\sim 80\gev$ become allowed. In the case of the contracted profile of table~\ref{tab:profiles}, thermal WIMPs with hadronic annihilations can be excluded for masses up to TeV.

We can compare our constraints to~\cite{Cuoco:2016eej,Cui:2016ppb} and find them to be in reasonable agreement despite very complementary approaches. Uncertainties present in~\cite{Cuoco:2016eej,Cui:2016ppb} related to the size of the diffusion halo where avoided in our analysis as we used positrons to limit $L$. We also note that inclusion of cross section uncertainties in the secondary flux did not considerably weaken the limits. While cross section mismodeling can lead to features in the residuals, these are typically smoother than those induced by primary signals.

It is also interesting to put our results into comparison with gamma ray searches for dark matter. An excess in the gamma ray flux from the galactic center was pointed out in~\cite{Goodenough:2009gk}. In figure~\ref{fig:dmlimits2} we depict the corresponding $2\,\sigma$ confidence regions for the case that it is interpreted in terms of dark matter. The preferred regions were derived using the spectrum~\cite{Krauss:2016cdi} which is, however, subject to seizable uncertainties~\cite{TheFermi-LAT:2017vmf}. It may seem intriguing that the gamma ray excess hints at a dark matter candidate with similar mass as the antiproton excess discussed earlier~\cite{Cuoco:2017rxb}. On the other hand, a consistent picture in terms of hadronic dark matter annihilations does not really emerge as cross sections required for the gamma ray excess are excluded by antiprotons. We also note that strong arguments for an astrophysical interpretation of the gamma ray excess in terms of point sources have been presented~\cite{Bartels:2015aea,Lee:2015fea,Fermi-LAT:2017yoi}. The strongest gamma ray constraints are set by the emission of dwarf galaxies~\cite{Ackermann:2015zua,Fermi-LAT:2016uux}. As shown in figure~\ref{fig:dmlimits2} antiproton limits in the considered channels are stronger by a factor 1.5-50 for the canonical NFW profile. Even for the very conservative cored profile, antiproton constraints dominate over a wide mass range. 

\section{Conclusion}

In this work we performed a systematic search for dark matter signals in the AMS-02 antiproton data. We included B/C and positron data in our analysis in order to narrow down uncertainties in the propagation of charged cosmic rays. A careful treatment of solar modulation including charge-sign dependent effects allowed us to reliably interpret spectra down to the lowest energies. Uncertainties in the secondary source terms of antiprotons and boron were rigorously modeled from the available accelerator data and embedded into a powerful spectral analysis. In particular, we investigated a reported antiproton excess at $\R\sim 10\gv$. The latter had been interpreted in terms of a WIMP with mass $m_{\text{DM}}\sim 80\gev$ and hadronic annihilations. Dark matter with similar properties had previously been considered as the explanation of the bright GeV gamma ray spectrum in the galactic center. We find that the boron and antiproton fluxes are consistent with pure secondary production. A mild antiproton excess corresponding to $m_{\text{DM}}\sim 80\gev$ is confirmed. But its significance hardly exceeds $1\,\sigma$, once all relevant uncertainties and the look-elsewhere effect are taken into account. Even if this tiny excess is taken serious, the required dark matter annihilation cross section does not fit to the galactic center excess. A consistent common explanation does not emerge, unless one assigns a fraction of the gamma ray excess to other sources. 

In the absence of a conclusive signal, we provided strong limits on dark matter annihilations into $b\bar{b}$ and $WW$ (see figures~\ref{fig:dmlimits} and~\ref{fig:dmlimits2}). For a standard NFW dark matter profile these exclude thermal WIMPs with masses up to $570\gev$ ($100-500\gev$) in the $b\bar{b}$-channel ($WW$-channel). The mass window at $m_{\text{DM}}\sim 80\gev$, however, opens up for slightly more conservative choices of the dark matter halo. Antiproton limits are significantly stronger than gamma ray limits from dwarf spheroidal galaxies over wide mass ranges, even for cored dark matter profiles.

Although antiprotons are already a very powerful channel, strong improvements in sensitivity are within reach. Due to the high precision of the data, uncertainties in the secondary spectra are currently still the most limiting element. In our analysis we were already facing some limitations in fully resolving the antiproton excess at $\R\sim 10\gv$. The astrophysical antiproton background at this energy is affected by the transition from near-threshold to scaling behavior in the secondary production at $\sqrt{s}\sim 10\gev$. The modeling of this energy regime still relies on accelerator data from the early 1970s with seizable uncertainties. 
Interpretation of the B/C spectrum faces similar challenges: even the most important reaction for boron production, the nucleon stripping of carbon, has only been measured with reasonable precision up to $T\simeq 4 \gev$. At the present stage, small residuals we observed in our fits to the antiproton and B/C data are consistent with uncertainties. However, they could turn into real features once the next level of precision is reached. It is encouraging that the variety of upcoming cosmic ray data will soon allow for new insights. In order to fully exploit the potential of cosmic ray observations, new measurements of particle physics cross sections are urgently needed. Not only is this important to further explore the parameter space of thermal WIMPs, but also to search for other cosmic ray sources and to develop the global picture of cosmic ray propagation.

\section*{Acknowledgments}
We would like to thank K.\ Blum, M.\ Boudaud, A.\ Cuoco, P.\ von Doetinchem, Y.\ Genolini, J.\ Lavalle, R.\ Kappl, P.\ Mertsch and S.\ Sarkar for helpful discussions on various topics covered in this work. Furthermore, we are grateful to T.\ Kamae for correspondence on the positron production cross sections.

\appendix

\section{Covariance Matrices of Source Term Uncertainties}\label{sec:appendix}

For the determination of $\Sigma^\text{B/C,source}$ we randomly generate a large number of tuples \{$\sigma_{1,i}$, $\xi_i$, $\Delta_i$, $f_i$, $\R_{l,i}$, $\alpha_i$, $\gamma_i$, $R_b$, $\Delta\gamma$, $s$\} from the probability distributions of the parameters. The probability distributions are derived from experimental data as described in section~\ref{sec:secondaryproduction}. Since $i$ runs over the 6 relevant boron progenitors the tuples are sets of 45 parameters which fix boron production completely. For each tuple we determine the corresponding B/C ratio in the 67 rigidity bins of AMS-02. The covariance between the $i$th and the $j$th bin of AMS-02 is then obtained as
\begin{equation}\label{eq:covariance}
\Sigma_{ij}^\text{B/C,source}=  \Big\langle(B/C)_i - \big\langle B/C\big\rangle_i\Big\rangle \Big\langle(B/C)_j - \big\langle B/C\big\rangle_j\Big\rangle\,.
\end{equation}
where $(B/C)_i$ denotes the predicted B/C in the $i$th bin. The averaging is performed over the B/C ratios corresponding to the different parameter tuples. A slight complication occurs as the covariance matrix is sensitive to the choice of propagation parameters. Since we want to avoid evaluating~\eqref{eq:covariance} for each set of propagation parameters, we define the relative covariance matrix
\begin{equation}\label{eq:relativecovariance}
 \tilde{\Sigma}_{ij}^\text{B/C,source} = \frac{\Sigma_{ij}^\text{B/C,source}}{(B/C)_i (B/C)_j}\,.
\end{equation}
The relative covariance matrix remains (nearly) constant under variations in the propagation. Therefore, in practice we just had to determine $\tilde{\Sigma}_{ij}^\text{B/C,source}$ for one set of propagation parameters\footnote{As the one set of propagation parameters we choose the configuration which minimizes $\chi_{\text{B/C}}^2$ in~\eqref{eq:chibc} if only experimental errors of AMS-02 are included.} and then obtained $\Sigma_{ij}^\text{B/C,source}$ by scaling it with the predicted B/C according to~\eqref{eq:relativecovariance}. The determination of $\Sigma^\text{$\bar{p}$,source}$ proceeds in complete analogy to $\Sigma^\text{B/C,source}$.

\pagebreak

\bibliography{antibib}

\begin{thebibliography}{100}

\bibitem{Adriani:2008zr}
PAMELA, O. Adriani et~al.,
\newblock Nature 458 (2009), 0810.4995.

\bibitem{Abdo:2010dk}
Fermi-LAT, A.A. Abdo et~al.,
\newblock JCAP 1004 (2010), 1002.4415.

\bibitem{Ackermann:2011wa}
Fermi-LAT, M. Ackermann et~al.,
\newblock Phys. Rev. Lett. 107 (2011), 1108.3546.

\bibitem{Galli:2009zc}
S. Galli et~al.,
\newblock Phys. Rev. D80 (2009), 0905.0003.

\bibitem{Slatyer:2009yq}
T.R. Slatyer, N. Padmanabhan and D.P. Finkbeiner,
\newblock Phys. Rev. D80 (2009), 0906.1197.

\bibitem{Huetsi:2009ex}
G. Huetsi, A. Hektor and M. Raidal,
\newblock Astron. Astrophys. 505 (2009), 0906.4550.

\bibitem{Boudaud:2016jvj}
M. Boudaud et~al.,
\newblock Astron. Astrophys. 605 (2017), 1612.03924.

\bibitem{tingtalk:2015}
S. Ting,
\newblock AMS Days at CERN  (2015).

\bibitem{Kappl:2015bqa}
R. Kappl, A. Reinert and M.W. Winkler,
\newblock JCAP 1510 (2015), 1506.04145.

\bibitem{Giesen:2015ufa}
G. Giesen et~al.,
\newblock JCAP 1509 (2015), 1504.04276.

\bibitem{Evoli:2015vaa}
C. Evoli, D. Gaggero and D. Grasso,
\newblock JCAP 1512 (2015), 1504.05175.

\bibitem{Kachelriess:2015wpa}
M. Kachelriess, I.V. Moskalenko and S.S. Ostapchenko,
\newblock Astrophys. J. 803 (2015), 1502.04158.

\bibitem{Winkler:2017xor}
M.W. Winkler,
\newblock JCAP 1702 (2017), 1701.04866.

\bibitem{Boschini:2017fxq}
M.J. Boschini et~al.,
\newblock Astrophys. J. 840 (2017), 1704.06337.

\bibitem{Aguilar:2014mma}
AMS, M. Aguilar et~al.,
\newblock Phys. Rev. Lett. 113 (2014).

\bibitem{Aguilar:2016vqr}
AMS, M. Aguilar et~al.,
\newblock Phys. Rev. Lett. 117 (2016).

\bibitem{Aguilar:2016kjl}
AMS, M. Aguilar et~al.,
\newblock Phys. Rev. Lett. 117 (2016).

\bibitem{Cuoco:2016eej}
A. Cuoco, M. Kr{\"{a}}mer and M. Korsmeier,
\newblock (2016), 1610.03071.

\bibitem{Cui:2016ppb}
M.Y. Cui et~al.,
\newblock (2016), 1610.03840.

\bibitem{Maurin:2001sj}
D. Maurin et~al.,
\newblock Astrophys. J. 555 (2001), astro-ph/0101231.

\bibitem{Donato:2001ms}
F. Donato et~al.,
\newblock Astrophys. J. 563 (2001), astro-ph/0103150.

\bibitem{Maurin:2002ua}
D. Maurin et~al.,
\newblock (2002), astro-ph/0212111.

\bibitem{Adriani:2012paa}
O. Adriani et~al.,
\newblock JETP Lett. 96 (2013).

\bibitem{Moskalenko:1997gh}
I.V. Moskalenko and A.W. Strong,
\newblock Astrophys. J. 493 (1998), astro-ph/9710124.

\bibitem{Strong:1998pw}
A.W. Strong and I.V. Moskalenko,
\newblock Astrophys. J. 509 (1998), astro-ph/9807150.

\bibitem{Strong:2001gh}
A.W. Strong and I.V. Moskalenko,
\newblock ICRC 2001, Proceedings  (2001), astro-ph/0106504.

\bibitem{Evoli:2008dv}
C. Evoli et~al.,
\newblock JCAP 0810 (2008), 0807.4730,
\newblock [Erratum: JCAP 1604 (2016)].

\bibitem{Evoli:2017vim}
C. Evoli et~al.,
\newblock (2017), 1711.09616.

\bibitem{Ptuskin:1997aa}
V.S. Ptuskin et~al.,
\newblock Astron. Astrophys. 321 (1997).

\bibitem{Delahaye:2008ua}
T. Delahaye et~al.,
\newblock Astron. Astrophys. 501 (2009), 0809.5268.

\bibitem{Tripathi:1999nw}
R.K. Tripathi, J.W. Wilson and F.A. Cucinotta,
\newblock Nucl. Instrum. Meth. B152 (1999).

\bibitem{Tripathi:1999}
R.K. Tripathi, F.A. Cucinotta and J.W. Wilson,
\newblock Nucl. Instrum. Meth. B155 (1999).

\bibitem{Protheroe:1981gj}
R.J. Protheroe,
\newblock Astrophys. J. 251 (1981).

\bibitem{Tan:1983de}
L.C. Tan and L.K. Ng,
\newblock J. Phys. G9 (1983).

\bibitem{Kappl:2011jw}
R. Kappl and M.W. Winkler,
\newblock Phys. Rev. D85 (2012), 1110.4376.

\bibitem{Putze:2010zn}
A. Putze, L. Derome and D. Maurin,
\newblock Astron. Astrophys. 516 (2010), 1001.0551.

\bibitem{Barrau:2001ev}
A. Barrau et~al.,
\newblock Astron. Astrophys. 388 (2002), astro-ph/0112486.

\bibitem{Donato:2003xg}
F. Donato et~al.,
\newblock Phys. Rev. D69 (2004), astro-ph/0306207.

\bibitem{Gleeson:1968zza}
L.J. Gleeson and W.I. Axford,
\newblock Astrophys. J. 154 (1968).

\bibitem{Kota:1979}
J. Kota,
\newblock International Cosmic Ray Conference 3 (1979).

\bibitem{Jokipii:1981}
J.R. Jokipii and B. Thomas,
\newblock Astrophys. J. 243 (1981).

\bibitem{Sun:2015}
X. Sun et~al.,
\newblock Astrophys. J. 798 (2015), 1410.8867.

\bibitem{Adriani:2016uhu}
O. Adriani et~al.,
\newblock Phys. Rev. Lett. 116 (2016), 1606.08626.

\bibitem{Cholis:2015gna}
I. Cholis, D. Hooper and T. Linden,
\newblock Phys. Rev. D93 (2016), 1511.01507.

\bibitem{Vos:2015}
E.E. Vos and M.S. Potgieter,
\newblock Astrophys. J. 815 (2015).

\bibitem{Ghelfi:2015tvu}
A. Ghelfi et~al.,
\newblock Astron. Astrophys. 591 (2016), 1511.08650,
\newblock [Erratum: Astron. Astrophys. 605 (2017)].

\bibitem{Corti:2015bqi}
C. Corti et~al.,
\newblock Astrophys. J. 829 (2016), 1511.08790.

\bibitem{Stone:2013}
E.C. Stone et~al.,
\newblock Science 341 (2013).

\bibitem{Aguilar:2015ooa}
AMS, M. Aguilar et~al.,
\newblock Phys. Rev. Lett. 114 (2015).

\bibitem{Boschini:2017gic}
M.J. Boschini et~al.,
\newblock (2017), 1704.03733.

\bibitem{Adriani:2013as}
O. Adriani et~al.,
\newblock Astrophys. J. 765 (2013), 1301.4108.

\bibitem{Webber:2003}
W.R. Webber et~al.,
\newblock Astrophys. J. Suppl. Series 144 (2003).

\bibitem{Silberberg:1973jxa}
R. Silberberg and C.H. Tsao,
\newblock Astrophys. J. Suppl. 25 (1973).

\bibitem{Silberberg:1998lxa}
R. Silberberg, C.H. Tsao and A.F. Barghouty,
\newblock Astrophys. J. 501 (1998).

\bibitem{Moskalenko:2003kp}
I.V. Moskalenko and S.G. Mashnik,
\newblock ICRC 2003, Proceedings  (2003), astro-ph/0306367.

\bibitem{Read:1984xme}
S.M. Read and V.E. Viola,
\newblock Atom. Data Nucl. Data Tabl. 31 (1984).

\bibitem{Genolini:2017dfb}
Y. Génolini et~al.,
\newblock (2017), 1706.09812.

\bibitem{Bodansky:1975}
D. Bodansky, W.W. Jacobs and D.L. Oberg,
\newblock Astrophys. J. 202 (1975).

\bibitem{Bodemann:1993}
R. Bodemann et~al.,
\newblock Nucl. Instrum. Meth.B 82 (1993).

\bibitem{Brechtmann:1988rs}
C. Brechtmann, W. Heinrich and E.V. Benton,
\newblock Phys. Rev. C39 (1989).

\bibitem{Davids:1970wh}
C.N. Davids, H. Laumer and S.M. Austin,
\newblock Phys. Rev. C1 (1970).

\bibitem{Epherre:1969qxr}
M. Epherre et~al.,
\newblock Nucl. Phys. A139 (1969).

\bibitem{Fontes:1977qq}
P. Fontes,
\newblock Phys. Rev. C15 (1977).

\bibitem{Fontes:1971rn}
P. Fontes et~al.,
\newblock Nucl. Phys. A165 (1971).

\bibitem{Goel:1969}
P.S. Goel,
\newblock Nature 223 (1969).

\bibitem{Jung:1970xr}
M. Jung et~al.,
\newblock Phys. Rev. C1 (1970).

\bibitem{Korejwo:2000pf}
A. Korejwo et~al.,
\newblock J. Phys. G26 (2000).

\bibitem{Korejwo:2002ts}
A. Korejwo et~al.,
\newblock J. Phys. G28 (2002).

\bibitem{Laumer:1974zza}
H. Laumer, S.M. Austin and L.M. Panggabean,
\newblock Phys. Rev. C10 (1974).

\bibitem{Laumer:1973zz}
H. Laumer et~al.,
\newblock Phys. Rev. C8 (1973).

\bibitem{Lindstrom:1975mi}
P.J. Lindstrom et~al.,
\newblock LBL-3650  (1975).

\bibitem{Moyle:1979zz}
R.A. Moyle et~al.,
\newblock Phys. Rev. C19 (1979).

\bibitem{Olson:1983jz}
D.l. Olson et~al.,
\newblock Phys. Rev. C28 (1983).

\bibitem{Raisbeck:1974zz}
G.M. Raisbeck and F. Yiou,
\newblock Phys. Rev. C9 (1974).

\bibitem{Raisbeck:1971ie}
G.M. Raisbeck and F. Yiou,
\newblock Phys. Rev. Lett. 27 (1971).

\bibitem{Roche:1976zz}
C.T. Roche et~al.,
\newblock Phys. Rev. C14 (1976).

\bibitem{Schiekel:1996}
T. Schiekel et~al.,
\newblock Nucl. Instrum. Meth.B 114 (1996).

\bibitem{Webber:1990kb}
W.R. Webber, J.C. Kish and D.A. Schrier,
\newblock Phys. Rev. C41, 533 (1990).

\bibitem{Webber:1990kc}
W.R. Webber, J.C. Kish and D.A. Schrier,
\newblock Phys. Rev. C41, 547 (1990).

\bibitem{Webber:1998}
W.R. Webber et~al.,
\newblock Astrophys. J. 508 (1998).

\bibitem{Yiou:1969}
F. Yiou, C. Seide and R. Bernas,
\newblock J. of Geophys. Research 74 (1969).

\bibitem{Zeitlin:2007sm}
C. Zeitlin et~al.,
\newblock Nucl. Phys. A784 (2007).

\bibitem{Zeitlin:2011qg}
C. Zeitlin et~al.,
\newblock Phys. Rev. C83 (2011), 1102.2848.

\bibitem{Zeitlin:2001ye}
C. Zeitlin et~al.,
\newblock Phys. Rev. C64 (2001).

\bibitem{Gupta:2013}
R. Gupta and A. Kumar,
\newblock Europ. Phys. J. A49 (2013).

\bibitem{Ferrando:1988tw}
P. Ferrando et~al.,
\newblock Phys. Rev. C37 (1988).

\bibitem{Kappl:2014hha}
R. Kappl and M.W. Winkler,
\newblock JCAP 1409 (2014), 1408.0299.

\bibitem{Tan:1982nc}
L.C. Tan and L.K. Ng,
\newblock Phys. Rev. D26 (1982).

\bibitem{diMauro:2014zea}
M. di~Mauro et~al.,
\newblock Phys. Rev. D90 (2014), 1408.0288.

\bibitem{Antinucci:1972ib}
M. Antinucci et~al.,
\newblock Lett. Nuovo Cim. 6 (1973).

\bibitem{Abelev:2008ab}
STAR, B.I. Abelev et~al.,
\newblock Phys. Rev. C79 (2009), 0808.2041.

\bibitem{Anticic:2009wd}
NA49, T. Anticic et~al.,
\newblock Eur. Phys. J. C65 (2010), 0904.2708.

\bibitem{Adare:2011vy}
PHENIX, A. Adare et~al.,
\newblock Phys. Rev. C83 (2011), 1102.0753.

\bibitem{Aamodt:2011zj}
ALICE, K. Aamodt et~al.,
\newblock Eur. Phys. J. C71 (2011), 1101.4110.

\bibitem{Chatrchyan:2012qb}
CMS, S. Chatrchyan et~al.,
\newblock Eur. Phys. J. C72 (2012), 1207.4724.

\bibitem{Baatar:2012fua}
NA49, B. Baatar et~al.,
\newblock Eur. Phys. J. C73 (2013), 1207.6520.

\bibitem{LHCb:2017tqz}
LHCb,
\newblock LHCb-CONF-2017-002  (2017).

\bibitem{Badhwar:1977}
G.D. Badhwar and S.A. Stephens,
\newblock ICRC, Conference Papers 1 (1977).

\bibitem{Tan:1984ha}
L.C. Tan and L.K. Ng,
\newblock J. Phys. G9 (1983).

\bibitem{Blum:2017iol}
K. Blum, R. Sato and M. Takimoto,
\newblock (2017), 1709.04953.

\bibitem{Kamae:2006bf}
T. Kamae et~al.,
\newblock Astrophys. J. 647 (2006), astro-ph/0605581,
\newblock [Erratum: Astrophys. J. 662 (2007)].

\bibitem{Tomassetti:2015doz}
N. Tomassetti,
\newblock PoS ICRC2015 (2016), 1510.09214.

\bibitem{Aguilar:2015ctt}
AMS, M. Aguilar et~al.,
\newblock Phys. Rev. Lett. 115 (2015).

\bibitem{talkXSCR}
AMS, Q. Yan,
\newblock Talk at XSCRC2017, CERN  (2017).

\bibitem{Engelmann:1990}
J.J. Engelmann et~al.,
\newblock Astrophys. J. 233 (1990).

\bibitem{Cirelli:2010xx}
M. Cirelli et~al.,
\newblock JCAP 1103 (2011), 1012.4515,
\newblock [Erratum: JCAP 1210 (2012)].

\bibitem{Drees:2009bi}
M. Drees, M. Kakizaki and S. Kulkarni,
\newblock Phys. Rev. D80 (2009), 0904.3046.

\bibitem{Laine:2006cp}
M. Laine and Y. Schroder,
\newblock Phys. Rev. D73 (2006), hep-ph/0603048.

\bibitem{Navarro:1995iw}
J.F. Navarro, C.S. Frenk and S.D.M. White,
\newblock Astrophys. J. 462 (1996), astro-ph/9508025.

\bibitem{Catena:2009mf}
R. Catena and P. Ullio,
\newblock JCAP 1008 (2010), 0907.0018.

\bibitem{Salucci:2010qr}
P. Salucci et~al.,
\newblock Astron. Astrophys. 523 (2010), 1003.3101.

\bibitem{Pato:2015dua}
M. Pato, F. Iocco and G. Bertone,
\newblock JCAP 1512 (2015), 1504.06324.

\bibitem{McMillan:2016}
P.J. McMillan,
\newblock Mon. Not. Roy. Astron. Soc. 465 (2017), 1608.00971.

\bibitem{Guedes:2011ux}
J. Guedes et~al.,
\newblock Astrophys. J. 742 (2011), 1103.6030.

\bibitem{Roca-Fabrega:2015gma}
S. Roca-Fàbrega et~al.,
\newblock Astrophys. J. 824 (2016), 1504.06261.

\bibitem{Zhu:2015jwa}
Q. Zhu et~al.,
\newblock Mon. Not. Roy. Astron. Soc. 458 (2016), 1506.05537.

\bibitem{Schaller:2015mua}
M. Schaller et~al.,
\newblock Mon. Not. Roy. Astron. Soc. 455 (2016), 1509.02166.

\bibitem{Mollitor:2014ara}
P. Mollitor, E. Nezri and R. Teyssier,
\newblock Mon. Not. Roy. Astron. Soc. 447 (2015), 1405.4318.

\bibitem{Serpico:2015caa}
P.D. Serpico,
\newblock PoS ICRC2015 (2016), 1509.04233.

\bibitem{Ptuskin:2012qu}
V. Ptuskin, V. Zirakashvili and E.S. Seo,
\newblock Astrophys. J. 763 (2013), 1212.0381.

\bibitem{Ohira:2015ega}
Y. Ohira, N. Kawanaka and K. Ioka,
\newblock Phys. Rev. D93 (2016), 1506.01196.

\bibitem{Blasi:2012yr}
P. Blasi, E. Amato and P.D. Serpico,
\newblock Phys. Rev. Lett. 109 (2012), 1207.3706.

\bibitem{Tomassetti:2012ga}
N. Tomassetti,
\newblock Astrophys. J. 752 (2012), 1204.4492.

\bibitem{Thoudam:2011aa}
S. Thoudam and J.R. Horandel,
\newblock Mon. Not. Roy. Astron. Soc. 421 (2012), 1112.3020.

\bibitem{Bernard:2012pia}
G. Bernard et~al.,
\newblock Astron. Astrophys. 555 (2013), 1207.4670.

\bibitem{Tomassetti:2015xem}
N. Tomassetti,
\newblock Astrophys. J. 815 (2015), 1511.04460.

\bibitem{Kachelriess:2017yzq}
M. Kachelrieß, A. Neronov and D.V. Semikoz,
\newblock (2017), 1710.02321.

\bibitem{Blasi:2011fm}
P. Blasi and E. Amato,
\newblock JCAP 1201 (2012), 1105.4529.

\bibitem{Salati:2016owk}
P. Salati et~al.,
\newblock {RICAP16, Proceedings}  (2016), 1611.06154.

\bibitem{Kraichnan:1965zz}
R.H. Kraichnan,
\newblock Phys. Fluids 8 (1965).

\bibitem{Kolmogorov:1941}
A. Kolmogorov,
\newblock Akademiia Nauk SSSR Doklady 30 (1941).

\bibitem{Hamaguchi:2015wga}
K. Hamaguchi, T. Moroi and K. Nakayama,
\newblock Phys. Lett. B747 (2015), 1504.05937.

\bibitem{Lin:2015taa}
S.J. Lin et~al.,
\newblock (2015), 1504.07230.

\bibitem{Chen:2015kla}
Y.H. Chen, K. Cheung and P.Y. Tseng,
\newblock Phys. Rev. D93 (2016), 1505.00134.

\bibitem{Katz:2009yd}
B. Katz, K. Blum and E. Waxman,
\newblock Mon. Not. Roy. Astron. Soc. 405 (2010), 0907.1686.

\bibitem{Blum:2013zsa}
K. Blum, B. Katz and E. Waxman,
\newblock Phys. Rev. Lett. 111 (2013), 1305.1324.

\bibitem{Lipari:2016vqk}
P. Lipari,
\newblock Phys. Rev. D95 (2017), 1608.02018.

\bibitem{Blum:2017iwq}
K. Blum, R. Sato and E. Waxman,
\newblock (2017), 1709.06507.

\bibitem{Cowsik:2013woa}
R. Cowsik, B. Burch and T. Madziwa-Nussinov,
\newblock Astrophys. J. 786 (2014), 1305.1242.

\bibitem{Kappl:2016qug}
R. Kappl and A. Reinert,
\newblock Phys. Dark Univ. 16 (2017), 1609.01300.

\bibitem{Blasi:2009hv}
P. Blasi,
\newblock Phys. Rev. Lett. 103 (2009), 0903.2794.

\bibitem{Mertsch:2009ph}
P. Mertsch and S. Sarkar,
\newblock Phys. Rev. Lett. 103 (2009), 0905.3152.

\bibitem{Mertsch:2014poa}
P. Mertsch and S. Sarkar,
\newblock Phys. Rev. D90 (2014), 1402.0855.

\bibitem{Aharonian:1995zz}
F.A. Aharonian, A.M. Atoyan and H.J. Volk,
\newblock Astron. Astrophys. 294 (1995).

\bibitem{Hooper:2008kg}
D. Hooper, P. Blasi and P.D. Serpico,
\newblock JCAP 0901 (2009), 0810.1527.

\bibitem{Abeysekara:2017old}
HAWC, A.U. Abeysekara et~al.,
\newblock Science 358 (2017), 1711.06223.

\bibitem{Lavalle:2014kca}
J. Lavalle, D. Maurin and A. Putze,
\newblock Phys. Rev. D90 (2014), 1407.2540.

\bibitem{Ackermann:2012pya}
Fermi-LAT, M. Ackermann et~al.,
\newblock Astrophys. J. 750 (2012), 1202.4039.

\bibitem{Bringmann:2011py}
T. Bringmann, F. Donato and R.A. Lineros,
\newblock JCAP 1201 (2012), 1106.4821.

\bibitem{DiBernardo:2012zu}
G. Di~Bernardo et~al.,
\newblock JCAP 1303 (2013), 1210.4546.

\bibitem{Orlando:2013ysa}
E. Orlando and A. Strong,
\newblock Mon. Not. Roy. Astron. Soc. 436 (2013), 1309.2947.

\bibitem{Davies:1987zz}
R.B. Davies,
\newblock Biometrika 74 (1987).

\bibitem{Gross:2010qma}
E. Gross and O. Vitells,
\newblock Eur. Phys. J. C70 (2010), 1005.1891.

\bibitem{Korsmeier:2017xzj}
M. Korsmeier, F. Donato and N. Fornengo,
\newblock (2017), 1711.08465.

\bibitem{Jones:2000qd}
F.C. Jones et~al.,
\newblock Astrophys. J. 547 (2001), astro-ph/0007293.

\bibitem{Ptuskin:2005ax}
V.S. Ptuskin et~al.,
\newblock Astrophys. J. 642 (2006), astro-ph/0510335.

\bibitem{Goodenough:2009gk}
L. Goodenough and D. Hooper,
\newblock (2009), 0910.2998.

\bibitem{Krauss:2016cdi}
M.E. Krauss et~al.,
\newblock Phys. Dark Univ. 14 (2016), 1605.05327.

\bibitem{TheFermi-LAT:2017vmf}
Fermi-LAT, M. Ackermann et~al.,
\newblock Astrophys. J. 840 (2017), 1704.03910.

\bibitem{Cuoco:2017rxb}
A. Cuoco et~al.,
\newblock JCAP 1710 (2017), 1704.08258.

\bibitem{Bartels:2015aea}
R. Bartels, S. Krishnamurthy and C. Weniger,
\newblock Phys. Rev. Lett. 116 (2016), 1506.05104.

\bibitem{Lee:2015fea}
S.K. Lee et~al.,
\newblock Phys. Rev. Lett. 116 (2016), 1506.05124.

\bibitem{Fermi-LAT:2017yoi}
Fermi-LAT, M. Ajello et~al.,
\newblock Submitted to: Astrophys. J.  (2017), 1705.00009.

\bibitem{Ackermann:2015zua}
Fermi-LAT, M. Ackermann et~al.,
\newblock Phys. Rev. Lett. 115 (2015), 1503.02641.

\bibitem{Fermi-LAT:2016uux}
DES, Fermi-LAT, A. Albert et~al.,
\newblock Astrophys. J. 834 (2017), 1611.03184.

\end{thebibliography}
\bibliographystyle{standard}

\end{document}